\documentclass[aps,asssymb,amsmath,notitlepage,superscriptaddress,pra,nofootinbib,twocolumn,balancelastpage,floatfix]{revtex4-1}
\usepackage{graphicx}
\usepackage{mathptmx}    %use Times fonts
\usepackage{braket}      %To use quantum computing macros
\usepackage{comment}     %To use comment block
\usepackage[colorlinks = true,
            linkcolor = cyan,
            urlcolor  = cyan,
            citecolor = red,
            anchorcolor = blue]{hyperref} %To link references inside text.
\usepackage{ragged2e}
\usepackage[format=hang,justification=RaggedRight]{caption}  %For captioning.
\usepackage{subcaption} %To use multiple figures in main figure, table, etc
\usepackage[T1]{fontenc}
\usepackage{amsmath,amsfonts,mathrsfs,amssymb} %Various font and typesetting packages.
\usepackage[usenames]{color}
\usepackage{morefloats}
\usepackage[pass]{geometry}
\makeatletter
\newcommand*{\balancecolsandclearpage}{%
  \close@column@grid
  \clearpage
  \twocolumngrid
}
\makeatother
\newcommand{\beq}{\begin{equation}}
\newcommand{\eeq}{\end{equation}}
\newcommand{\beqnn}{\begin{equation*}}
\newcommand{\eeqnn}{\end{equation*}}
\newcommand{\bea}{\begin{eqnarray}}
\newcommand{\eea}{\end{eqnarray}}
\newcommand{\beann}{\begin{eqnarray*}}
\newcommand{\eeann}{\end{eqnarray*}}
\newcommand{\bes} {\begin{subequations}}
\newcommand{\ees} {\end{subequations}}
\date{\today}
\begin{document}
\title{Performance of two different quantum annealing correction codes}

\author{Anurag Mishra}
\affiliation{Department of Physics and Astronomy, University of Southern California, Los Angeles, California 90089, USA}
\affiliation{Center for Quantum Information Science \& Technology, University of Southern California, Los Angeles, California 90089, USA}

\author{Tameem Albash}
\affiliation{Department of Physics and Astronomy, University of Southern California, Los Angeles, California 90089, USA}
\affiliation{Center for Quantum Information Science \& Technology, University of Southern California, Los Angeles, California 90089, USA}
\affiliation{Information Sciences Institute, University of Southern California, Marina del Rey, CA 90292}

\author{Daniel A. Lidar}
\affiliation{Department of Physics and Astronomy, University of Southern California, Los Angeles, California 90089, USA}
\affiliation{Center for Quantum Information Science \& Technology, University of Southern California, Los Angeles, California 90089, USA}
\affiliation{Department of Electrical Engineering, University of Southern California, Los Angeles, California 90089, USA}
\affiliation{Department of Chemistry, University of Southern California, Los Angeles, California 90089, USA}

\begin{abstract}
Quantum annealing is a promising approach for solving optimization problems, but like all other quantum information processing methods, it requires error correction to ensure scalability. In this work we experimentally compare two quantum annealing correction codes in the setting of antiferromagnetic chains, using two different quantum annealing processors. The lower temperature processor gives rise to higher success probabilities.
The two codes differ in a number of interesting and important ways, but both require four physical qubits per encoded qubit. We find significant performance differences, which we explain in terms of the effective energy boost provided by the respective redundantly encoded logical operators of the two codes. The code with the higher energy boost results in improved performance, at the expense of a lower degree encoded graph.
Therefore, we find that there exists an important tradeoff between encoded connectivity and performance for quantum annealing correction codes.
\end{abstract}
\maketitle
%
%%%%%%%%%%%%%%%%%%%%%%%%%%%%%%%
\section{Introduction}
\label{sec:Introduction}
%%%%%%%%%%%%%%%%%%%%%%%%%%%%%%%
%
Steady progress is being made towards the realization of a universal fault-tolerant quantum computer (e.g., \cite{Kelly:2015fv,Corcoles:2015zl}), yet the development of a scalable architecture remains a tenacious obstacle.  This has spurred the arrival of alternative quantum computing devices that sacrifice universality in order to allow for more rapid progress and thus hopefully usher in the era of quantum computing, albeit for a limited set of computational tasks, such as quantum simulation \cite{Lloyd:96,Buluta:2009fk,barreiro2011open}. Another example is quantum annealing \cite{finnila_quantum_1994,kadowaki_quantum_1998,Brooke1999,brooke_tunable_2001,2002quant.ph.11152K,Dwave}, an analog quantum approach for solving optimization problems \cite{farhi_quantum_2000,farhi_quantum_2001}, that may offer a quicker route than the standard circuit model towards the demonstration of the utility of quantum computation. The task addressed by quantum annealing is the well-known NP-hard problem of solving for the ground state of a classical Ising Hamiltonian \cite{Barahona1982},
\begin{equation}
  \label{eq:isingHam}
  H_{I}=\sum_{i \in V(G)} h_{i}Z_{i}+\sum_{ \left\{ (i, j) \right\} \in E(G) }J_{ij}Z_{i}Z_{j} \ ,
\end{equation}
where $\{h_{i}\}$ are the local fields on the vertices $V$ of the connectivity graph $G$, $\{J_{ij} \}$ are the couplings along the edges $E$ of $G$, and $Z_i$ is the Pauli-$Z$ operator on the $i$th qubit. This is done, ideally, by evolving the system for a total annealing time $t_f$ according to the time-dependent Hamiltonian
\beq
H(t) = -A(t) \sum_{i} X_i + B(t) H_{I} \ ,
\eeq
where $X_i$ is the Pauli-$X$ operator on the $i{\textrm{th}}$ qubit, and where $A(t)$ and $B(t)$ are the annealing schedules, satisfying $A(0) \gg B(0)$ and $A(t_f) \ll B(t_f)$.  In the closed system setting, starting in the ground state of $H(0)$ and evolving adiabatically, the system is guaranteed to reach the ground state of $H_I$ with high probability \cite{Kato:50,Jansen:07,lidar:102106}.

Although adiabatic dynamics is robust against certain forms of decoherence appearing in the more realistic open system setting \cite{2002quant.ph.11152K,childs_robustness_2001,PhysRevLett.95.250503,Aberg:2005rt,PhysRevA.71.032330,amin_decoherence_2009,Albash:2015nx}, it remains susceptible to thermal noise and specification errors \cite{Young:2013fk}, which can jeopardize the efficiency of the quantum computation.  Therefore, any scalable quantum annealing architecture will require quantum error correction \cite{Lidar-Brun:book}.  Unfortunately, theoretical progress in quantum error correction for adiabatic quantum computing and quantum annealing has not enjoyed the same success as that of other quantum computing paradigms, in spite of recent advances \cite{jordan2006error,PhysRevLett.100.160506,PhysRevA.86.042333,Ganti:13,Bookatz:2014uq}.  Physical constraints, such as locality of the interaction terms in the Hamiltonian \cite{Young:13,Sarovar:2013kx}, and a no-go theorem constraining what can be achieved with commuting two-local interactions \cite{Marvian:2014nr}, remain stubborn hurdles. An accuracy-threshold theorem rivaling that of the circuit model (e.g., \cite{Aliferis:05,Knill:05}) remains elusive despite recent attempts \cite{Mizel:2014sp}.

Still, the absence of such a theorem has not stopped the development of analog quantum computing devices, and quantum annealers are now commercially available \cite{Johnson:2010ys,Berkley:2010zr,Harris:2010kx}. Can such devices benefit from some form of error correction? Here we address this question by comparing two codes and two quantum annealing devices in the setting of the toy optimization problem of finding the ground states of antiferromagnetic chains.

Work on the first generation D-Wave 1 (DW1) ``Rainier'' and second generation D-Wave 2 (DW2) ``Vesuvius'' processors has already demonstrated that error correction can substantially benefit quantum annealing \cite{PAL:13,PAL:14,Vinci:2015jt}. Namely, it was shown that even a relatively simple quantum repetition code incorporating energy penalty terms and a decoding procedure, can significantly improve the success probability of finding ground states, as well as overcome precision issues in the specification of the Ising Hamiltonian. This technique is known as quantum annealing correction (QAC).  QAC is designed to work within the present technological restrictions of available quantum annealers whereby only the problem Hamiltonian ($H_I$) and not the transverse field Hamiltonian ($\sum_{i} X_i$) can be encoded.  This differs from the standard scheme of error-suppression via energy penalties for AQC wherein the entire Hamiltonian is encoded \cite{jordan2006error,Bookatz:2014uq}. This type of encoding in principle allows for the suppression of arbitrary errors, but requires $k$-local interactions with $k\geq 3$ \cite{Young:13}. QAC provides a pragmatic compromise, in that it only suppresses errors that do not commute with the repetition code stabilizers (i.e., it does not suppress pure dephasing errors), but is directly implementable since it only requires $2$-local $Z_i Z_j$ interactions.

\begin{figure*}[t]
  \centering
  \begin{subfigure}[b]{0.45\textwidth}
    \includegraphics[scale=0.35]{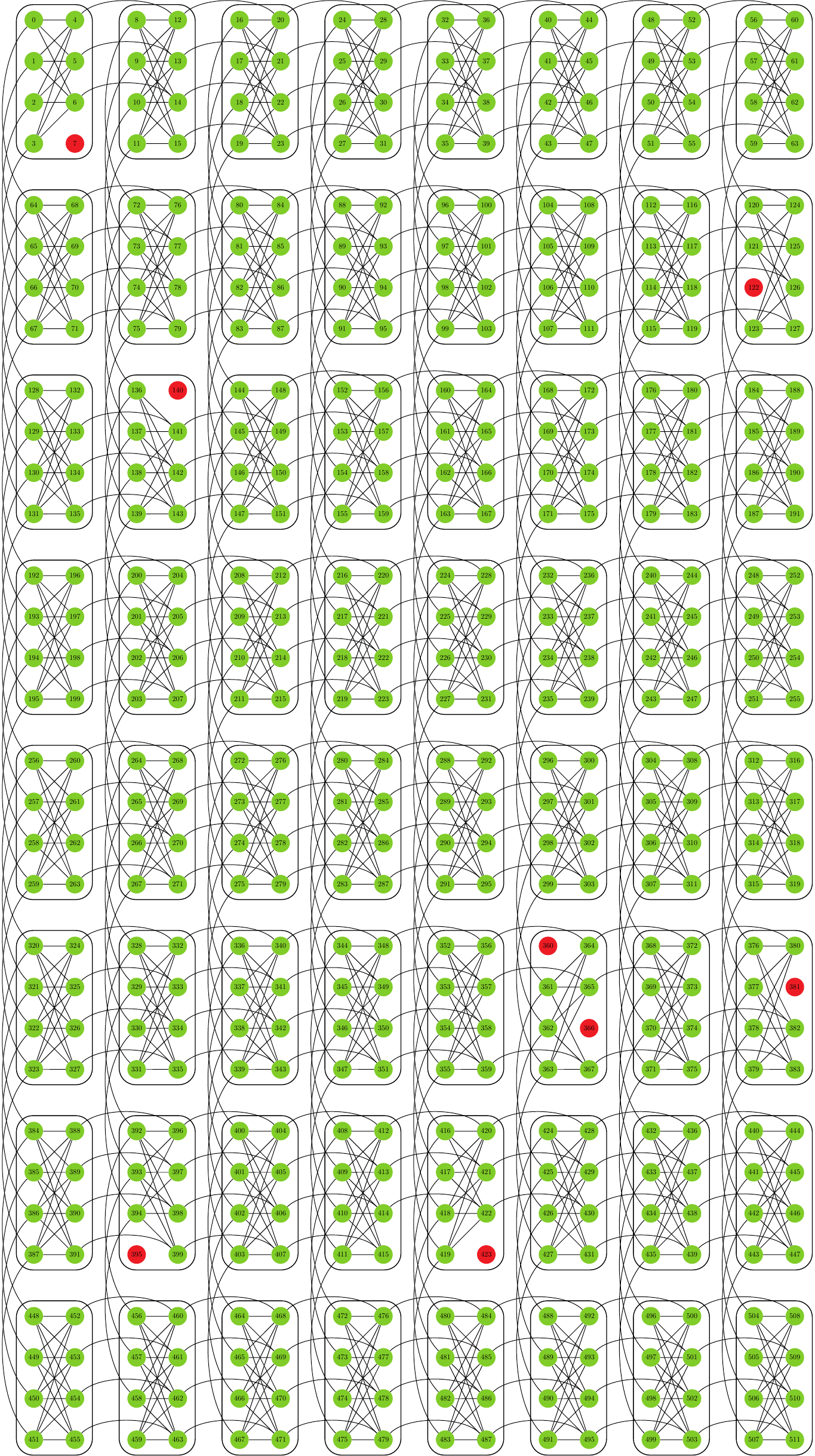}
    \caption{DW2-ISI hardware graph}
  \end{subfigure}
  \begin{subfigure}[b]{0.45\textwidth}
    \includegraphics[scale=0.35]{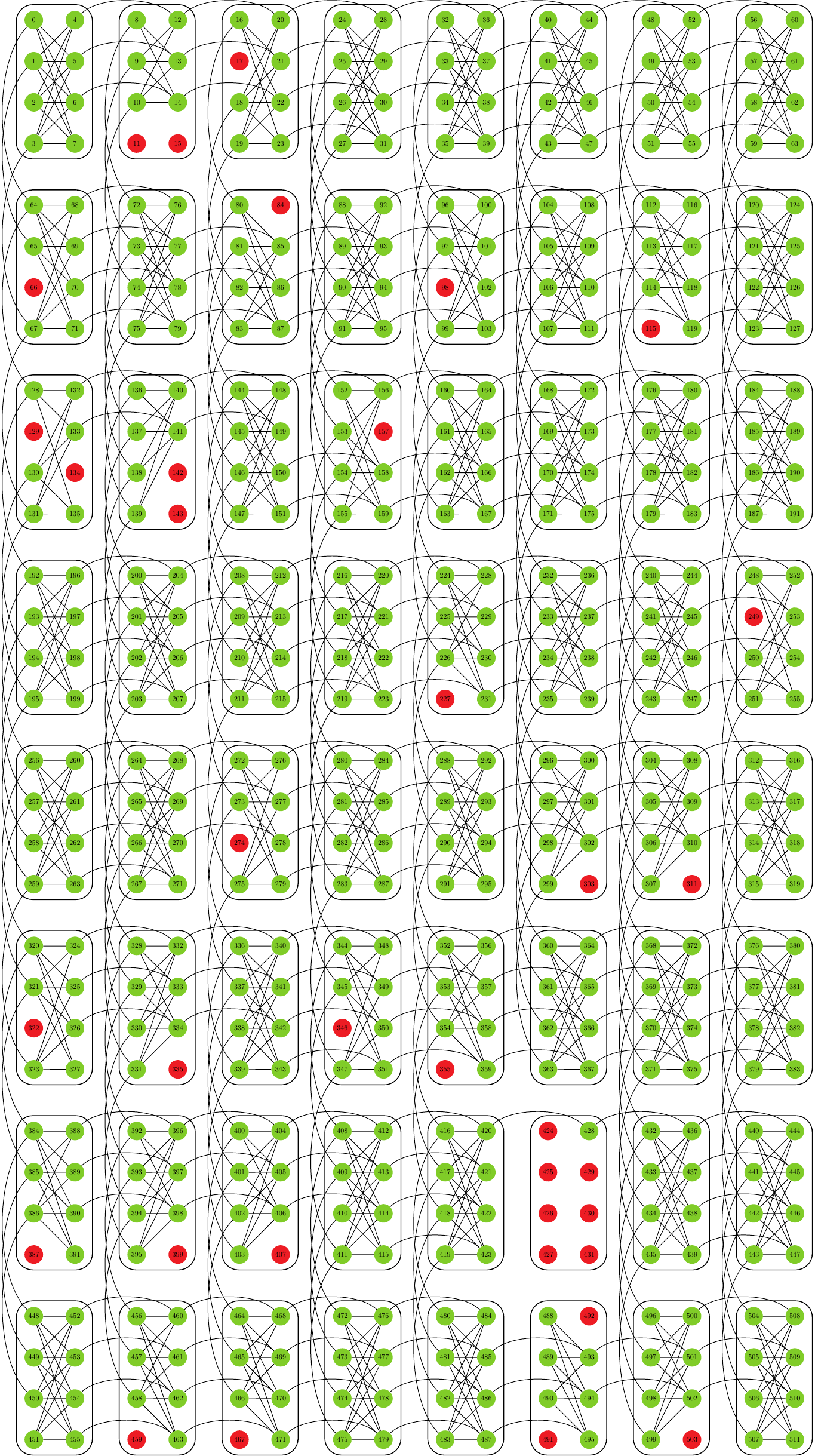}
    \caption{S6 hardware graph}
  \end{subfigure}
  \caption{The DW2 processors have unit cells arranged in an $8\times8$ grid, each containing $8$ qubits forming a $K_{4,4}$ bipartite graph. The active (inactive) qubits are shown in green (red), and active couplers are shown as solid black lines. Out of the $512$ qubits on the full DW2 Chimera graph, $504$ and $476$ were functional on the DW2-ISI and S6 devices, respectively. A comparison of the physical parameters of the two devices is given in Table~\ref{tab:annealerComp}.}
  \label{fig:dw2Vesuvius}
\end{figure*}

We use the notation ${[n,k,d]}_{n_P}$ to denote a distance $d$ code that uses $n$ physical data qubits and $n_{P}$ dedicated penalty qubits to encode $k$ qubits \cite{Vinci:2015jt}. The original QAC code introduced in Ref.~\cite{PAL:13} is a ${[3,1,3]}_{1}$ code, designed to be compatible with the D-Wave Chimera hardware graph (shown in Fig.~\ref{fig:dw2Vesuvius}).
Recently, Ref.~\cite{Vinci:2015jt} introduced a new ${[4,1,4]}_{0}$ code that uses the same physical resources as the ${[3,1,3]}_{1}$ code, without a dedicated penalty qubit. The two key benefits of the new code are that (i) the encoded hardware graph corresponding to the ${[4,1,4]}_{0}$ code has a higher degree (of connectivity), and (ii) it can be concatenated to give higher distance codes. Given that the ${[3,1,3]}_{1}$ and ${[4,1,4]}_{0}$ codes consume the same physical resources, it is natural to ask for a comparison between the two. Here we address this by testing the two codes on uniform antiferromagnetic chains, the same problem first studied in Ref.~\cite{PAL:13}. These problems are simple---their ground state can be trivially written down---but they are instructive since they are particularly error-prone in quantum annealers because of the existence of numerous low energy excitations (domain walls).  We demonstrate that for sufficiently long chains and sufficiently high noise rates, the ${[4,1,4]}_{0}$ code outperforms the classical strategy of running four chains in parallel and selecting the best, which also consumes the same physical resources. The ${[4,1,4]}_{0}$ code was shown in Ref.~\cite{Vinci:2015jt} to significantly improve the performance of quantum annealing in a minor embedding setting, a technique enabling the embedding of a given graph of interactions $G$ into one of a smaller degree, by using several physical qubits to represent a single logical qubit. This is crucial for applications, where one often starts from a logical problem defined on a high-degree (even complete) graph  \cite{2013arXiv1302.5843L}.  In particular, the $[4,1,4]_0$ code can be viewed as a minor embedding on the Chimera graph of two interconnected square graphs, as shown in Figure~\ref{fig:sqCode-latticeSection}. Here we find that the $[4,1,4]_0$ code is bested by the ${[3,1,3]}_{1}$ code in the setting of chains. We provide and verify an explanation for this performance difference in terms of the different effective energy scales generated by the two codes.

A novel aspect of this work is that we compare two different quantum annealing devices, namely two DW2 devices with somewhat different operating characteristics. This allows us to observe the role of temperature effects, among others.

This paper is organized as follows. In Sec.~\ref{sec:Methods}, we briefly review QAC in the context of the ${[4,1,4]}_{0}$ and ${[3,1,3]}_{1}$ codes, including decoding strategies. In Sec.~\ref{sec:benchmarking-AFM-chains} we describe our benchmarking procedure and define the strategies that QAC is compared against.  Section~\ref{sec:Results} presents our experimental results. Theoretical analysis is provided in Sec.~\ref{sec:theory}. We conclude in Sec.~\ref{sec:conclusions}. Additional details are provided in the Appendix.

%%%%%%%%%%%%%%%%%%%%%%%%%%%%%%%
\section{QAC using the ${[4,1,4]}_{0}$ and ${[3,1,3]}_{1}$ codes}
\label{sec:Methods}
%%%%%%%%%%%%%%%%%%%%%%%%%%%%%%%
%
\subsection{Layout}
We first briefly review the layout of the ${[4,1,4]}_{0}$ and ${[3,1,3]}_{1}$ codes, which are both quantum repetition codes against bit-flip errors.
The qubits on the Chimera graph of the D-Wave device (depicted in Fig.~\ref{fig:dw2Vesuvius}) are arranged in a square grid of unit cells, where each unit cell forms a complete $K_{4,4}$ bipartite graph.  This graph supports a number of QAC codes wherein each encoded qubit is represented by several physical qubits, which we call an ``encoded group.''  Figure~\ref{fig:squareCodeLattice} describes the ${[4,1,4]}_{0}$ code. We split the unit cell horizontally into two halves. The top and the bottom halves separately form two encoded qubits, where each of the four physical qubits are maximally connected via intra-cell ferromagnetic penalty couplings. The encoded qubit connects to the encoded qubits on each side via inter-cell problem couplings and also connects to the other encoded qubit in its unit cell via intra-cell problem couplings. On the Chimera graph, this generates an $8\times8\times2$ lattice, as shown in Fig.~\ref{fig:squareCodeGraph}. Incidentally, this is the two-level-grid (2LG) used in the original proof of the NP-hardness of the Ising model \cite{Barahona1982}.

The ${[3,1,3]}_{1}$ code splits a unit cell vertically into two halves. As shown in Fig.~\ref{fig:pudenzCodeLattice}, the three qubits on either half of the unit cell are combined with one qubit on the opposite half of the cell, which plays the role of a dedicated penalty qubit, to form an encoded qubit. Each unit cell thus contains two encoded qubits.  This construction gives rise to another encoded graph, which is shown in Fig.~\ref{fig:pudenzCodeGraph}. While both the ${[4,1,4]}_{0}$ code and the ${[3,1,3]}_{1}$ code use the same number of physical qubits and generate a non-planar encoded graph, their encoded graphs differ in connectivity, with the ${[4,1,4]}_{0}$ code having the advantage of degree $5$ over the degree $3$ of the ${[3,1,3]}_{1}$ code.
%
%%%%%%%%%%%%%%%%%%%%%%%%%%%%%
\subsection{Encoding}
\label{sec:quant-anneal-corr}
%%%%%%%%%%%%%%%%%%%%%%%%%%%%%
%
Encoding is achieved by replacing the Pauli-$Z$ operators in the Ising Hamiltonian in Eq.~\eqref{eq:isingHam} by their encoded counterparts.  Thus, the encoded Ising Hamiltonian can be written as
\begin{equation}
  \label{eq:encodedIsingHam}
  \overline{H_{I}}=\sum_{i \in V(\overline{G})}h_{i}\overline{Z_{i}}+\sum_{ \left\{ (i, j) \right\} \in E(\overline{G}) }J_{ij}\overline{{Z}_{i} Z_{j}} \ ,
\end{equation}
where the $h_i$ and $J_{ij}$ values are inherited from the original problem Hamiltonian, Eq.~\eqref{eq:isingHam}, and where $\overline{G}\subset G$ is the encoded graph. $\overline{G}$ is also a minor of $G$, i.e., it is formed by collapsing vertices and removing certain edges. To tie the encoded group together, we introduce an energy penalty Hamiltonian $H_P$, which is the sum of the stabilizer generators of the code. The energy penalty Hamiltonian serves to energetically penalize differences among the physical qubits in the encoded group, which helps to suppress bit flip errors. With the energy penalty term included, the overall time-dependent Hamiltonian during the evolution is
\begin{equation}
  \label{eq:fullEvolution}
  H(t) = A(t)H_{X} + B(t) ( \alpha \overline{H_{I}}+\gamma H_{P} ) \ ,
\end{equation}
where $H_{X}=-\sum_{i}X_{i}$ is the original (unencoded) transverse field, and $(\alpha,\gamma)$ are two controllable experimental parameters that can be varied in the range $[0,1]$ to control the relative strength of the problem and the penalty Hamiltonians. Because $H_{P}$ is a part of physical problem Hamiltonian $H_I$ it inherits the latter's time-dependence, i.e., is turned on via the annealing schedule $B(t)$. This aspect of QAC differs from standard error suppression~\cite{jordan2006error}.

Note that $H_{X}$ is itself a sum of bit-flip operators, so it plays a dual role: it is used to prepare the initial superposition state (its ground state), and is an ``error'' from the perspective of the penalty Hamiltonian. This is unavoidable in the setting of the D-Wave device, which (also unlike Ref.~\cite{jordan2006error}) prevents $H_{X}$ from being encoded, as this would require many-body $X^{\otimes n}$ terms, which are experimentally unavailable. Because of this tension there is an optimal penalty value $\gamma$ that depends on $\alpha$, the problem instance, and other variables. In particular, the optimal $\gamma$ keeps decodable states lower in the energy spectrum. We shall return to this point later.

\subsubsection{${[4,1,4]}_{0}$ code}
\begin{figure*}
  \centering
  \begin{subfigure}[b]{0.45\textwidth}
    \includegraphics[scale=0.35]{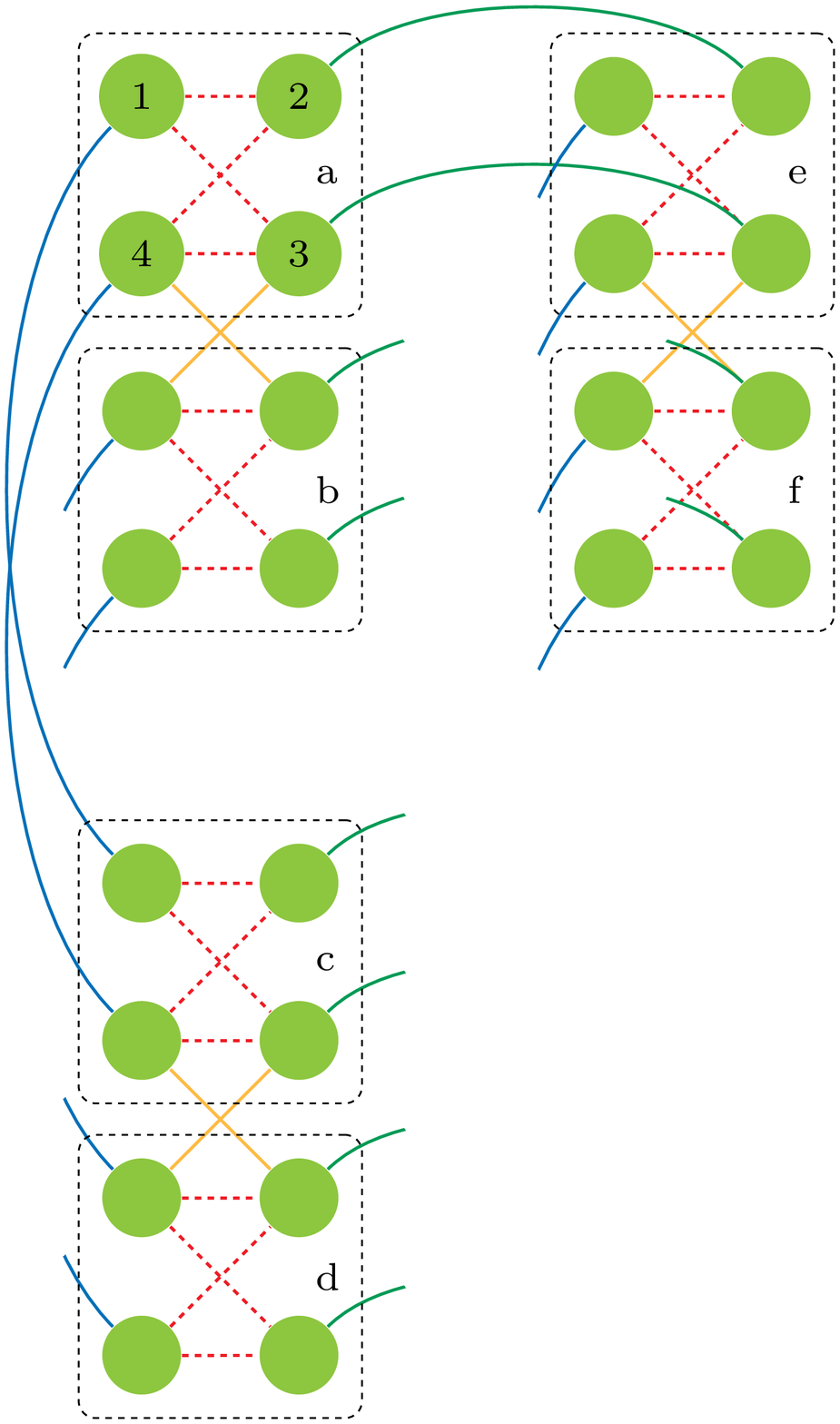}
    \caption{Arrangement of physical qubits}
    \label{fig:sqCode-phyQubits}
  \end{subfigure}
  \qquad
  \begin{subfigure}[b]{0.45\textwidth}
    \raisebox{3cm}{\includegraphics[scale=0.35]{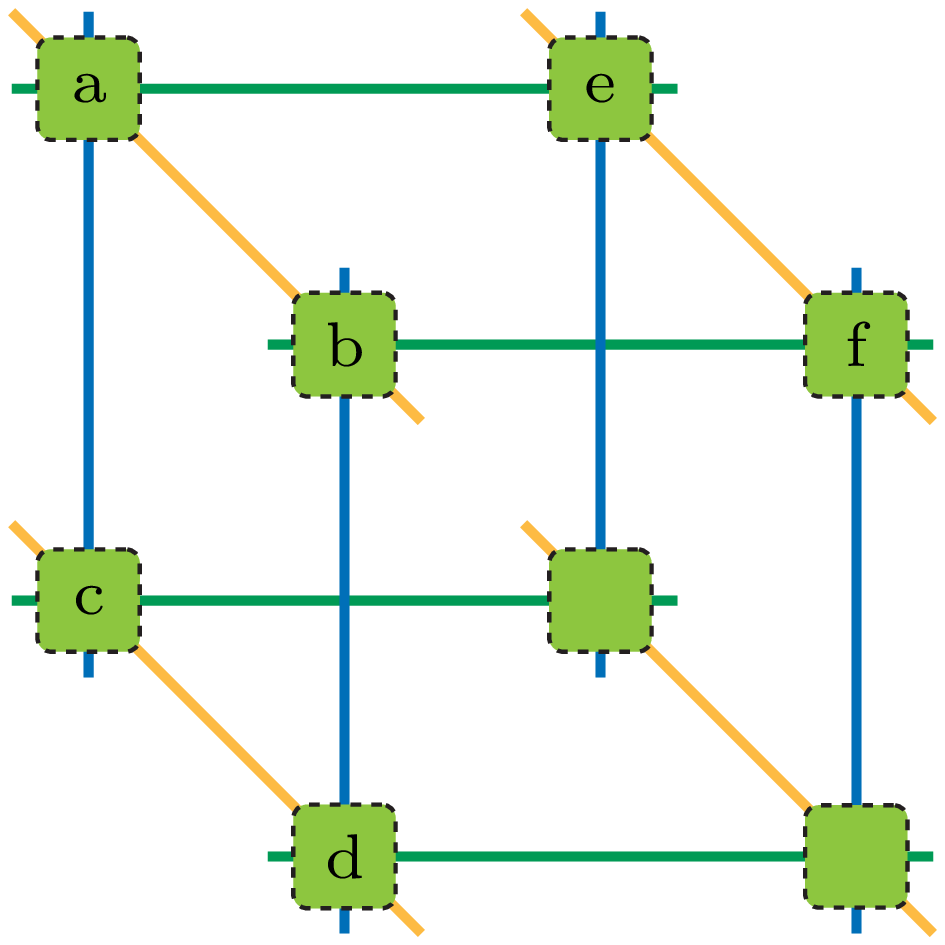}}
    \caption{Logical qubit lattice}
    \label{fig:sqCode-latticeSection}
  \end{subfigure}
  \caption{\textbf{Construction of the ${[4,1,4]}_{0}$ code.} Two encoded qubits are constructed using the four physical qubits in the upper and the lower halves of the unit cell. In \subref{fig:sqCode-phyQubits}, the dotted lines represent the  penalty terms. The solid lines form the encoded Hamiltonian couplings. Since each encoded coupling is formed by $2$ physical couplings, the energy scale of the Ising problem is boosted by factor of $2$.  In \subref{fig:sqCode-latticeSection} we show the section of the encoded graph formed by \subref{fig:sqCode-phyQubits}, with the same color scheme for the couplings. Roman letters labels the same encoded qubits in \subref{fig:sqCode-phyQubits} and \subref{fig:sqCode-latticeSection}.}
  \label{fig:squareCodeLattice}
\end{figure*}

\begin{figure*}
 \centering
 \begin{subfigure}[b]{0.45\textwidth}
   \includegraphics[scale=0.4]{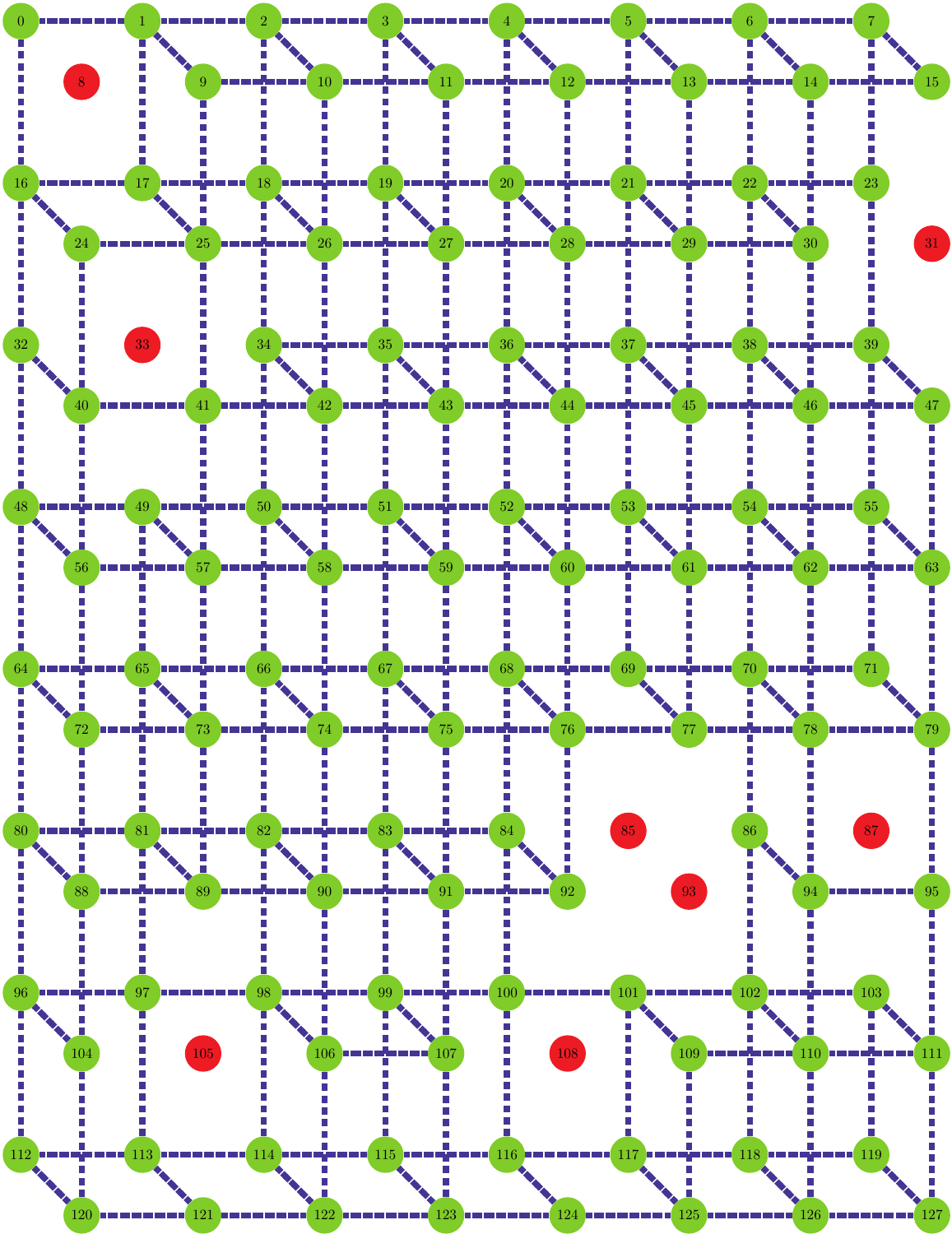}
   \caption{DW2-ISI device}
   \label{fig:squareCodeGraph-a}
 \end{subfigure}
 \begin{subfigure}[b]{0.45\textwidth}
   \includegraphics[scale=0.4]{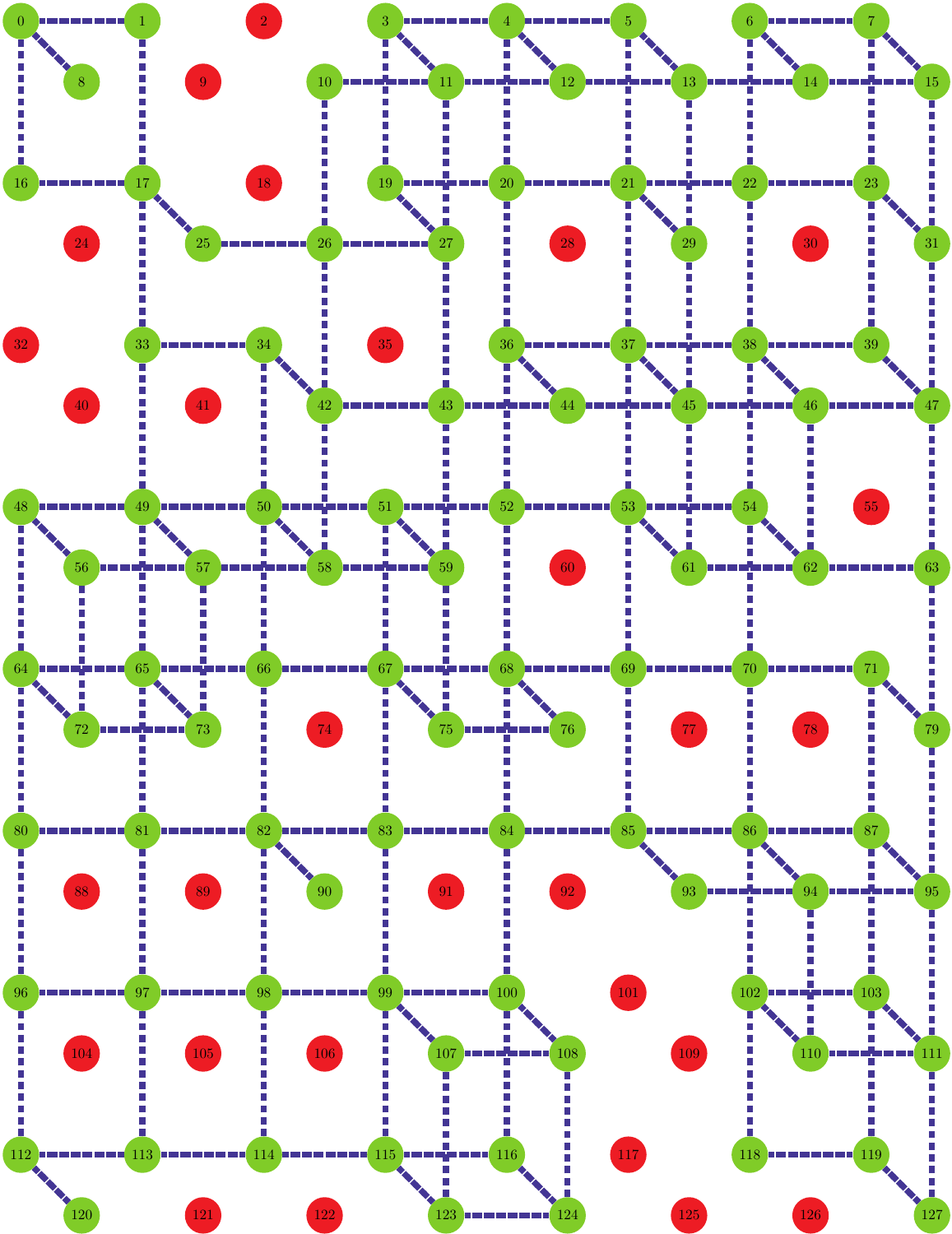}
   \caption{S6 device}
   \label{fig:squareCodeGraph-b}
 \end{subfigure}
 \caption{\textbf{The ${[4,1,4]}_{0}$ encoded graph.} Each encoded qubit is composed of four physical qubits, and the encoded couplings are formed from two physical couplers.  The green (red) circles denote functional (inactive) qubits. Out of 128 possible encoded qubits on the complete graph, $120$ were functional on the DW2-ISI device \subref{fig:squareCodeGraph-a} and $99$ on the S6 device \subref{fig:squareCodeGraph-b}. The encoded graph is a $2$-level grid \cite{Barahona1982} and has degree $5$.}
  \label{fig:squareCodeGraph}
\end{figure*}

\begin{figure*}
  \centering
  \begin{subfigure}[b]{0.45\textwidth}
    \includegraphics[scale=0.35]{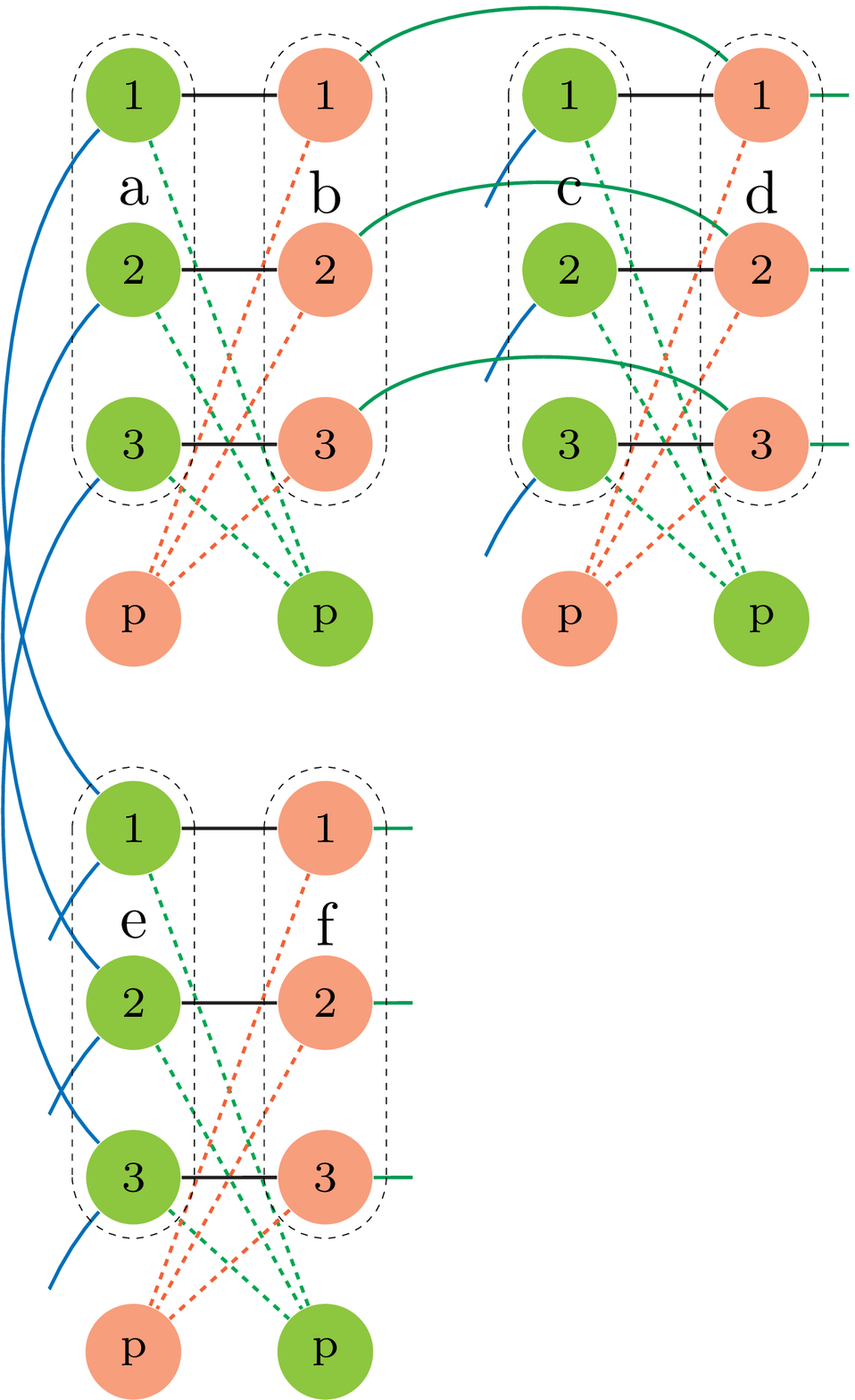}
    \caption{Arrangement of physical qubits}
    \label{fig:PudCode-phyQubits}
  \end{subfigure}
  \qquad
  \begin{subfigure}[b]{0.45\textwidth}
     \raisebox{3cm}{\includegraphics[scale=0.35]{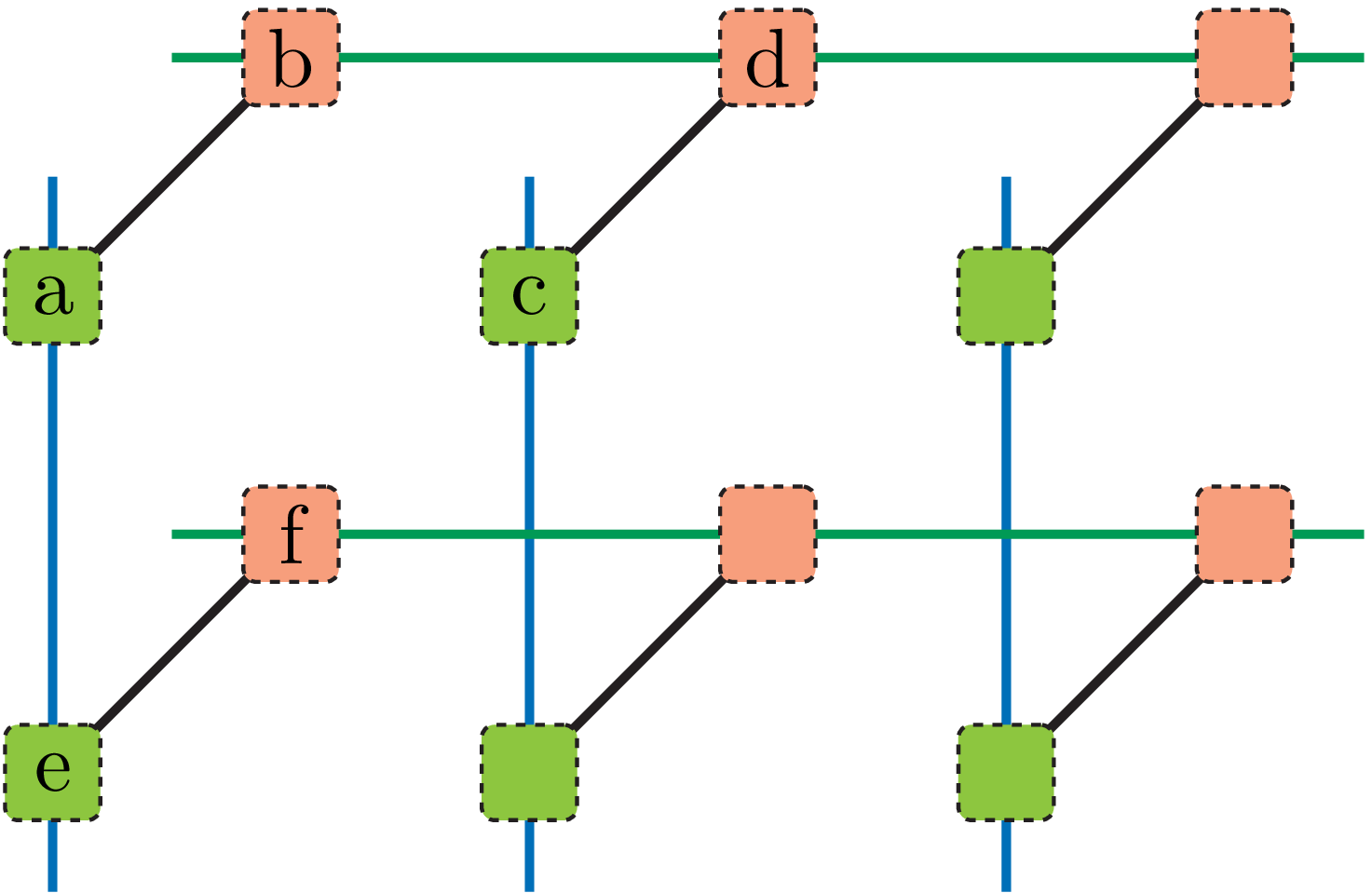}}
    \caption{Logical qubit lattice}
    \label{fig:PudCode-latticeSection}
  \end{subfigure}
  \caption{\textbf{Construction of the ${[3,1,3]}_{1}$ code.} An encoded qubit is constructed using three data qubits from each vertical half of the unit cell and a penalty qubit from the opposite half. In \subref{fig:PudCode-phyQubits}, the four physical qubits forming the encoded group are shown in the same color, and the dashed lines represent the stabilizer couplings. The solid lines form the encoded Hamiltonian coupling. Since each encoded coupling comprises $3$ physical couplings, the energy scale of the encoded problem is boosted by a factor of $3$. In \subref{fig:PudCode-latticeSection} we show the section of the encoded graph formed by \subref{fig:PudCode-phyQubits}, with the same color scheme for the couplings.  In both \subref{fig:PudCode-phyQubits} and \subref{fig:PudCode-latticeSection}, the Roman alphabet labels the respective encoded qubits.}
  \label{fig:pudenzCodeLattice}
\end{figure*}

\begin{figure*}
 \centering
  \begin{subfigure}[b]{0.45\textwidth}
    \centering
    \includegraphics[scale=0.3]{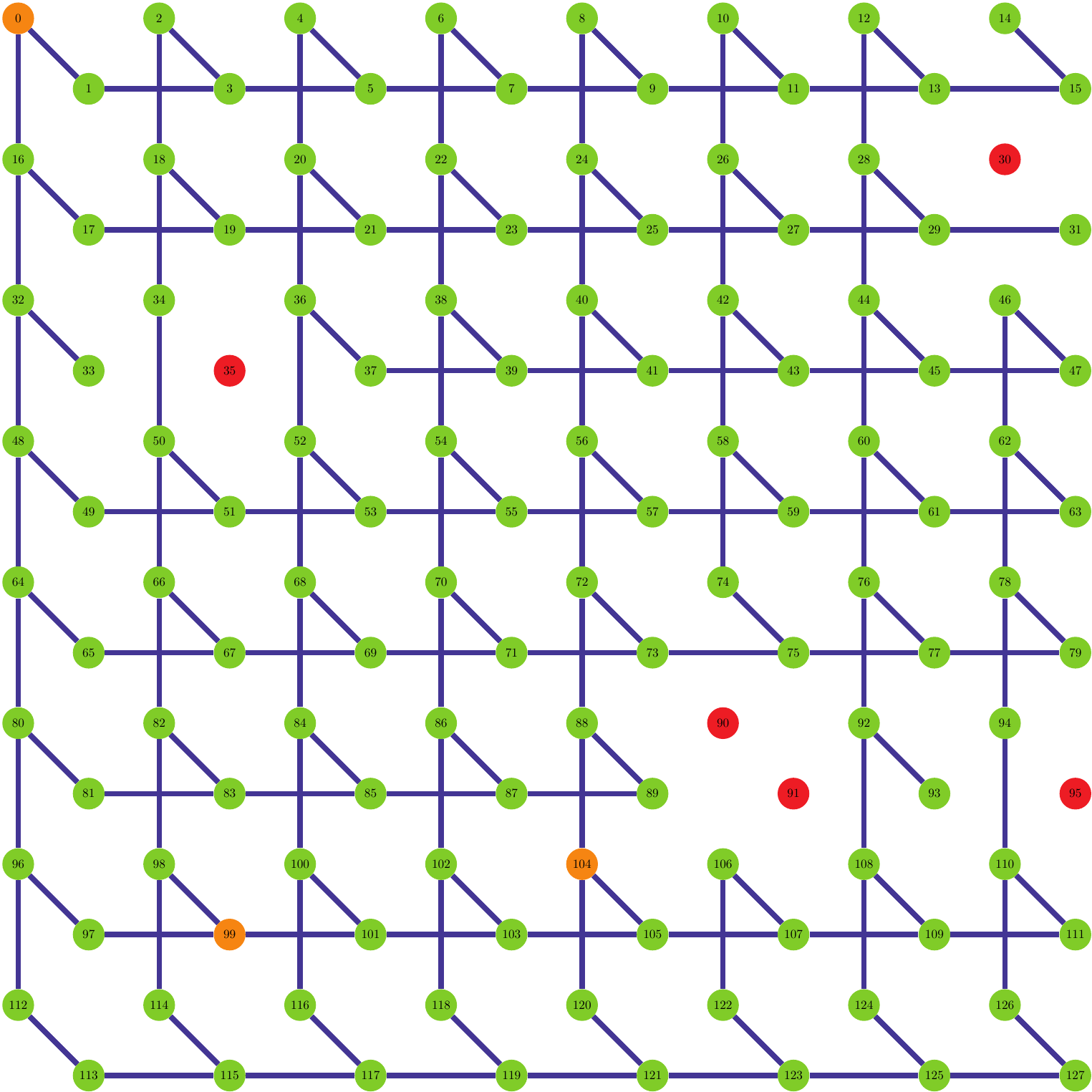}
    \caption{DW2-ISI device}
    \label{fig:pudenzCodeGraph-a}
  \end{subfigure}
  \begin{subfigure}[b]{0.45\textwidth}
    \centering
    \includegraphics[scale=0.3]{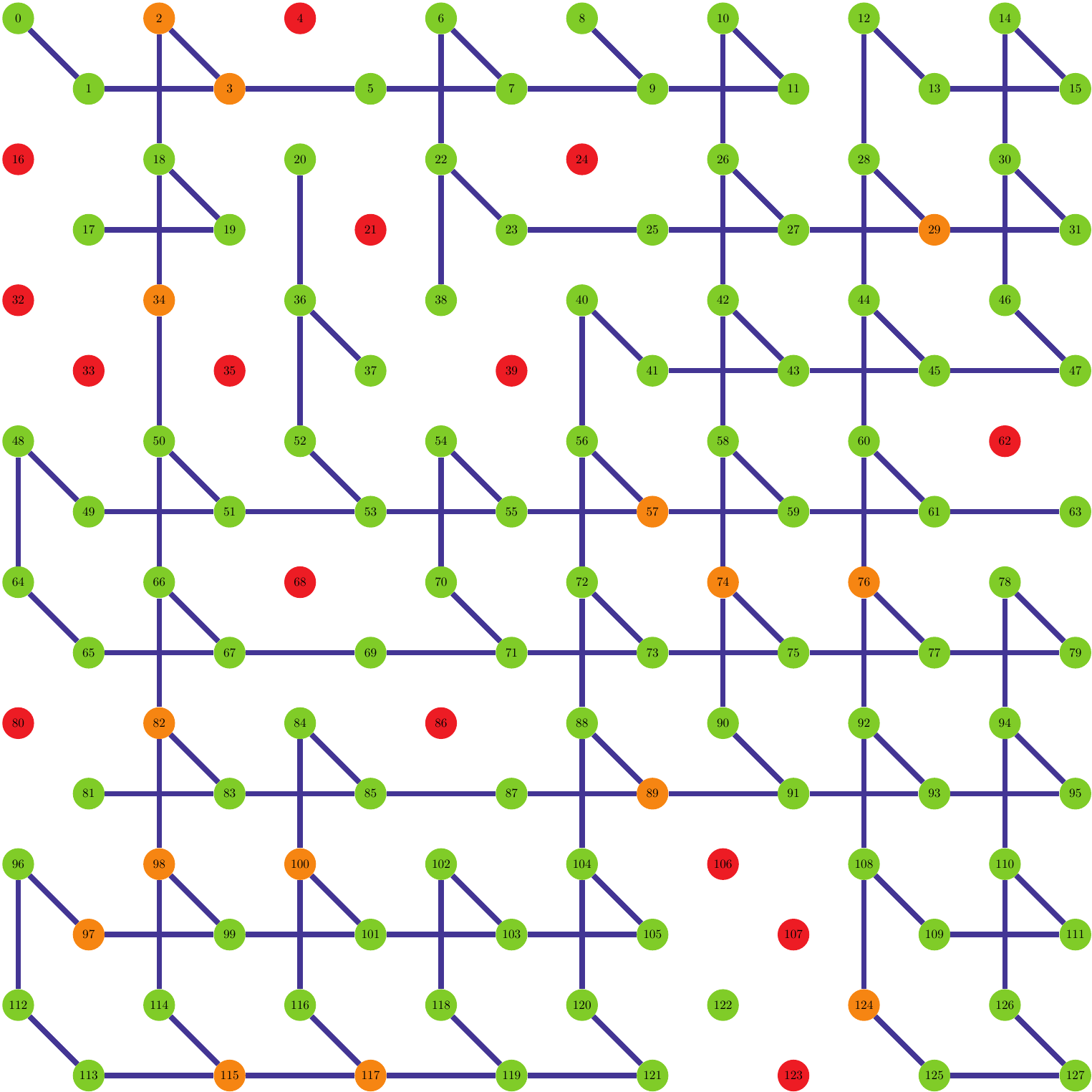}
    \caption{S6 device}
    \label{fig:pudenzCodeGraph-b}
  \end{subfigure}
  \caption{\textbf{The ${[3,1,3]}_{1}$ code encoded graph.} Each encoded qubit is composed of four physical qubits, and the encoded couplings are formed from three physical couplers. The green (red) circles denote functional (inactive) qubits. Orange circles indicate encoded qubits that have all three data qubits but are missing their penalty qubits. Out of $128$ possible encoded qubits on the complete graph, $120$ were fully functional while $3$ were missing penalty qubits on the DW2-ISI device \subref{fig:pudenzCodeGraph-a}; $95$ were fully functional and $18$ were missing penalty qubit on the S6 device \subref{fig:pudenzCodeGraph-b}. We only used fully functional encoded qubits in our experiments. The encoded graph has degree $3$.}
  \label{fig:pudenzCodeGraph}
\end{figure*}

For the ${[4,1,4]}_{0}$ code, the encoded Pauli-$Z$ operators can be constructed from physical operators as follows:
\bes
\begin{align}
\label{eq:5a}
Z_i &\mapsto \overline{Z_{i}} = \frac{1}{2} (Z^{1}_{i}+Z^{2}_{i}+Z^{3}_{i}+Z^{4}_{i}) \ , \\
Z_{i} Z_{j} &\mapsto \overline{Z_{i} Z_{j}} = \sum_{k}Z_{i}^{k}Z_{j}^{k}\ ,
\end{align}
\ees
where $k$ runs over two of the four physical qubits depicted in Fig.~\ref{fig:sqCode-phyQubits}, i.e., solid lines of the same color in that figure. By encoding in this fashion, we boost the Ising problem energy scale uniformly by a factor of two. The penalty Hamiltonian is chosen as indicated by the dotted couplings in Fig.~\ref{fig:sqCode-phyQubits}, i.e.:
\begin{equation}
  \label{eq:penaltyHamSquareCode}
  H_{P} = -\sum_{i=1}^{\overline{N}} (Z_{i}^{1}Z_{i}^{2} + Z_{i}^{1}Z_{i}^{3} + Z_{i}^{2}Z_{i}^{4} + Z_{i}^{3}Z_{i}^{4}) \ ,
\end{equation}
where henceforth $\overline{N} = |V(\overline{G})|$ denotes the number of encoded qubits.

\subsubsection{${[3,1,3]}_{1}$ code}
The ${[3,1,3]}_{1}$ code uses a similar construction, except a distinction is made between the four physical qubits in the encoded group.  They are categorized into a single ``penalty qubit'' and three ``data qubits,'' depicted in Fig~\ref{fig:PudCode-phyQubits}. Now the encoded Pauli-$Z$ operators are constructed from physical operators as follows:
\bes
\begin{align}
\label{eq:7a}
{Z_{i}} &\mapsto \overline{Z_{i}} = Z^{1}_{i}+Z^{2}_{i}+Z^{3}_{i} \\
Z_{i} {Z_{j}} &\mapsto \overline{Z_{i} {Z_{j}}} = \sum_{k}Z_{i}^{k}Z_{j}^{k}
\end{align}
\ees
where $k$ runs over the three data qubits. This encoding boosts the energy scale by a factor of three, which is more than the boost provided by the ${[4,1,4]}_{0}$ code. The importance of this difference is discussed in detail below. The penalty Hamiltonian, which is again the sum of the stabilizer generators of the code, is formed by coupling the data qubits to the penalty qubit, i.e.:
\begin{equation}
 \label{eq:penaltyHamPudenzCode}
 H_{P} = - \sum_{i=1}^{\overline{N}} (Z_{i}^{1}Z_{i}^{P} + Z_{i}^{2}Z_{i}^{P} + Z_{i}^{3}Z_{i}^{P}) \ .
\end{equation} \\
%
%%%%%%%%%%%%%%%%%%%%%%%%%%%%%
\subsection{Decoding strategies}
\label{sec:decodingStrategies}
%%%%%%%%%%%%%%%%%%%%%%%%%%%%%
%

The encoded state is decoded via a majority vote on the physical qubits in an encoded group.  Since the number of qubits in the ${[4,1,4]}_{0}$ code is even, a majority vote alone does not suffice since ties are possible. To decode in such cases we follow two different decoding schemes (see also Ref.~\cite{Vinci:2015jt}):
\begin{itemize}
\label{list:decoding-schemes}
\item \textit{Coin tossing} (CT).  We flip an unbiased coin to break each tie, i.e., we assign a random $\pm 1$ value to each encoded qubit. This random decoding strategy serves as a baseline against which we can compare other strategies.
\item  \textit{Energy Minimization} (EM).  The tied qubits can be treated as an Ising system with effective local fields (due to the now fixed decoded qubits) and couplings to other tied qubits. This (hopefully small) system is then solved exactly by explicitly checking the energy of all possible configurations. EM is guaranteed to give the lowest possible energy from decoding, and it remains a feasible decoding scheme as long as the size of the tied clusters does not scale with the size of the problem\footnote{It was shown in Ref.~\cite{Vinci:2015jt} that in general this is related to the per-site percolation threshold of the encoded graph, though this is not relevant in the case of chains.}.
\end{itemize}
We call an excited physical state ``decodable'' if, upon applying either of the decoding procedures described above, the decoded state is an encoded ground state. When this happens, we declare a success.
For a given decoding scheme at a given problem energy scale $\alpha$, we always locate the optimal energy penalty strength $\gamma_{\textrm{opt}}$ that maximizes the success probability.

%%%%%%%%%%%%%%%%%%%%%%%%%%%%%
\section{Benchmarking using antiferromagnetic chains}
\label{sec:benchmarking-AFM-chains}
%%%%%%%%%%%%%%%%%%%%%%%%%%%%%

We implement an (unencoded) $N$-qubit antiferromagnetic chain with the following Ising Hamiltonian:
\begin{equation}
  \label{eq:AFM-chain-hamiltonian}
  H_{I} = \sum_{i=1}^{N-1}Z_{i}Z_{i+1} \ .
\end{equation}
In order to quantify the performance of the QAC scheme, we also test two classical strategies and one additional quantum strategy \cite{PAL:13,PAL:14,Vinci:2015jt}.

\begin{itemize}
\item \textit{Unprotected} (U): In this case we directly embed the Hamiltonian in Eq.~\eqref{eq:AFM-chain-hamiltonian} on the device hardware graph. Each run in which either one of the two degenerate ground states of the chain is found is then declared a success.
\item \textit{Classical} (C): Here we use simple classical repetition, i.e., we run four copies of the chain in parallel. This uses an equal number of physical qubits as our QAC scheme. A run is then considered a success if at least one of the four copies is in one of the two ground states. If $p$ is the success probability for the U case, then, assuming the chains running in parallel are independent, the success probability for the C strategy is $1-{(1-p)}^{4}$ (i.e., at least one chain is correct).
\item \textit{Energy Penalty} (EP):  Here we encode into either the ${[4,1,4]}_{0}$ or the ${[3,1,3]}_{1}$ code, including the energy penalty, but we do not decode. That is, we declare a success only when the physical ground state of the physical graph obtained after encoding is observed (note that this graph is not a chain). We refer to this as the EP strategy since it relies only on the energy penalty, but not on decoding, to increase the success probability.
\end{itemize}

The comparison of the U, C, EP and QAC strategies allows us to isolate different aspects of our overall error correction strategy. The U case is the baseline against which all other strategies are measured. EP informs us about whether the energy penalty is helping. In order for QAC to be considered successful, it should exhibit better performance than the C strategy.

The experiments detailed next were performed on two different programmable quantum annealing devices.  The DW2 processor at the USC Information Sciences Institute (DW2-ISI) has $504$ functional qubits with an operating temperature of $16\pm1$ mK. Another DW2 processor, at D-Wave Inc. in Burnaby (S6) had $476$ operational qubits and operated at $11\pm 1$ mK\footnote{Both processors have meanwhile been dismantled.}. These devices and the underlying technology have been described before in detail in various publications (e.g., Refs.~\cite{Johnson:2010ys,Berkley:2010zr,Harris:2010kx,speedup}).

All experiments were performed with $30$ instances of randomly placed chains on the hardware graph. The error bars in all figures below are the standard error of the means calculated over these $30$ instances. Each chain instance was run $1000$ times (in a single programming cycle), and the fraction of successful runs was taken to be the success probability of an instance.
%\red{no gauges?} \blue{No gauges. The idea was to keep the methodology as close to Kristen's paper.}

\section{Experimental Results}
\label{sec:Results}
In this section we present our success probability results for the various strategies and the two devices tested, and we analyze these results from a number of different angles. All our results use optimized penalty values.
\subsection{Success Probability Comparison}
\label{sec:success-prob-comp}
\begin{figure*}
 \centering
 \null\hfill
 \begin{subfigure}[t]{0.45\textwidth}
  \includegraphics[width=\textwidth]{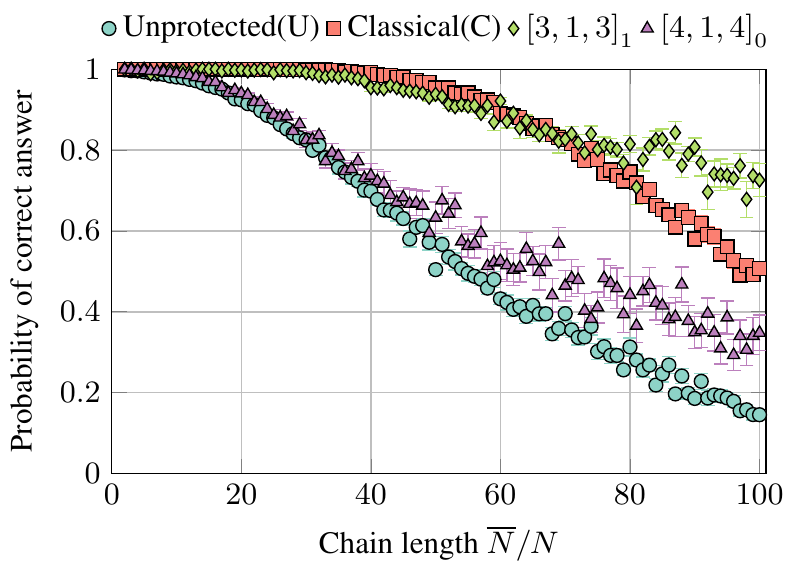}
  \caption{$\alpha=1$}
  \label{subfig:ISI-allPlots-alpha100}
 \end{subfigure}
 \hfill
 \begin{subfigure}[t]{0.45\textwidth}
  \includegraphics[width=\textwidth]{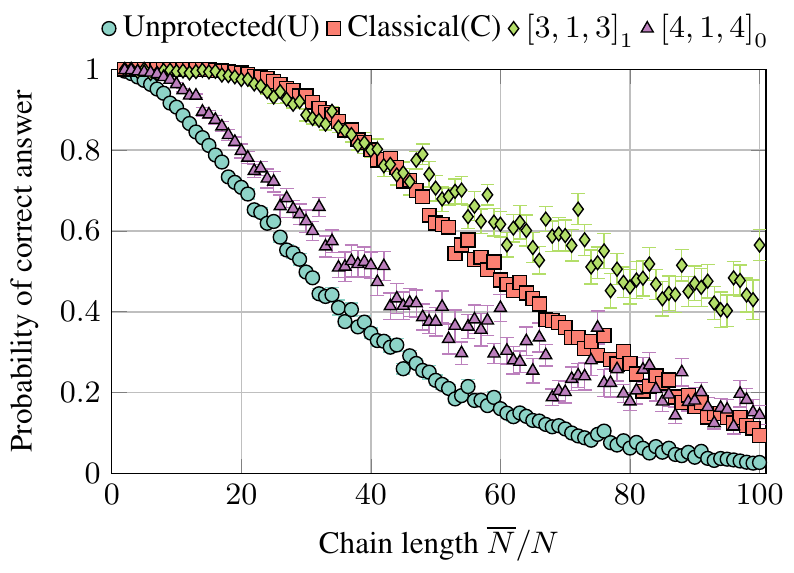}
  \caption{$\alpha=0.5$}
   \label{subfig:ISI-allPlots-alpha050}
 \end{subfigure}
 \hfill\null \\
 \null\hfill
\begin{subfigure}[t]{0.45\textwidth}
  \includegraphics[width=\textwidth]{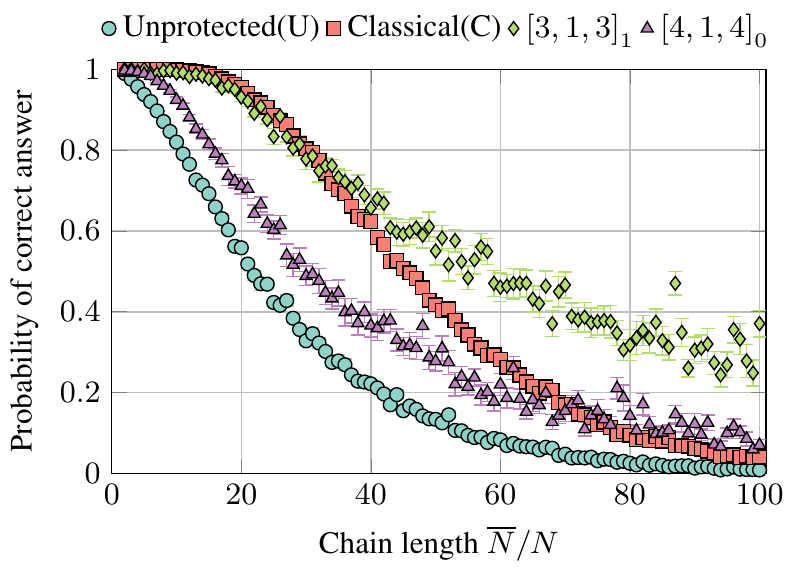}
  \caption{$\alpha=0.4$}
  \label{subfig:ISI-allPlots-alpha040}
 \hfill
 \end{subfigure}
\begin{subfigure}[t]{0.45\textwidth}
  \raisebox{1.3mm}{\includegraphics[width=\textwidth]{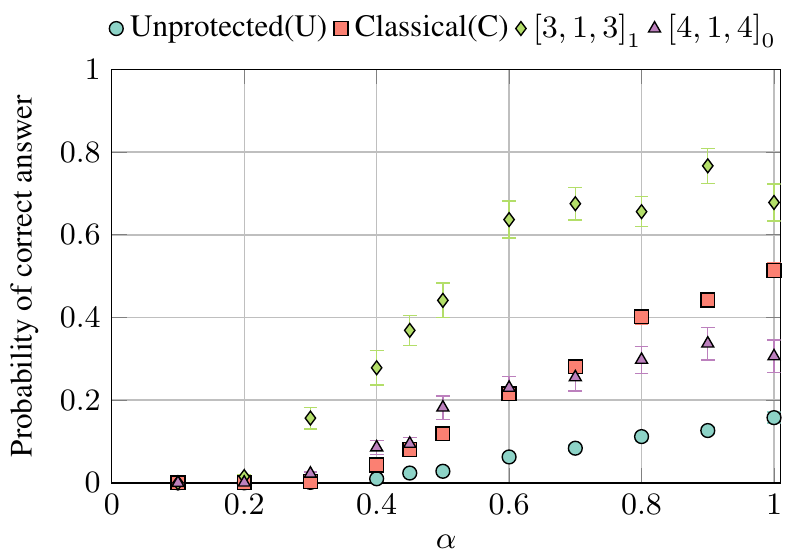}}
  \caption{$\overline{N}=98$}
  \label{subfig:ISI-alphaComparison}
 \end{subfigure}
 \hfill\null
 \caption{\textbf{Results for the DW2-ISI device.} Panels~\subref{subfig:ISI-allPlots-alpha100},~\subref{subfig:ISI-allPlots-alpha050} and~\subref{subfig:ISI-allPlots-alpha040} compare the results for chains using the U, C, ${[3,1,3]}_{1}$ code and $[4,1,4]$ code at scaling parameters $\alpha=1$, $0.5$ and $0.4$ respectively. Panel~\subref{subfig:ISI-alphaComparison} shows a comparison at a fixed chain length of $\overline{N}=98$ as $\alpha$ is varied. For $\alpha \lesssim 0.5$ and sufficiently long chain lengths, the ${[4,1,4]}_{0}$ code starts to outperform the C strategy. The ${[3,1,3]}_{1}$ code outperforms all other strategies at all values of $\alpha$, for sufficiently long chain lengths.}
 \label{fig:ISImainChainComparison}
\end{figure*}

Figure~\ref{fig:ISImainChainComparison} displays the DW2-ISI results. It shows that for the highest chain length studied ($\bar{N} = 98$) the ${[4,1,4]}_{0}$ code (with EM) is bested by the C strategy at higher $\alpha$, but outperforms the C strategy at lower $\alpha$, corresponding to a transition from a regimes of low to high error rates. The cross-over point occurs around $\alpha=0.6$. The ${[3,1,3]}_{1}$ code provides superior error correction at all scales for this largest size, confirming and extending the results of Ref.~\cite{PAL:13}.

\begin{figure*}
 \centering
 \null\hfill
 \begin{subfigure}[t]{0.45\textwidth}
  \includegraphics[width=\textwidth]{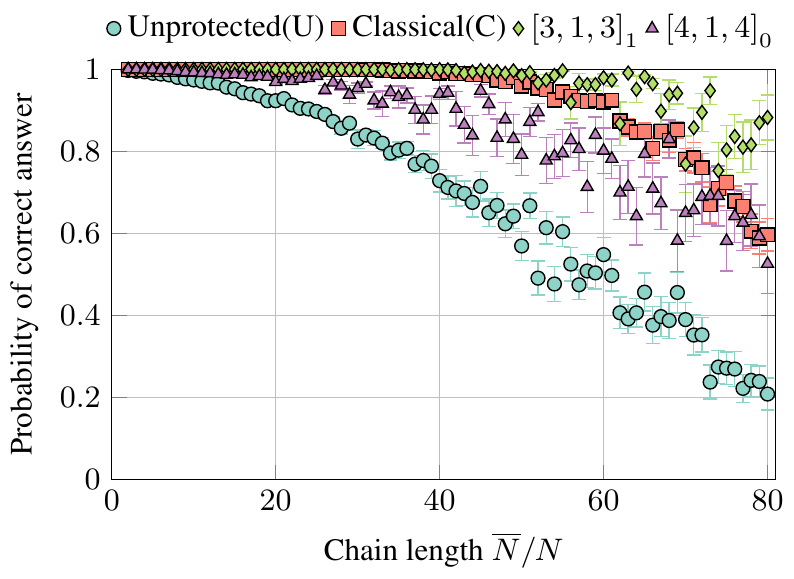}
  \caption{$\alpha = 1$}
  \label{subfig:S6-allPlots-alpha100}
 \end{subfigure}
 \hfill
 \begin{subfigure}[t]{0.45\textwidth}
  \includegraphics[width=\textwidth]{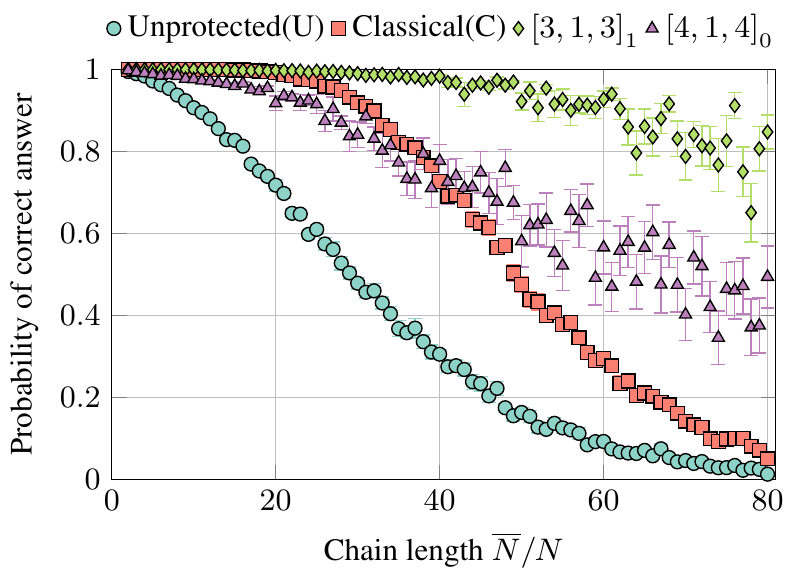}
  \caption{$\alpha = 0.5$}
   \label{subfig:S6-allPlots-alpha050}
 \end{subfigure}
 \hfill\null \\
 \null\hfill
\begin{subfigure}[t]{0.45\textwidth}
  \includegraphics[width=\textwidth]{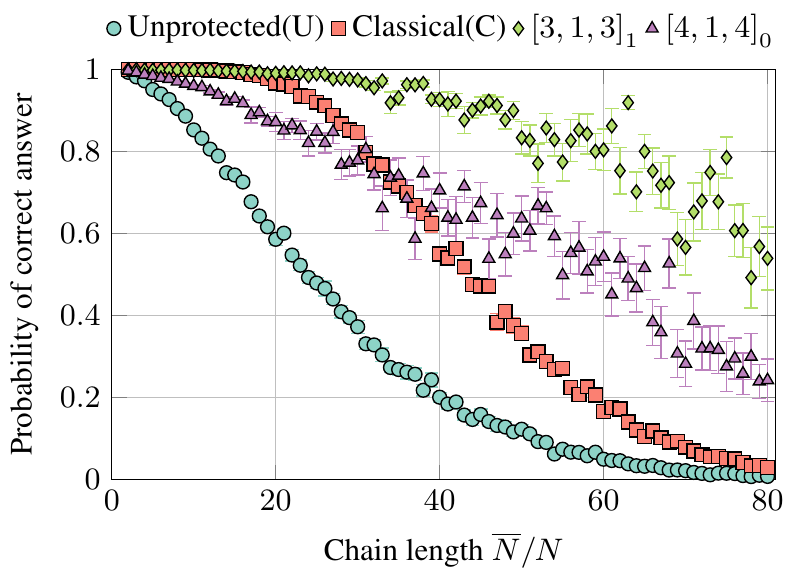}
  \caption{$\alpha = 0.4$}
  \label{subfig:S6-allPlots-alpha040}
 \hfill
 \end{subfigure}
 \begin{subfigure}[t]{0.45\textwidth}
  \raisebox{1.95mm}{\includegraphics[width=\textwidth]{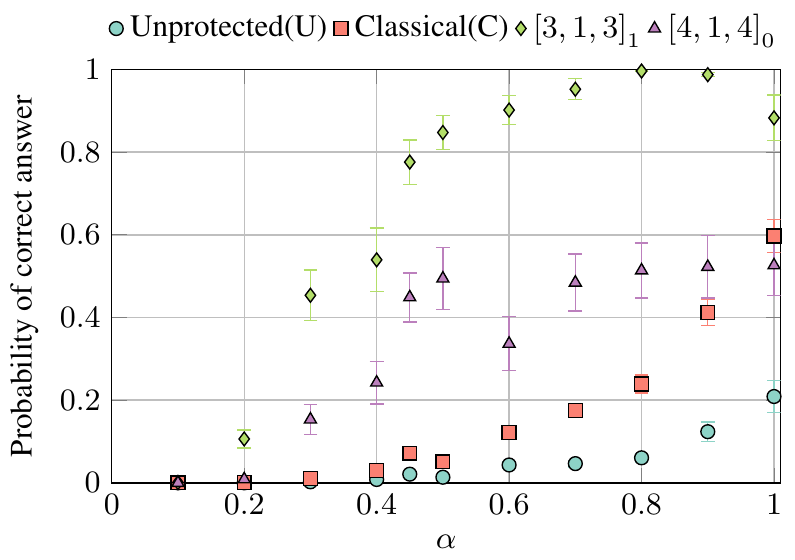}}
  \caption{$\overline{N}=80$}
  \label{subfig:S6-alphaComparison}
 \end{subfigure}
 \hfill\null
 \caption{\textbf{Results for the S6 device.} Panels~\subref{subfig:S6-allPlots-alpha100},~\subref{subfig:S6-allPlots-alpha050} and~\subref{subfig:S6-allPlots-alpha040} compare the results for chains using the U, C, ${[3,1,3]}_{1}$ code and $[4,1,4]$ code at scaling parameters $\alpha=1$, $0.5$ and $0.4$ respectively. Panel~\subref{subfig:S6-alphaComparison} shows a comparison at a fixed chain length of $\overline{N}=80$ as $\alpha$ is varied. For $\alpha \lesssim 0.90$ and for sufficiently long chains, the ${[4,1,4]}_{0}$ code starts to outperform the C strategy. The ${[3,1,3]}_{1}$ code outperforms all other strategies at all values of $\alpha$, for sufficiently long chain lengths.}
 \label{fig:S6mainChainComparison}
\end{figure*}

Figure~\ref{fig:S6mainChainComparison} displays the same for the S6 device. Since this device operated at a lower temperature than the DW2-ISI device, we expect it to have a lower thermal excitation error rate. This means that the probability of multiple bit-flips per encoded qubit will decrease, i.e., more errors will be decodable, and so we can expect that---all else being equal---QAC will be more effective.
This explains why the $\alpha$ cross-over point for the ${[4,1,4]}_{0}$ code shifts to higher values; it is now closer to $\alpha=0.9$.

\begin{figure*}
  \centering
  \includegraphics[width=0.5\textwidth]{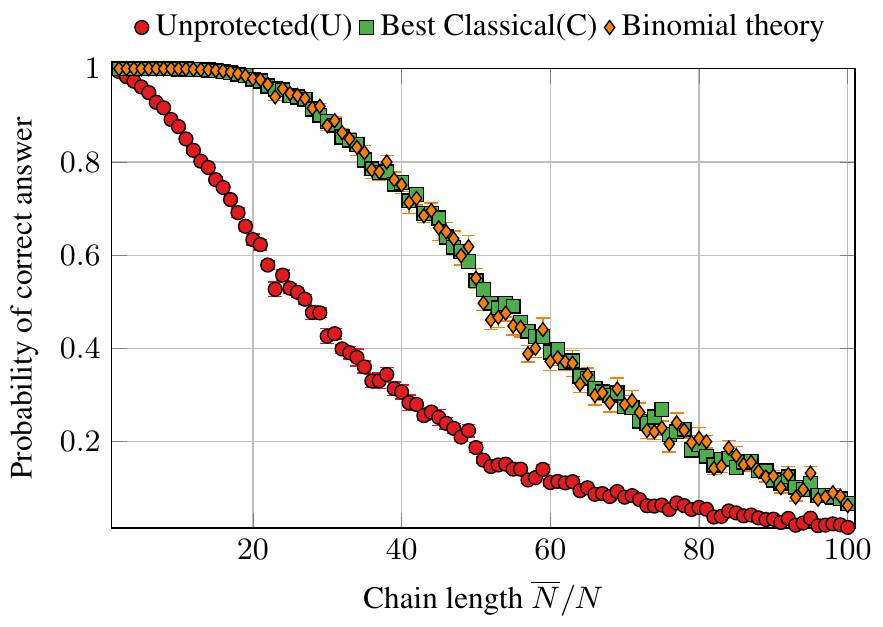}
  \caption{\textbf{Independence test}. Using the data from the {DW2-ISI device}, we compare the performance of the U strategy to the C strategy at $\alpha=0.45$. The C strategy agrees with the prediction from binomial theory that assumes independent chains.}
  \label{fig:UCPlot}
\end{figure*}

In order to demonstrate the independence of the four copies in the C strategy, we compare the performance of the U and C strategies in Fig~\ref{fig:UCPlot}. The C strategy's performance is close to the one predicted for independent runs, indicating that chains indeed behave independently. The success probabilities are high for small chains, but rapidly drop as we increase the chain length. The same conclusion holds across the range of scaling parameter $\alpha$.

\begin{figure*}
 \centering
 \includegraphics[width=0.5\textwidth]{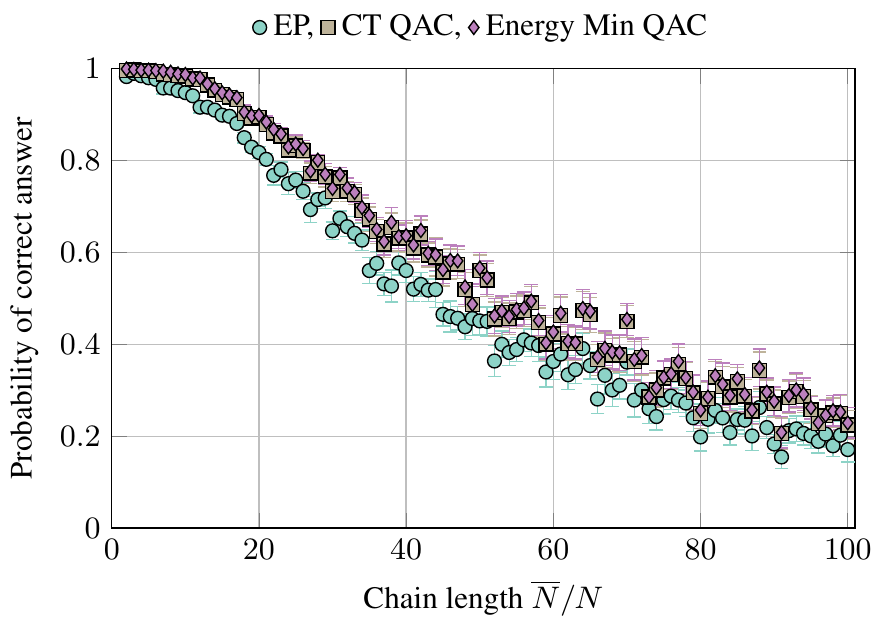}
 \caption{\textbf{Decoding strategies for the ${[4,1,4]}_{0}$ code}.  DW2-ISI device at $\alpha=0.70$. The EP strategy is marginally improved upon by the use of decoding. Ties are broken by either coin tossing or energy minimization; the latter performs slightly better at all chain lengths.}
 \label{fig:squareCodeComparison}
\end{figure*}

In Fig~\ref{fig:squareCodeComparison}, we compare the results for the EP, CT, and EM decoding strategies.  As expected, the EM strategy outperforms the EP and CT strategies when decoding chains. The small enhancement in the success probability of the EM strategy over CT indicates that the number of ties is correspondingly small. This is confirmed, along with additional results compare the various decoding strategies at other $\alpha$ values, in Appendix~\ref{sec:comp-decod-strat}.

%
%%%%%%%%%%%%%%%%%%%%%%%%%%
\subsection{The role of energy scaling}
\label{sec:energyscaling}
%%%%%%%%%%%%%%%%%%%%%%%%%%
%
We now address the performance mismatch between the two QAC codes.  Recall that QAC boosts the problem scale via a redundant representation of the $\overline{Z}_i$ and $\overline{Z_i Z_j}$ operators. One may expect this energy boost to reduce errors due to the combination of two effects: thermal excitations are suppressed via the Boltzmann factor, and raising the overall problem energy scale reduces diabatic transitions by increasing the minimum gap during the evolution, though the latter effect is difficult to quantify without diagonalizing the full Hamiltonian $H(t)$\footnote{See Ref.~\cite{PAL:14} for an analytically solvable model that exhibits an increased gap via this mechanism.}.  As noted above, the ${[3,1,3]}_{1}$ code boosts the problem energy scale by a factor of three, while the ${[4,1,4]}_{0}$ code boosts the problem energy scale by a factor of two, so we might expect the ${[3,1,3]}_{1}$ code to outperform the ${[4,1,4]}_{0}$ code on this basis alone.

\begin{figure*}
 \centering
 \begin{subfigure}[b]{0.45\textwidth}
  \includegraphics[width=\textwidth]{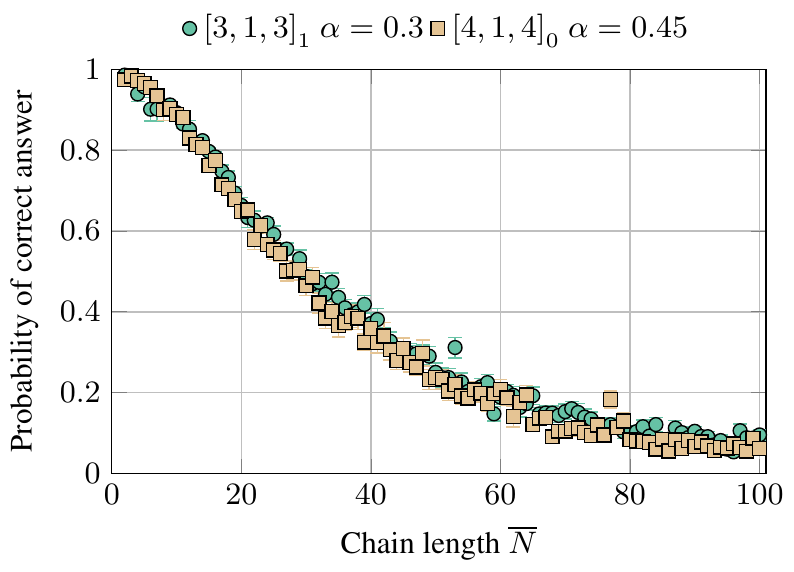}
  \caption{EP}
  \label{subfig:ISI-compareEP}
 \end{subfigure}
 \begin{subfigure}[b]{0.45\textwidth}
  \includegraphics[width=\textwidth]{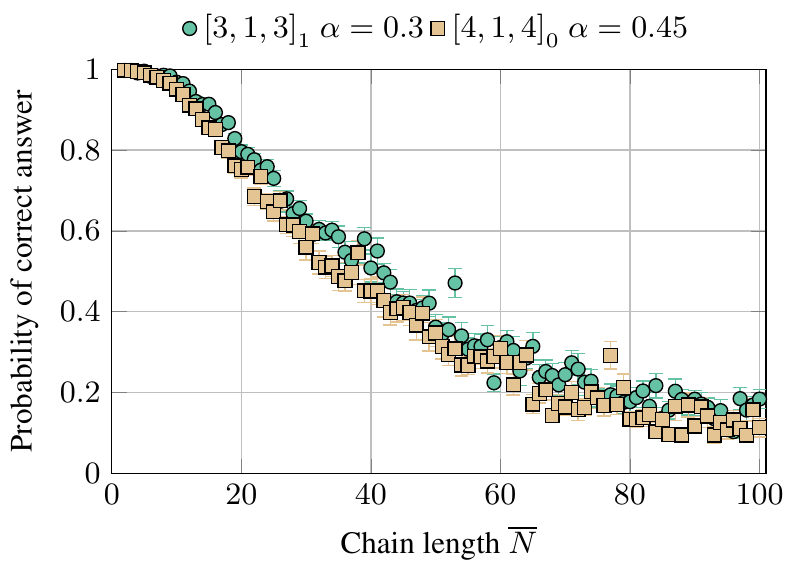}
  \caption{QAC}
  \label{subfig:ISI-compareDecodingPerformance}
 \end{subfigure}
 \caption{ \textbf{Code comparison at equalized effective energy scales for the DW2-ISI device.} Panel~\subref{subfig:ISI-compareEP} compares the EP performance of the two codes at equivalent effective energy scales: $3\times 0.3$ and $2\times 0.45$ for the ${[3,1,3]}_{1}$ and ${[4,1,4]}_{0}$ code, respectively. Panel~\subref{subfig:ISI-compareDecodingPerformance} compares the two codes after decoding. Code performance is essentially indistinguishable in the EP case, indicating that the effective energy scale is the dominant performance-determining factor. The ${[3,1,3]}_{1}$ code exhibits a slight advantage across all chain lengths after decoding.}
 \label{fig:energyProtection}
\end{figure*}

To compare the performance of the two codes we can equalize their effective problem energy scales, defined as $\alpha$ times the boost factor due to the redundancy in representing the $\overline{Z}_i$ and $\overline{Z_i Z_j}$ operators. We set $\alpha=0.3$ for the ${[3,1,3]}_{1}$ code and $\alpha=0.45$ for the ${[4,1,4]}_{0}$, so that, after the energy boost is accounted for, the effective scale of both is $0.9$. We first compare the two using the EP strategy in order to eliminate the role of decoding.  Figure~\ref{subfig:ISI-compareEP} reveals that the two codes perform almost identically when tested at the same effective energy scale, indicating that the protection offered by both codes is quantitatively determined by this scale.  Figure~\ref{subfig:ISI-compareDecodingPerformance} shows the results after decoding, with a slight advantage for the majority vote decoding of the ${[3,1,3]}_{1}$ code over the energy minimization of the ${[4,1,4]}_{0}$ code.

\subsection{Decodable states: energy \textit{vs} Hamming weight}
\label{app:states-decoded}
We expect both codes to enable correct decoding of physical states that are a small Hamming distance away from the physical ground state, but not when the physical state is far away in Hamming distance from the physical ground state. Figure~11 confirms this intuition. Additionally, we note that for both codes we can correct highly excited states, as long as they are within a small Hamming distance from the physical ground state. Thus, the requirement of remaining in the ground state throughout the evolution, which is impractical for non-zero temperature quantum annealers but is typically the condition imposed by adiabaticity in closed-system AQC, is seen to be overly restrictive in the present setting, since QAC is able to tolerate certain excitations out of the ground state. The observation that error correction for AQC or quantum annealing is designed to tolerate excitations has of course been made before, e.g., in Refs.~\cite{Young:2013fk,Young:13,PAL:13}.

\begin{figure*}
  \centering
  \null\hfill
  \begin{subfigure}[t]{0.45\textwidth}
      \includegraphics[width=\textwidth]{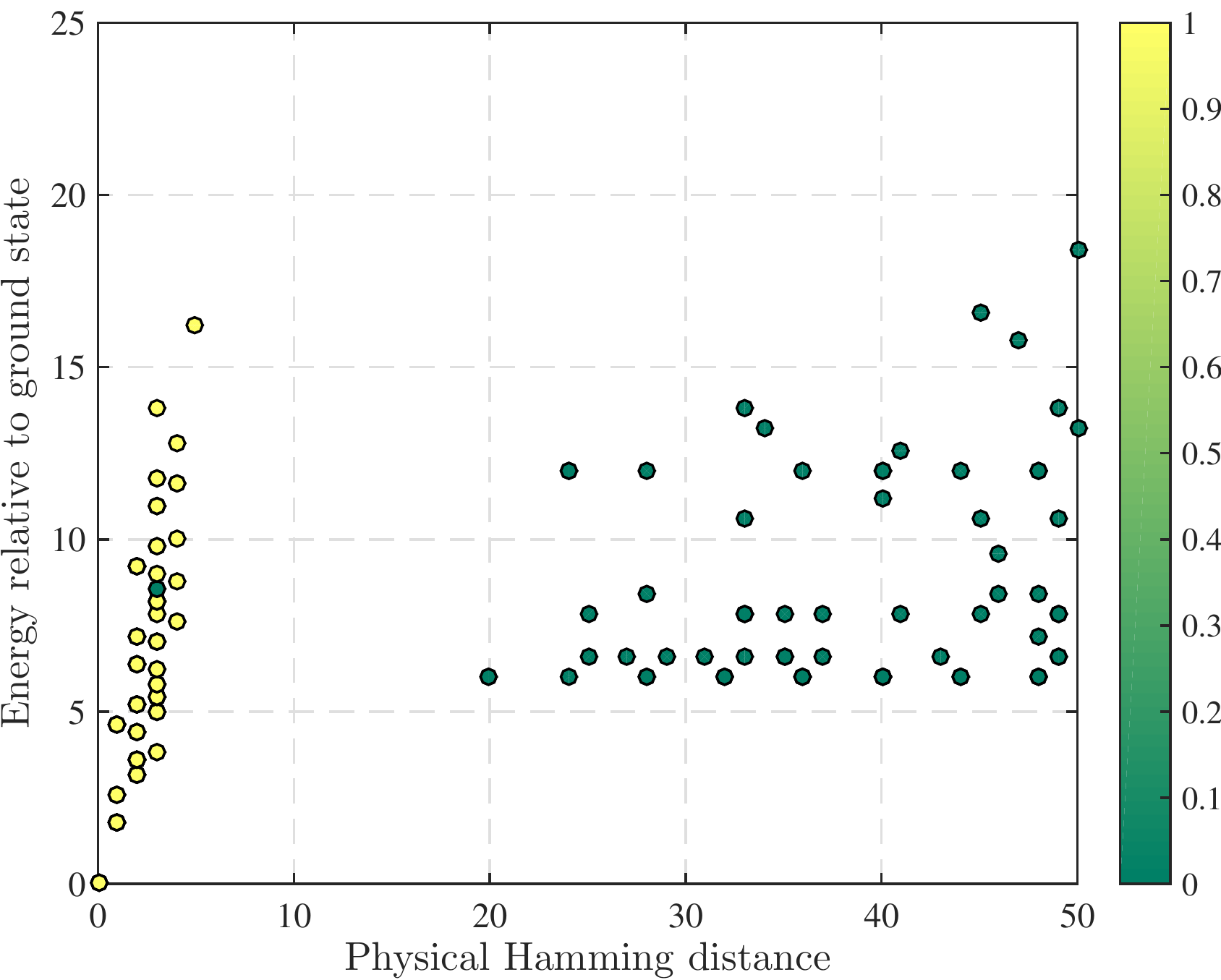}
      \caption{${[3,1,3]}_{1}$}
      \label{subfig:pudenz-hamD}
  \end{subfigure}
  \hfill
  \begin{subfigure}[t]{0.45\textwidth}
      \includegraphics[width=\textwidth]{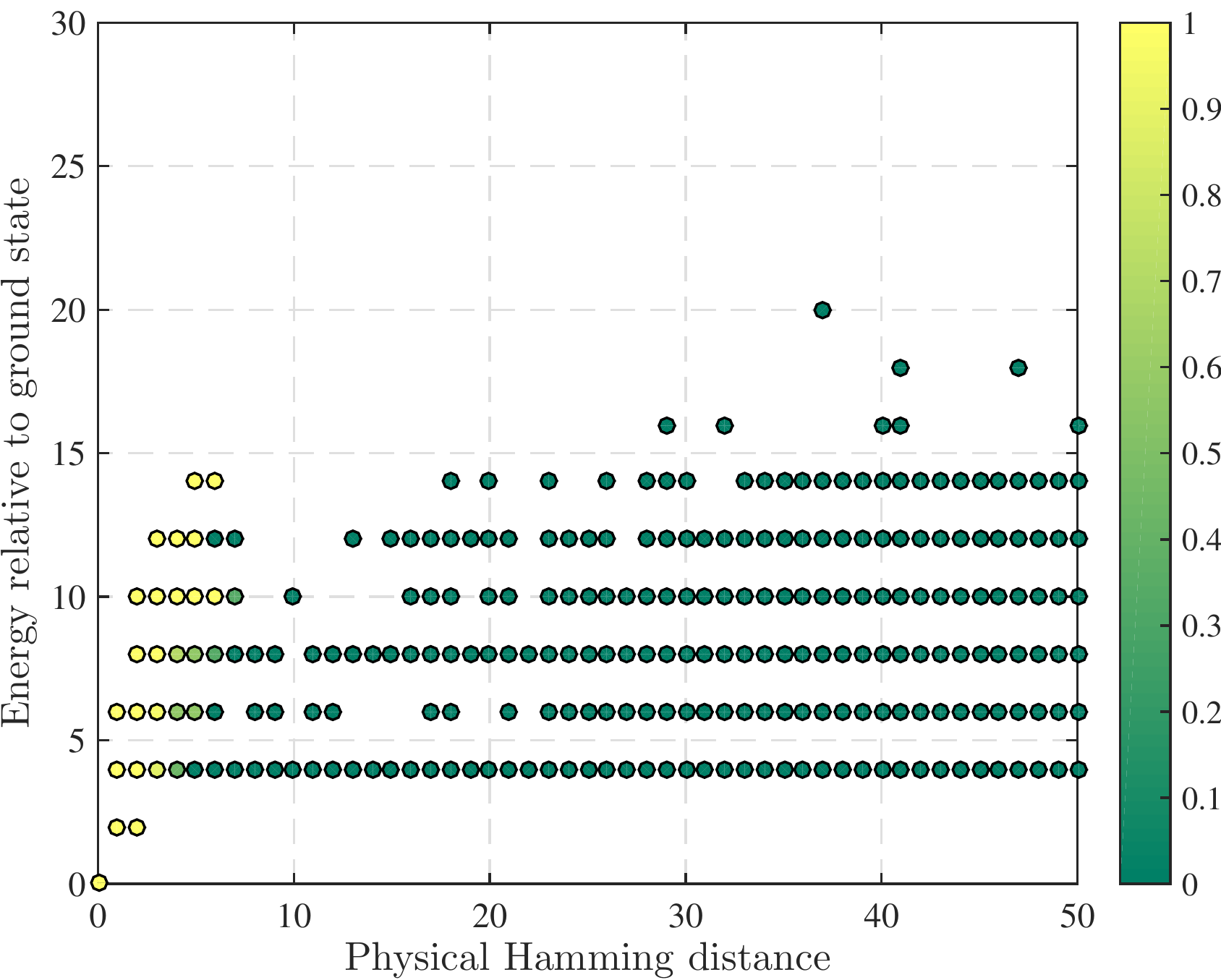}
      \caption{$[4,1,4]_{0}$}
      \label{subfig:square-hamD}
  \end{subfigure}
  \hfill\null
  \caption{\textbf{Decodable states.} Panels~\subref{subfig:pudenz-hamD} and \subref{subfig:square-hamD} display the performance of the ${[3,1,3]}_{1}$ and ${[4,1,4]}_{0}$ codes, respectively, for the longest chain of length $100$, $\alpha=1$, and optimal $\gamma$ observed on the DW2-ISI device. We show the energies of all the states observed relative to the encoded ground state along the vertical axis, while the horizontal axis is the Hamming distance from the encoded ground state. We did not find any decodable states of Hamming distance higher than $50$. Color indicates the fraction of decodable states at each observed energy and Hamming distance. States with a small Hamming distance are mostly decodable. The ${[3,1,3]}_{1}$ code is decoded via majority vote, while the ${[4,1,4]}_{0}$ code  uses the EM strategy.}
  \label{fig:hamming-distance-vs-energy}
\end{figure*}

Figure~\ref{fig:hamming-distance-vs-energy} reveals a striking difference between the ${[3,1,3]}_{1}$ and ${[4,1,4]}_{0}$ codes. The latter exhibits many undecodable states over the entire range of Hamming distances from the encoded ground state, starting from the second excited state. The ${[3,1,3]}_{1}$ code, on the other hand, exhibits a large Hamming distance separation between decodable and undecodable states, with the latter appearing only for relatively high excited states. This reflects the higher effectiveness of the penalty term in the ${[3,1,3]}_{1}$ code, and at the same time gives a detailed view of the different failure mechanisms of both codes.

\subsection{The role of temperature}
\begin{figure*}
  \begin{subfigure}[t]{0.45\textwidth}
    \includegraphics[width=\textwidth]{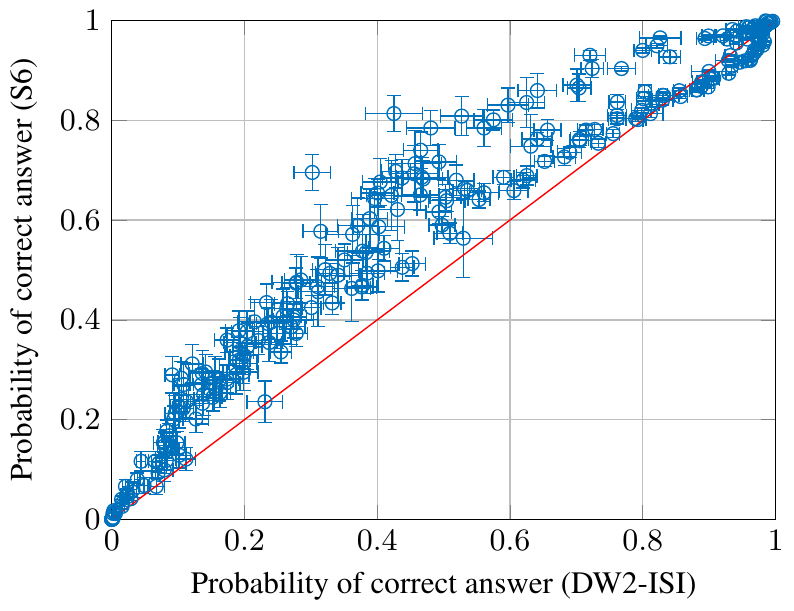}
    \caption{EP}
    \label{fig:ComparisonEP}
  \end{subfigure}\;\;\qquad
  \begin{subfigure}[t]{0.45\textwidth}
    \includegraphics[width=\textwidth]{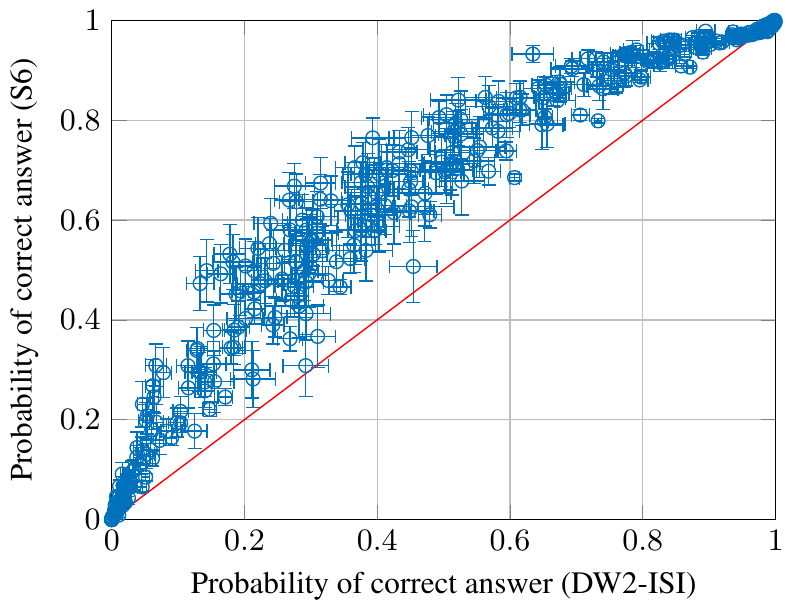}
    \caption{EM}
    \label{fig:ComparisonEM}
  \end{subfigure}
\caption{\textbf{Comparison of the performance of the ISI and S6 devices.}  Correlation plot of the success probability on the S6 and ISI devices for instances that have the same $\gamma_{\mathrm{opt}}$ (for the same $\alpha$ and chain length) using the EP strategy [panel~\subref{fig:ComparisonEP}, with $264$ instances] and the QAC with EM strategy [panel~\subref{fig:ComparisonEM}, with $376$ instances].  Instances to the left (right) of the diagonal line have a higher success probability on the S6 (DW2-ISI) device. Virtually all instances were solved with a higher success probability on the S6 device.}
\label{fig:S6ISIComparison}
\end{figure*}

Having access to two quantum annealers operating at two different temperatures and with different characteristics (see Table~\ref{tab:annealerComp}), we can compare the performance of the two devices at equivalent programming parameters.  We show in Fig.~\ref{fig:S6ISIComparison} a correlation plot for instances encoded using the ${[4,1,4]}_{0}$ code, with equal $\gamma_{\textrm{opt}}$ for a given $\alpha$.  We observe a clear advantage for the S6 device, which we attribute to its lower operating temperature.

\subsection{Behavior of the optimal energy penalty for the ${[4,1,4]}_{0}$ code}
\begin{figure*}
  \centering
  \begin{subfigure}[t]{0.45\textwidth}
  \includegraphics[width=\textwidth]{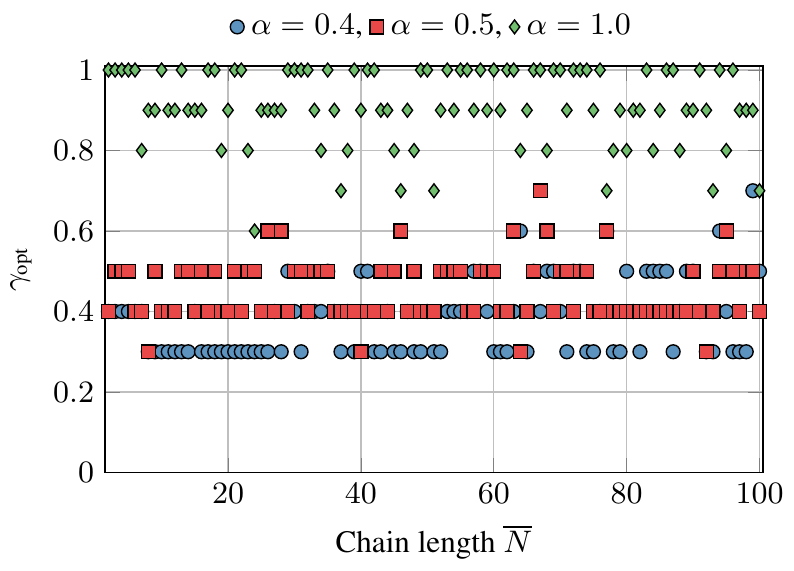}
  \caption{{DW2-ISI device} --- EP}
  \label{subfig:ISI-optimalPenalty-squareCode-EP}
  \end{subfigure}\;\;
  \begin{subfigure}[t]{0.45\textwidth}
  \includegraphics[width=\textwidth]{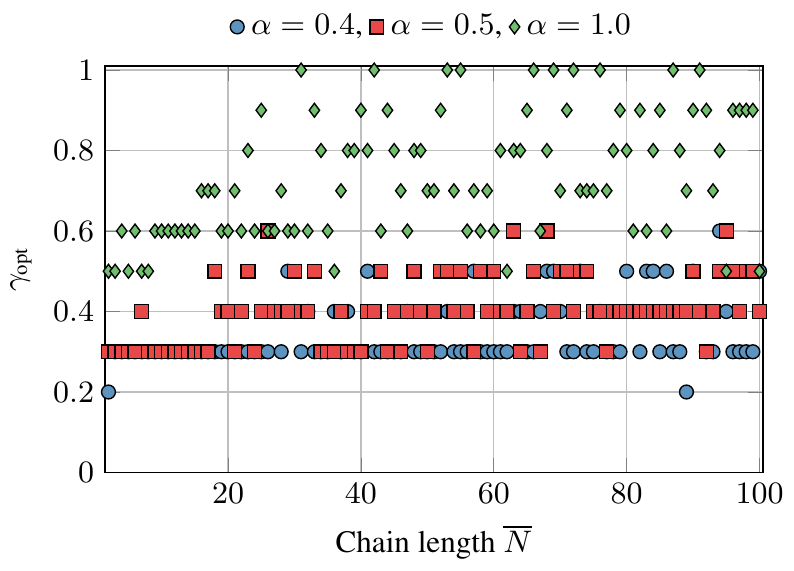}
  \caption{{DW2-ISI device} --- QAC with EM decoding}
  \label{subfig:ISI-optimalPenalty-squareCode-energyMin}
  \end{subfigure}\\
  \begin{subfigure}[t]{0.45\textwidth}
  \includegraphics[width=\textwidth]{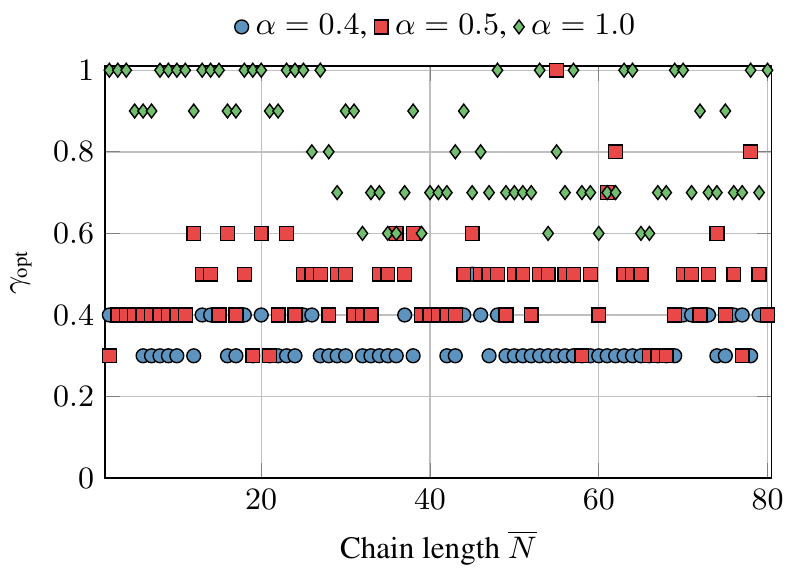}
  \caption{{S6 device} --- EP}
  \label{subfig:S6-optimalPenalty-squareCode-EP}
  \end{subfigure}\;\;
  \begin{subfigure}[t]{0.45\textwidth}
  \includegraphics[width=\textwidth]{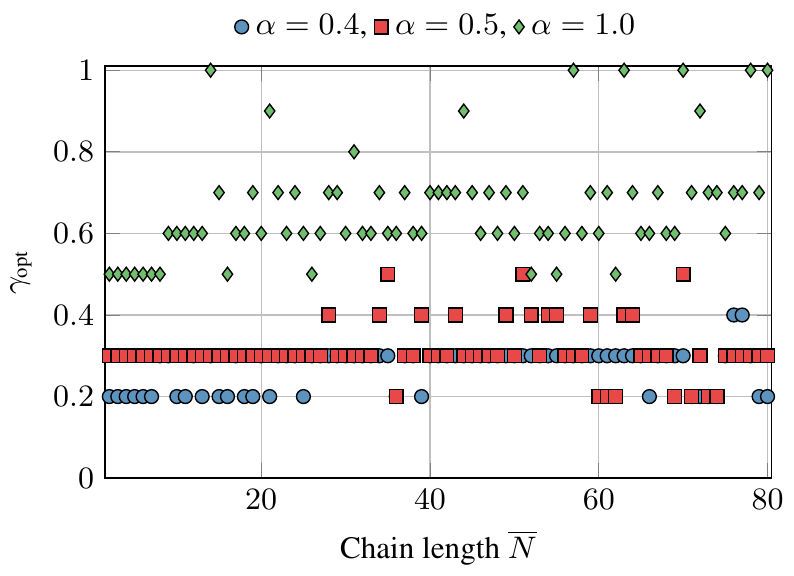}
  \caption{{S6 device} --- QAC with EM decoding}
    \label{subfig:S6-optimalPenalty-squareCode-energyMin}
  \end{subfigure}
  \caption{\textbf{Optimal $\boldsymbol{\gamma}$ for the ${[4,1,4]}_{0}$ code.}  Panels~\subref{subfig:ISI-optimalPenalty-squareCode-EP} and~\subref{subfig:ISI-optimalPenalty-squareCode-energyMin} show the optimal $\gamma$ values for the {DW2-ISI device}. Panels~\subref{subfig:S6-optimalPenalty-squareCode-EP} and~\subref{subfig:S6-optimalPenalty-squareCode-energyMin} show the same for the S6 device. For three representative values of the scaling parameter $\alpha$, we note that the EP strategy consistently requires a higher value for the optimal $\gamma$. There is a slight tendency for longer chains to have a larger optimal penalty. Additionally, since the DW2-ISI device operates at a higher temperature and hence is more prone to errors, it requires a higher value for the optimal $\gamma$ than the S6 device. The difference is most prominent in the QAC case, i.e., when comparing panels \subref{subfig:ISI-optimalPenalty-squareCode-energyMin} and \subref{subfig:S6-optimalPenalty-squareCode-energyMin}.}
  \label{fig:optimalBeta}
\end{figure*}

It is instructive to study the dependence of the optimal penalty value on $\alpha$ and chain length $\overline{N}$.  Figure~\ref{fig:optimalBeta} shows the results of the optimization of success probabilities for the ${[4,1,4]}_{0}$ code on the two quantum annealing devices. The optimal penalty value scales with $\alpha$, i.e., $\gamma_{\text{opt}}\propto \alpha$, which is quite unlike the behavior of the ${[3,1,3]}_{1}$ code reported in Ref.~\cite{PAL:13}, for which $\gamma_{\textrm{opt}}$ was found to be essentially constant (this is reproduced in Appendix~\ref{app:optimising-beta}). We may perhaps attribute this difference to the fact that the ${[3,1,3]}_{1}$ code has a dedicated penalty qubit which is therefore not as sensitive to values of the problem couplings as are the data qubits of the ${[4,1,4]}_{0}$ code, which participate simultaneously in the penalty and problem Hamiltonians.

Figure~\ref{fig:optimalBeta} also shows that lower values of $\gamma_{\text{opt}}$ were required on the {S6 device}, which can again be attributed to its lower operating temperature.

Appendix~\ref{app:optimising-beta} provides additional results, showing the full dependence of the success probabilities on the penalty values, $\alpha$, and chain length.
\section{Theoretical Analysis}
\label{sec:theory}

In this section we provide a theoretical analysis of some of our results. In particular, we provide a simple thermodynamic explanation of the performance difference between the ${[4,1,4]}_{0}$ and ${[3,1,3]}_{1}$ codes, which we can attribute primarily to the effective energy scale. In addition, we explain the decodability of the ${[4,1,4]}_{0}$ code in terms of an intuitively appealing criterion of the ordering of decodable \textit{vs} undecodable excited states.

\subsection{Thermodynamic comparison}
%Code comparison tables.
\begin{table*}
  \subcaptionbox{${[3,1,3]}_{1}$ code. The first three columns are the values of the data qubits; the fourth column is the penalty qubit; the fifth is the energy penalty counted as twice the number of violated couplings $v$; the sixth is the magnetization $m=- \sum_i s_i$; the seventh is the multiplicity, which is the number of states that are topologically equivalent to the state shown, up to relabelling of qubits; and the last is the decodability of the state.\label{tab:PudenzCodeAnalysis}}[0.45\textwidth]
     {\begin{tabular}{|c|c|c|c|c|r|c|c|}
         \hline
         $1$ & $2$ & $3$ & $p$ & $2v$ & $m$ & mult & dec \\
         \hline
         0 & 0 & 0 & 0 & 0 & -4 & 1 & y \\
         0 & 0 & 1 & 0 & 2 & -2 & 3 & y \\
         0 & 1 & 1 & 0 & 4 & 0 & 3 & n \\
         1 & 1 & 1 & 0 & 6 & 2 & 1 & n \\
         0 & 0 & 0 & 1 & 6 & -2 & 1 & y \\
         0 & 0 & 1 & 1 & 4 & 0 & 3 & y \\
         0 & 1 & 1 & 1 & 2 & 2 & 3 & n \\
         1 & 1 & 1 & 1 & 0 & 4 & 1 & n \\
         \hline
     \end{tabular}}
\qquad
  \subcaptionbox{${[4,1,4]}_{0}$ code. The first four columns are the values of the data qubits; the fifth is the energy penalty; the sixth is the magnetization $m=- \sum_i s_i$; the seventh is the multiplicity, which is number of states that are topologically equivalent to the state shown, up to relabelling of qubits; and the last is the decodability, where ``t'' denotes a tie. Rows 3--5 denote three distinct ways of placing the two plus and two minus states, where the qubits are numbered 1--4 clockwise, starting from top-left [as in Fig.~\ref{fig:sqCode-phyQubits}].\label{tab:SquareCodeAnalysis}}[0.45\textwidth]
     {\begin{tabular}{|c|c|c|c|c|c|c|c|}
\hline
$1$ & $2$ & $3$ & $4$ & $2v$ & $m$ & mult & dec \\
\hline
0 & 0 & 0 & 0 & 0 & -4 & 1 & y \\
0 & 0 & 0 & 1 & 4 & -2 & 4 & y \\
0 & 0 & 1 & 1 & 4 & 0 & 2 & t \\
0 & 1 & 0 & 1 & 4 & 0 & 2 & t \\
0 & 1 & 1 & 0 & 8 & 0 & 2 & t \\
0 & 1 & 1 & 1 & 4 & 2 & 4 & n \\
1 & 1 & 1 & 1 & 0 & 4 & 1 & n \\
\hline
     \end{tabular}}
\caption{Comparison of the two codes for a $1$-qubit encoded problem with a local field.}
\end{table*}

To gain a better understanding, we compare the theoretical performance of the two codes for decoding a single encoded qubit using a simple thermodynamic argument (see Ref.~\cite{matsuura_mean_2015} for a much more detailed analysis of the QAC partition function along the annealing evolution, for a fully connected ferromagnetic transverse field Ising model). We assume the presence of a local field of strength $h$ acting on the encoded qubits; this would translate to a local field of $\tilde{h} = h/2$ on all physical qubits of the ${[4,1,4]}_{0}$ code [by Eq.~\eqref{eq:5a}], and a local field of $\tilde{h} = h$ on the three data qubits of the ${[3,1,3]}_{1}$ code [by Eq.~\eqref{eq:7a}]. If $h<0$, then the state $\ket{0000}$ is the ground state of the system. Success would be declared if the evolution takes the system to this state, or if of the final state is correctly decodable to $\ket{\overline{0}}$.  Tables~\ref{tab:PudenzCodeAnalysis} and~\ref{tab:SquareCodeAnalysis} enumerate all $16$ cases along with their energy penalty and decodability for the two codes.

Let us now assume that the state at the end of the anneal is thermal. The Boltzmann weight of any of the $16$ states is given by $e^{-\beta(2v\gamma-\tilde{h} m)}/Z$ where $m=-\sum_{i} s_{i}$ is the magnetization, $v$ counts the number of violated couplings, $\beta=1/kT$ is the inverse temperature, and $Z$ is the partition function.  The probability $p_{\textrm{err}}$ for an error in the ${[3,1,3]}_{1}$ code case is the sum of the Boltzmann factors of the undecodable states, while for the ${[4,1,4]}_{0}$ code we must also include the tied cases with a factor of $1/2$, assuming that these cases are decoded by coin tossing. We write $Z=W+W'$ where $W$ is the sum of the unnormalized Boltzmann factors for the encoded error cases (rows with `n' in the decodability column of tables~\ref{tab:PudenzCodeAnalysis} and~\ref{tab:SquareCodeAnalysis}, and half of the `t' cases in table~\ref{tab:SquareCodeAnalysis}) and $W'$ is the sum of the decodable cases (rows with `y' in the decodability column of tables~\ref{tab:PudenzCodeAnalysis} and~\ref{tab:SquareCodeAnalysis}, and the other half of the `t' cases in table~\ref{tab:SquareCodeAnalysis}). The error probabilities $p_{\text{err}}^{[3]/[4]}$ are functions of $\gamma$ and $h$, where the labels $[3]/[4]$ label the ${[3,1,3]}_{1}/{[4,1,4]}_{0}$ code respectively:

\bes
\begin{align}
p^{\textrm{[3]}}_{\textrm{err}} & = \frac{W_{\textrm{[3]}}}{W_{\textrm{[3]}}+W'_{\textrm{[3]}}}   \ , \qquad p^{\textrm{[4]}}_{\textrm{err}}  = \frac{W_{\textrm{[4]}}}{W_{\textrm{[4]}}+W'_{\textrm{[4]}}} \ , \\
W_{\textrm{[3]}}   &= 3e^{-\beta(4\gamma+h)} + e^{-\beta(6\gamma+3h)} +3e^{-\beta(2\gamma+2h)} + e^{-{3\beta h}} \ , \\
 W'_{\textrm{[3]}} &=  3e^{-\beta(4\gamma-h)} + e^{-\beta(6\gamma-3h)} +  3e^{-\beta(2\gamma-2h)} + e^{{3\beta h}} \ ,  \\
W_{\textrm{[4]}}   &=  \frac{1}{2}\left( 4e^{-{4\beta\gamma}} + 2e^{-{8\beta\gamma}}\right) +4e^{-\beta(4\gamma+h)} + e^{-{2\beta h}} \ , \\
W'_{\textrm{[4]}}  &=  \frac{1}{2}\left( 4e^{-{4\beta\gamma}} + 2e^{-{8\beta\gamma}}\right)  +4e^{-\beta(4\gamma-h)}  +e^{{2\beta h}} \ .
\end{align}
\ees

\begin{figure*}
  \begin{subfigure}[t]{0.45\textwidth}
    \includegraphics[width=\textwidth]{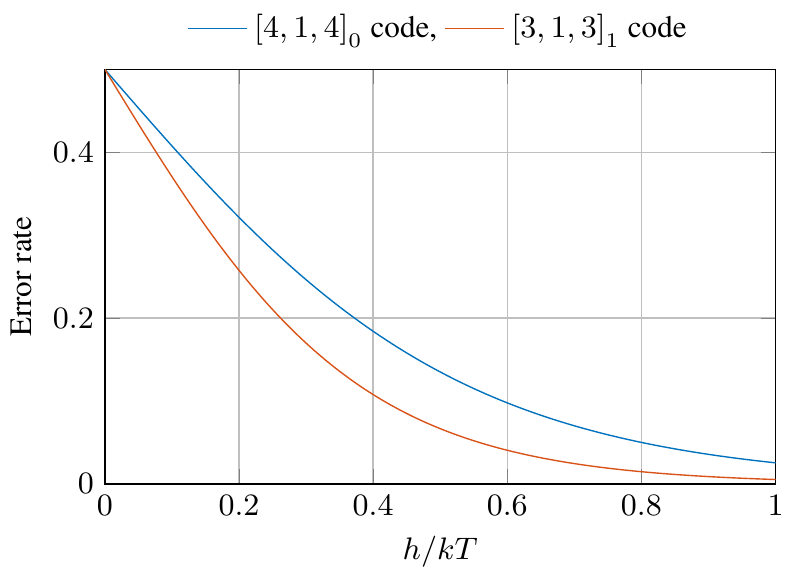}
    \caption{}
    \label{fig:thermoCodeComparison}
  \end{subfigure}\;\;\qquad
  \begin{subfigure}[t]{0.45\textwidth}
    \includegraphics[width=\textwidth]{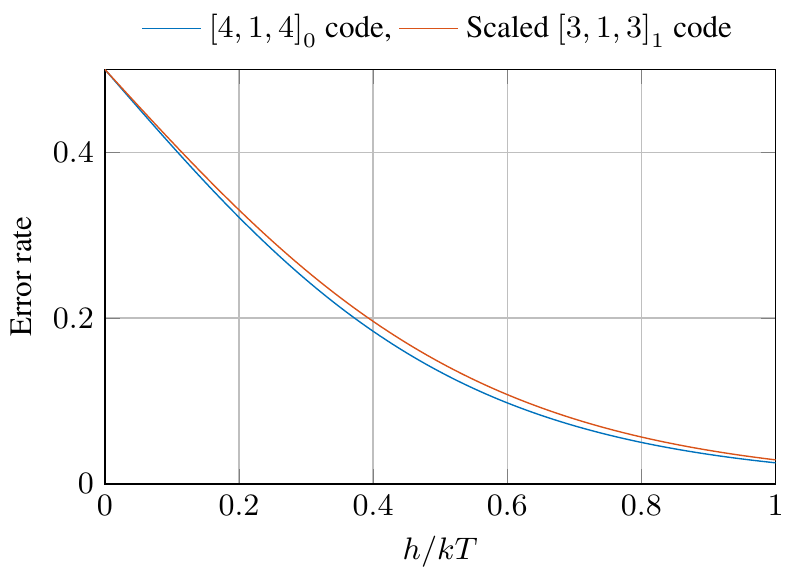}
    \caption{}
    \label{fig:thermoScaledCodeComparison}
  \end{subfigure}
\caption{\textbf{Thermodynamic comparison of codes.} Panel~\subref{fig:thermoCodeComparison} shows the thermal error rates of the two codes at the same encoded energy scale $h$. Panel~\subref{fig:thermoScaledCodeComparison} compares the two codes at equivalent effective energy scales, i.e., $h$ for the ${[4,1,4]}_{0}$ code and $2h/3$ for the ${[3,1,3]}_{1}$ code.}
\end{figure*}

We minimize the error probabilities with respect to $\gamma$ for each value of $h$, noting that the optimal $\gamma$ value is different for the two codes.  Figure~\ref{fig:thermoCodeComparison} shows the error rates of the two codes as $\beta h$ is varied. We note that the ${[3,1,3]}_{1}$ code exhibits a lower error rate than the ${[4,1,4]}_{0}$ code. This agrees with our experimental findings and is a simple consequence of the ${[3,1,3]}_{1}$ code operating at a higher boosted energy scale than the ${[4,1,4]}_{0}$ code.

We also compare the error rates at equivalent effective energy scales, i.e., $2h/3$ for the ${[3,1,3]}_{1}$ code and $h$ for the ${[4,1,4]}_{0}$ code. Figure~\ref{fig:thermoScaledCodeComparison} shows that at equivalent effective energy scales the two codes have similar error rates, with the ${[3,1,3]}_{1}$ code performing slightly worse for all $\beta h$ values.  This is the opposite of the experimental findings presented in Fig.~\ref{fig:energyProtection} and suggests that the output of the D-Wave devices for these problems is not fully captured by the thermal model. Nevertheless, this analysis confirms that the error rate due to a thermal bath would be similar for the two codes, when operated at equivalent effective energy scales.
\subsection{Decodability of the ${[4,1,4]}_{0}$ code}
\label{app:two-qubit-afm}
\begin{figure*}
  \centering
  \begin{subfigure}[t]{0.5\textwidth}
    \includegraphics[width=\textwidth]{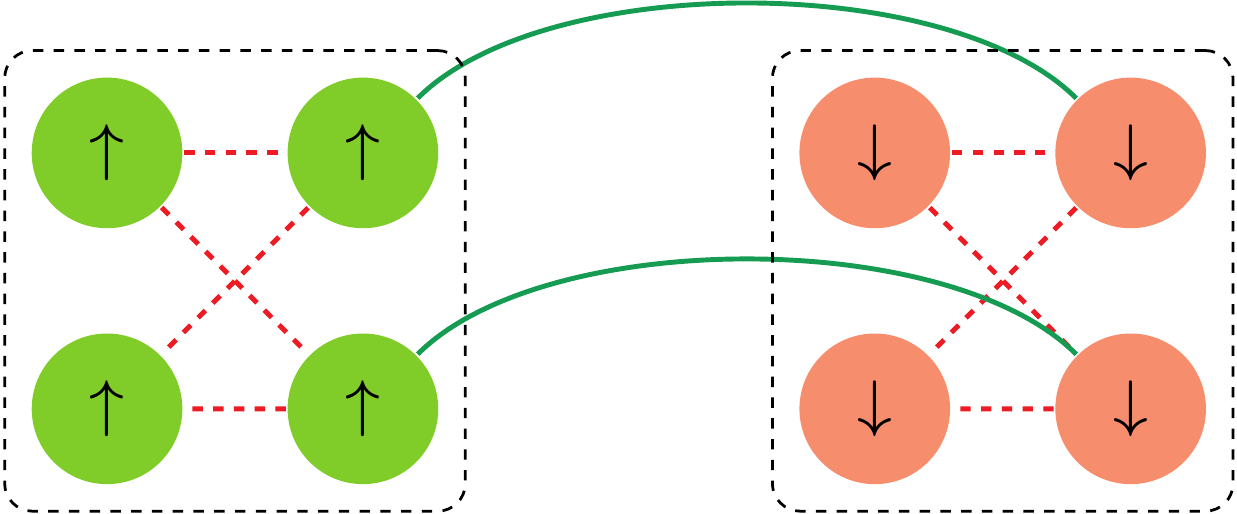}
    \caption{The ground state.}
  \end{subfigure}\\
  \begin{subfigure}[t]{0.25\textwidth}
    \includegraphics[width=\textwidth]{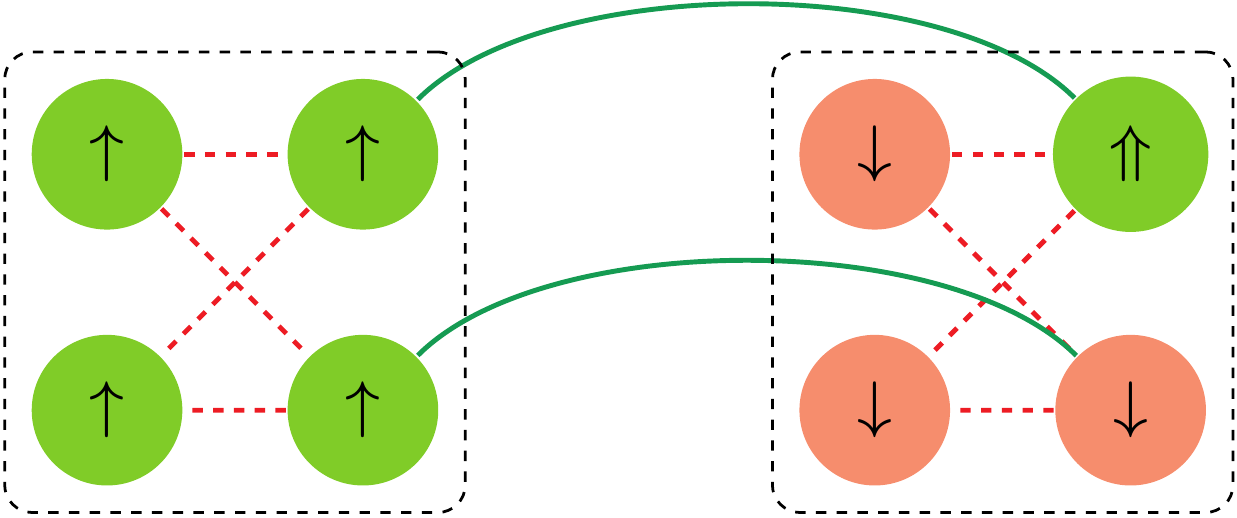}
    \caption{One flip ($\times2$). $2\alpha+4\gamma$}
  \end{subfigure}\qquad\qquad
  \begin{subfigure}[t]{0.25\textwidth}
    \includegraphics[width=\textwidth]{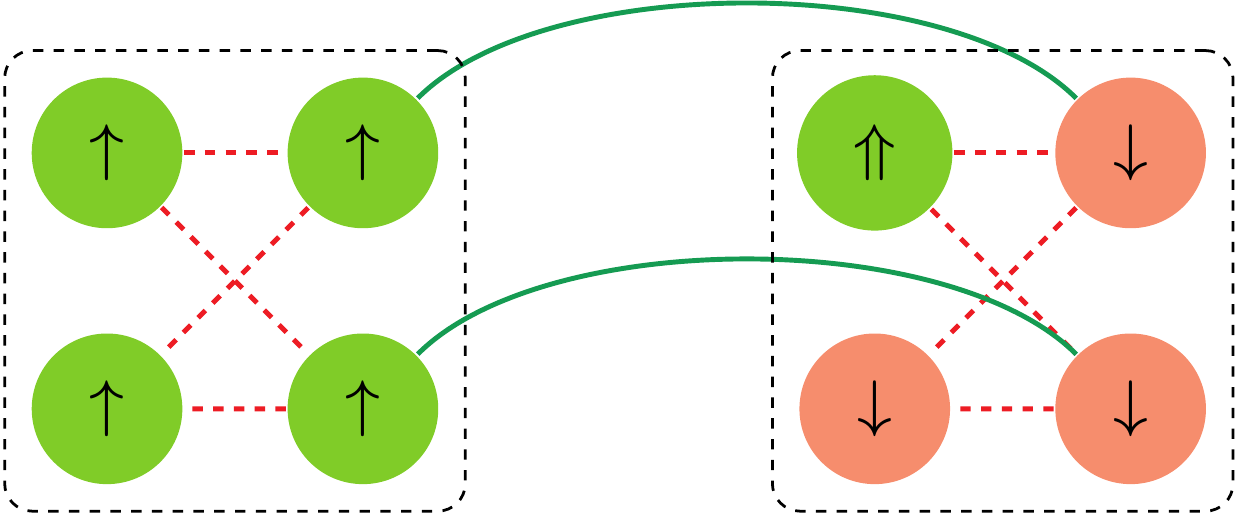}
    \caption{One flip ($\times2$). $4\gamma$}
  \end{subfigure}\\
  \begin{subfigure}[t]{0.25\textwidth}
    \includegraphics[width=\textwidth]{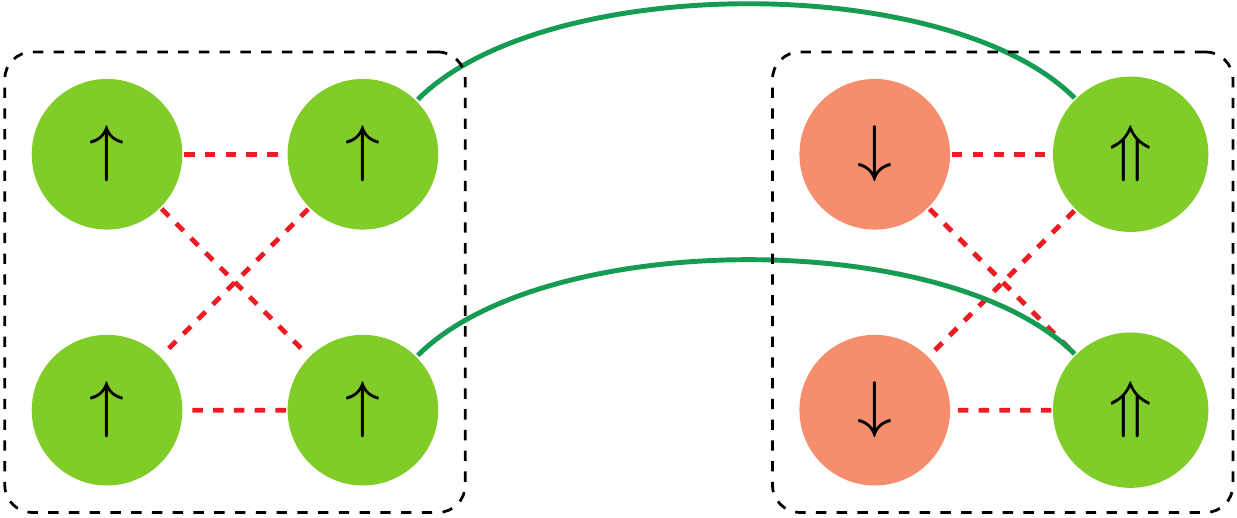}
    \caption{Two flips ($\times1$). $4\alpha+8\gamma$}
  \end{subfigure}\qquad\qquad
  \begin{subfigure}[t]{0.25\textwidth}
    \includegraphics[width=\textwidth]{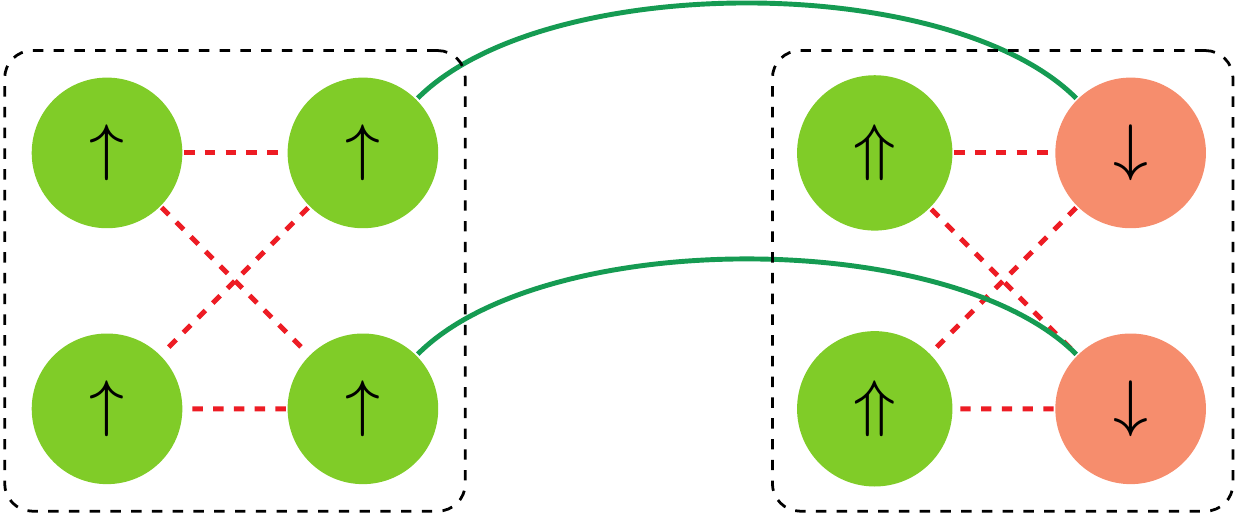}
    \caption{Two flips ($\times1$). $8\gamma$ }
  \end{subfigure}\qquad\qquad
  \begin{subfigure}[t]{0.25\textwidth}
    \includegraphics[width=\textwidth]{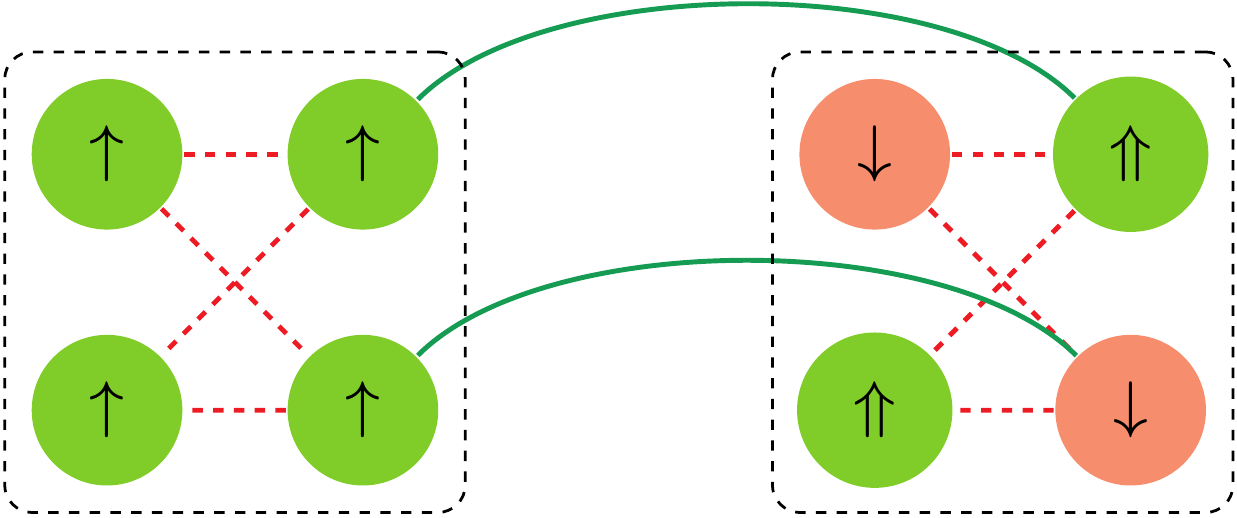}
    \caption{Two flips ($\times4$). $2\alpha+4\gamma$}
  \end{subfigure}\\
  \begin{subfigure}[t]{0.25\textwidth}
    \includegraphics[width=\textwidth]{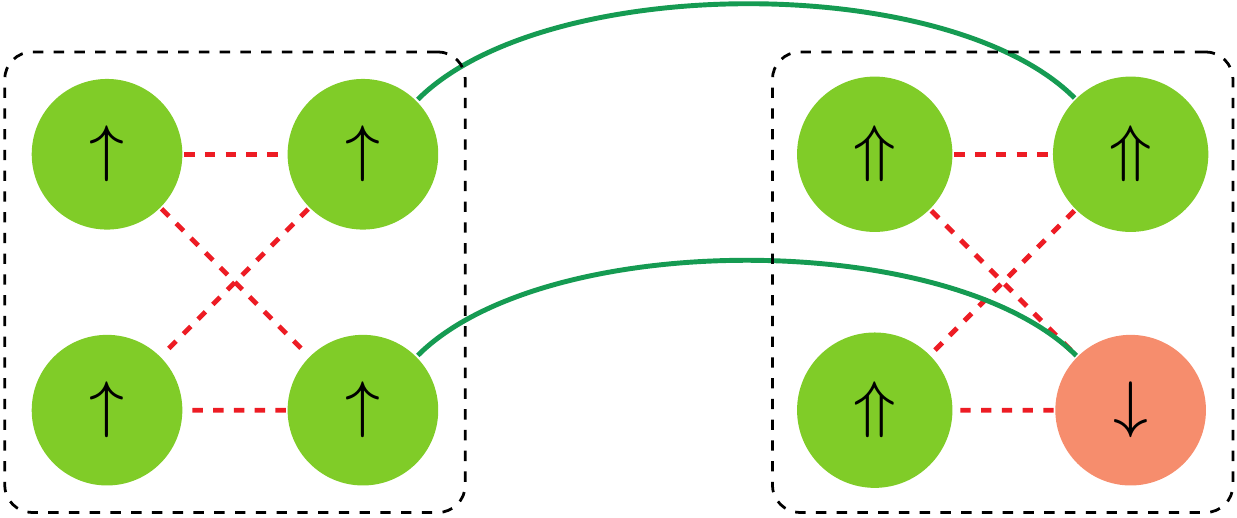}
    \caption{Three flips ($\times2$). $2\alpha+4\gamma$}
  \end{subfigure}\qquad\qquad
  \begin{subfigure}[t]{0.25\textwidth}
    \includegraphics[width=\textwidth]{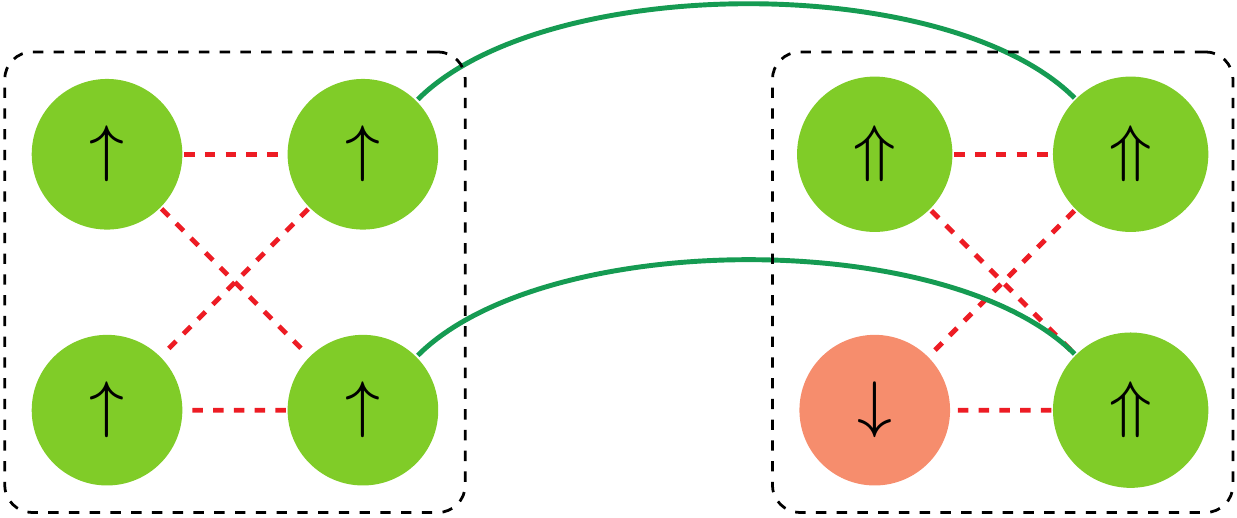}
    \caption{Three flips ($\times2$). $4\alpha+4\gamma$}
  \end{subfigure}\\
  \begin{subfigure}[t]{0.28\textwidth}
    \includegraphics[width=\textwidth]{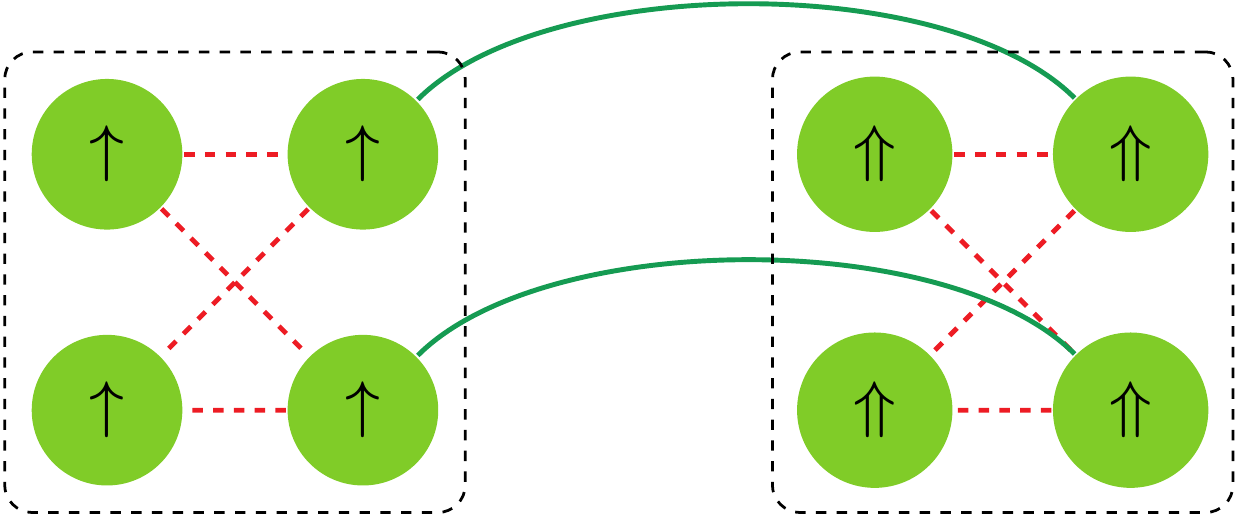}
    \caption{Four flips ($\times1$). $4\alpha$}
  \end{subfigure}
  \caption{\textbf{Two coupled encoded ${[4,1,4]}_{0}$ code qubits.} The two encoded ${[4,1,4]}_{0}$ code qubits consists of $8$ physical qubits. We show one of the two degenerate ground states, and all inequivalent ways in which bit-flip errors might accumulate on one of the encoded qubits. Each flip leads to an excited state. The number in parentheses denotes the multiplicity of such states, followed by the energy separation from the shown ground state. The $\uparrow$ symbol denotes the correct state of the qubit while the $\Uparrow$ symbol denotes the occurrence of a flip.}
  \label{fig:square-code-two-qubits}
\end{figure*}

\begin{figure*}
  \centering
  \begin{subfigure}[t]{0.45\textwidth}
      \includegraphics[width=\textwidth]{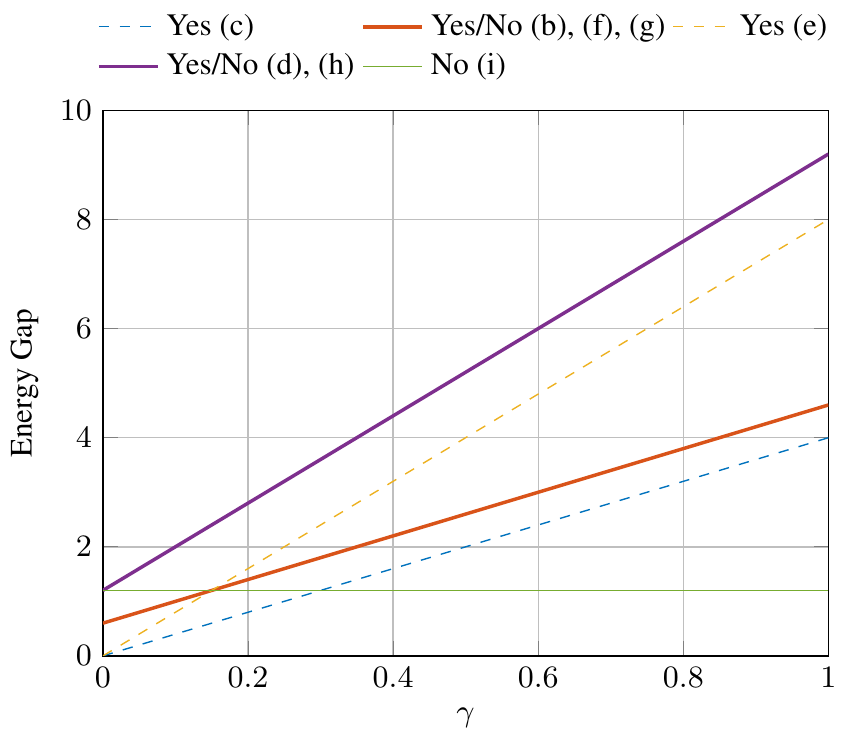}
      \caption{}
      \label{subfig:decode-excitedStates}
  \end{subfigure}\;\;
  \begin{subfigure}[t]{0.46\textwidth}
     \includegraphics[width=\textwidth]{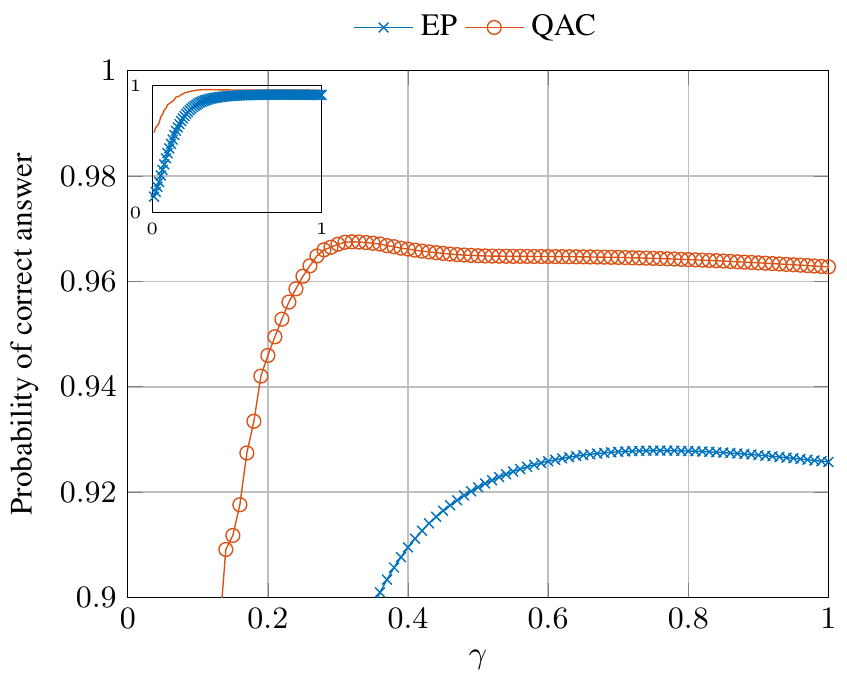}
     \caption{}
     \label{subfig:decode-ME}
  \end{subfigure}
  \caption{\textbf{Decodability analysis for $\boldsymbol{\alpha = 0.3}$.} \subref{subfig:decode-excitedStates} The first few excited states at the final time for two coupled ${[4,1,4]}_{0}$ code qubits. The legend labels refer to the states in Fig.~\ref{fig:square-code-two-qubits}, indicating their decodability. The optimal $\gamma$ value occurs right before the first excited state goes from being decodable to undecodable. Logical error states are represented by solid lines, decodable states by dashed lines. Thick lines indicate degenerate excited states, some of which are decodable, and some of which are encoded errors. \subref{subfig:decode-ME} Success probability calculated using the adiabatic master equation \cite{ABLZ:12-SI}. Inset: a zoomed out version. The success probability of the QAC strategy is maximized near $\gamma\approx0.3$, which agrees with the value of $\gamma$ in (a) where an undecodable state becomes the first excited state.  Population lost to this state in the simulations cannot be recovered after decoding.  This explains the comment made in Sec.~\ref{sec:quant-anneal-corr} that the optimal $\gamma$ keeps the decodable states lower in the energy spectrum.  The EP strategy is optimized at a larger value of $\gamma$.}
  \label{fig:decodabilityAnalysis}
\end{figure*}

In order to study the decodability of the ${[4,1,4]}_{0}$ code, and in particular the effect of varying the penalty strength $\gamma$, we consider two antiferromagnetically coupled encoded qubits, decoded via EM.  In Fig.~\ref{fig:square-code-two-qubits}, we show how different bit-flip errors can accumulate on a two qubit chain, pushing the system into one of the excited states, and the energy gap of these excited state from one of the ground state. We show in Fig.~\ref{subfig:decode-excitedStates} the spectrum of these excited states and whether they can or cannot be decoded at $\alpha=0.3$, as a function of $\gamma$. For sufficiently high $\gamma$ a non-decodable state becomes lower in energy than a decodable state.  This coincides with the optimal $\gamma$ value from a quantum adiabatic master equation simulation~\cite{ABLZ:12-SI}. When the EP strategy is used instead of EM, the optimal $\gamma$ occurs at a larger value as shown in Fig.~\ref{subfig:decode-ME}.\\
%
%
%%%%%%%%%%%%%%%%%%%%%%%%%%
\section{Conclusions}
\label{sec:conclusions}
%%%%%%%%%%%%%%%%%%%%%%%%%%
%
Quantum annealing will require error correction in order to become a scalable form of quantum information processing. While our results depend heavily on the Chimera architecture of the D-Wave devices, it is only possible to make progress in the field of experimental quantum error correction by studying specific devices  that provide snapshots of evolving technologies (e.g., Refs.~\cite{Kelly:2015fv,Corcoles:2015zl,2012Natur.482..382R}). With this caveat in mind, our study contains several valuable long-term lessons.

Specifically, in this work we studied two quantum annealing correction codes---the ${[4,1,4]}_{0}$ and ${[3,1,3]}_{1}$ codes---and compared their performance using the simple case of antiferromagnetic chains, on two different experimental platforms belonging to the same generation of D-Wave Two devices. The two codes differ in the energy boost they provide by redundantly encoding their logical operators, and the two quantum annealers differ in operating temperature. We have shown that these differences translate into performance gains as expected, i.e., both a higher energy boost and a lower operating temperature result in improved success probabilities. This conclusion has immediate implications for the design of future quantum annealing devices: despite results indicating that thermal effects can assist AQC \cite{TAQC}, and our (and previous \cite{Young:2013fk,Young:13,PAL:13}) results supporting the notion that error correction can tolerate thermal excitations, significant performance gains are to be realized via the straightforward mechanisms of cooling and increasing the energy scale.

Despite delivering lower success probabilities, the ${[4,1,4]}_{0}$ code with the smaller energy boost is  interesting, since it gives rise to an encoded graph with higher degree than the ${[3,1,3]}_{1}$ code, and physical implementations of quantum annealers are likely to be subject in general to constraints that reduce connectivity.  The tradeoff between code performance and the degree of the encoded graph may thus be worthwhile, as long as an improvement over purely classical error correction strategies is achieved, as we have demonstrated here for sufficiently high noise levels and problem sizes. An intriguing question is whether this tradeoff is necessary. Our work will hopefully inspire the design of quantum annealing architectures with higher connectivity and of codes that better leverage encoded graph degree and energy boosts.

\acknowledgements
Access to the D-Wave Two quantum annealers was made available by the USC-Lockheed Martin Quantum Computing Center and D-Wave Systems Inc. This work was supported under ARO grant number W911NF-12-1-0523, ARO MURI Grant No. W911NF-11-1-0268, NSF Grant No. CCF-1551064, and Fermilab Grant No. 622302. A.M. was also supported by the USC Provost Ph.D. fellowship.
\balancecolsandclearpage
\onecolumngrid
\appendix

\centerline
{\large Supplementary Material}
\vspace{.5cm}
In this appendix we present some additional results to complement the main text.

\section{Optimizing ${\pmb\gamma}$}
\label{app:optimising-beta}

\begin{figure*}
  \centering
  \begin{subfigure}[t]{\textwidth}
  \centering
  \includegraphics[width=0.28\textwidth]{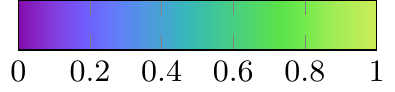}
  \end{subfigure}\\
  \vspace{0.2cm}
  \begin{subfigure}[t]{0.28\textwidth}
      \includegraphics[width=\textwidth]{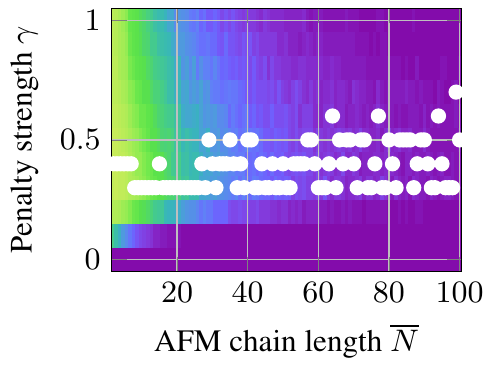}
      \caption{$\alpha=0.4$ EP}
  \end{subfigure}\;\;
  \begin{subfigure}[t]{0.28\textwidth}
      \includegraphics[width=\textwidth]{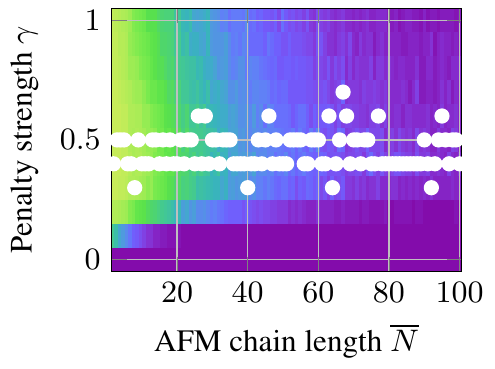}
      \caption{$\alpha=0.5$ EP}
  \end{subfigure}\;\;
  \begin{subfigure}[t]{0.28\textwidth}
      \includegraphics[width=\textwidth]{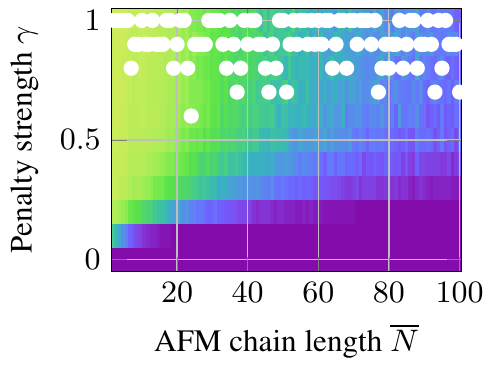}
      \caption{$\alpha=1.0$ EP}
  \end{subfigure}
  \begin{subfigure}[t]{0.28\textwidth}
      \includegraphics[width=\textwidth]{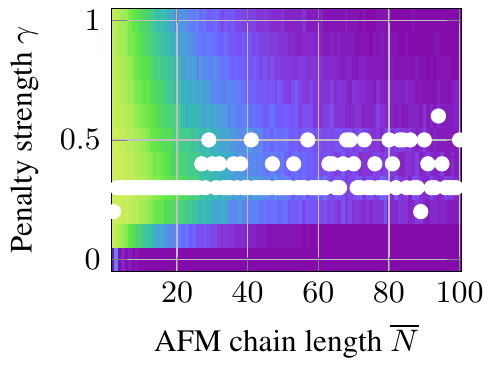}
      \caption{$\alpha=0.4$ QAC (EM)}
  \end{subfigure}\;\;
  \begin{subfigure}[t]{0.28\textwidth}
      \includegraphics[width=\textwidth]{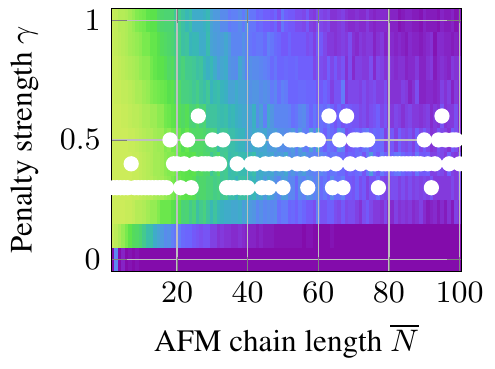}
      \caption{$\alpha=0.5$ QAC (EM)}
  \end{subfigure}\;\;
  \begin{subfigure}[t]{0.28\textwidth}
      \includegraphics[width=\textwidth]{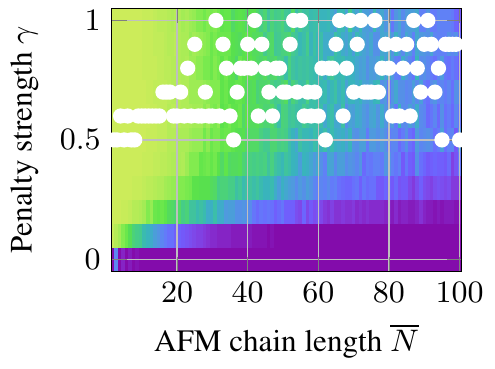}
      \caption{$\alpha=1.0$ QAC (EM)}
  \end{subfigure}
  \caption{\textbf{Optimal $\boldsymbol{\gamma}$ for the ${[4,1,4]}_{0}$ code on the {DW2-ISI device}.} The color scale represents the success probability, while the white circles indicate the optimal penalty value for a given chain length. The top and bottom three panels show the EP and QAC with EM strategies, respectively. The optimal $\gamma$ increases proportionally to the problem scale $\alpha$. A higher $\gamma_{\textrm{opt}}$ is required for the EP case, where we perform no decoding. The success probability depends strongly on $\overline{N}$, $\gamma$, and $\alpha$.}
\label{fig:ISI-squareCode-opt-beta}
\end{figure*}

\begin{figure*}
  \centering
  \begin{subfigure}[t]{0.28\textwidth}
      \includegraphics[width=\textwidth]{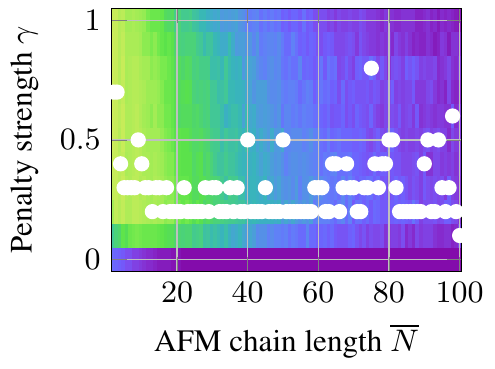}
      \caption{$\alpha=0.4$ EP}
  \end{subfigure}\;\;
  \begin{subfigure}[t]{0.28\textwidth}
      \includegraphics[width=\textwidth]{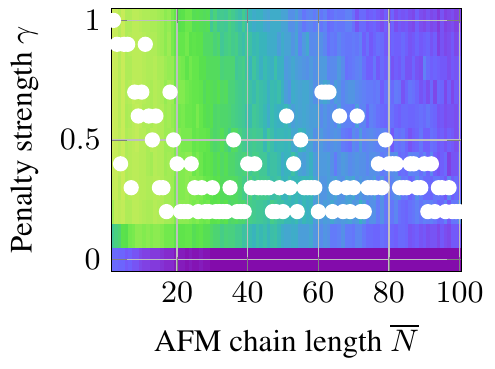}
      \caption{$\alpha=0.5$ EP}
  \end{subfigure}\;\;
  \begin{subfigure}[t]{0.28\textwidth}
      \includegraphics[width=\textwidth]{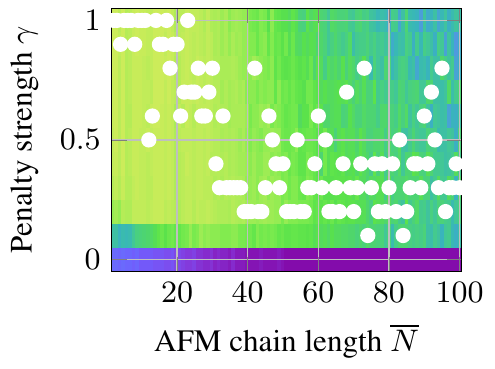}
      \caption{$\alpha=1.0$ EP}
  \end{subfigure}
  \begin{subfigure}[t]{0.28\textwidth}
      \includegraphics[width=\textwidth]{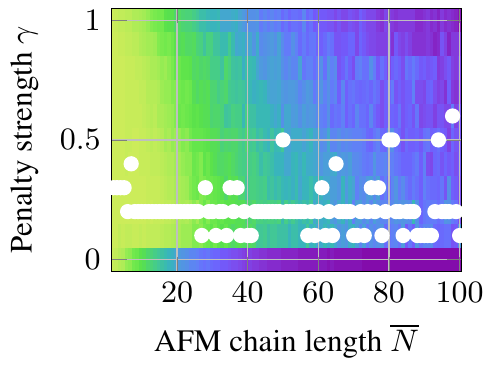}
      \caption{$\alpha=0.4$ QAC}
  \end{subfigure}\;\;
  \begin{subfigure}[t]{0.28\textwidth}
      \includegraphics[width=\textwidth]{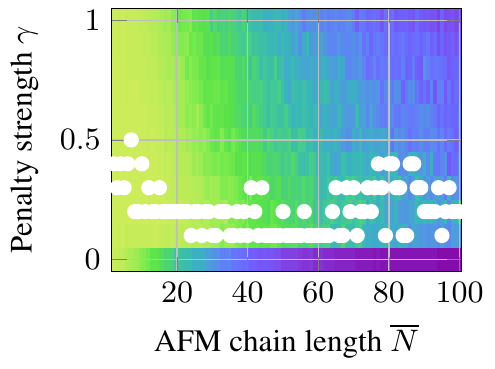}
      \caption{$\alpha=0.5$ QAC}
  \end{subfigure}\;\;
  \begin{subfigure}[t]{0.28\textwidth}
      \includegraphics[width=\textwidth]{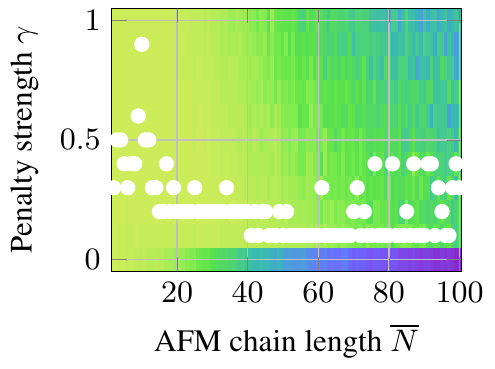}
      \caption{$\alpha=1.0$ QAC}
  \end{subfigure}
  \caption{\textbf{Optimal $\gamma$ for the ${[3,1,3]}_{1}$ code on the DW2-ISI device.} The optimal $\gamma$ increases proportionally to the problem scale $\alpha$ in the EP case, but remains fairly constant for the QAC case, in agreement with Ref.~\cite{PAL:13}. The success probability depends strongly on $\overline{N}$ and $\alpha$, but not as strongly on $\gamma$.}
\label{fig:ISI-pudenzCode-opt-beta}
\end{figure*}

\begin{figure*}
  \centering
  \begin{subfigure}[t]{\textwidth}
  \centering
  \includegraphics[width=0.28\textwidth]{common-colorbar.pdf}
  \end{subfigure}\\
  \vspace{0.2cm}
  \begin{subfigure}[t]{0.28\textwidth}
      \includegraphics[width=\textwidth]{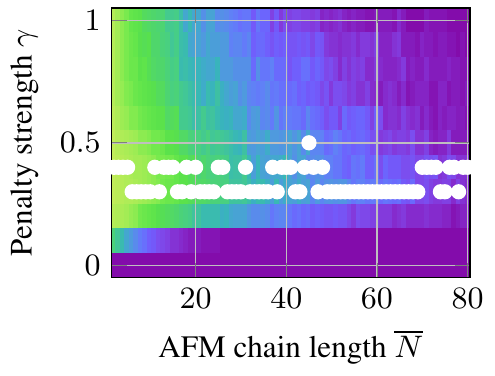}
      \caption{$\alpha=0.4$ EP}
  \end{subfigure}\;\;
  \begin{subfigure}[t]{0.28\textwidth}
      \includegraphics[width=\textwidth]{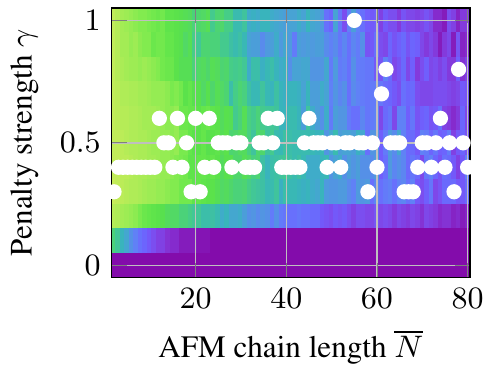}
      \caption{$\alpha=0.5$ EP}
  \end{subfigure}\;\;
  \begin{subfigure}[t]{0.28\textwidth}
      \includegraphics[width=\textwidth]{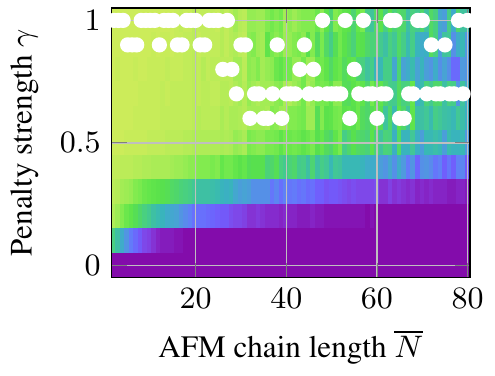}
      \caption{$\alpha=1.0$ EP}
  \end{subfigure}
  \begin{subfigure}[t]{0.28\textwidth}
      \includegraphics[width=\textwidth]{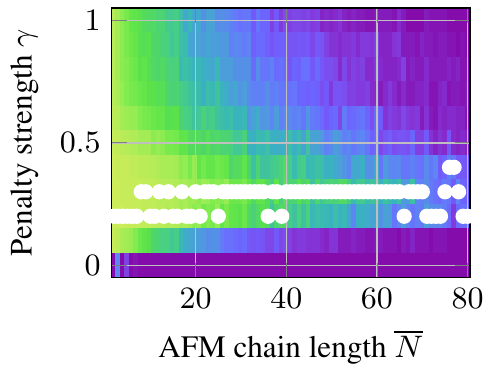}
      \caption{$\alpha=0.4$ QAC (EM)}
  \end{subfigure}\;\;
  \begin{subfigure}[t]{0.28\textwidth}
      \includegraphics[width=\textwidth]{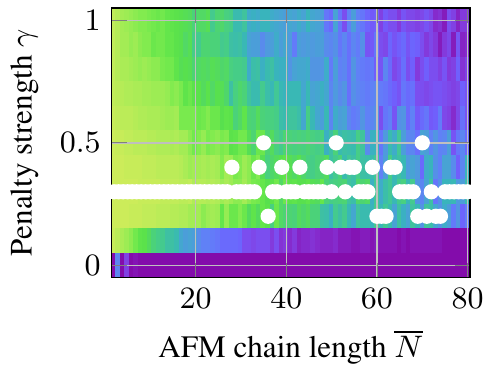}
      \caption{$\alpha=0.5$ QAC (EM)}
  \end{subfigure}\;\;
  \begin{subfigure}[t]{0.28\textwidth}
      \includegraphics[width=\textwidth]{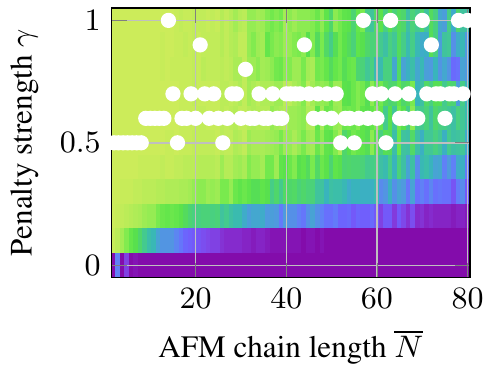}
      \caption{$\alpha=1.0$ QAC (EM)}
  \end{subfigure}
  \caption{\textbf{Optimal $\boldsymbol{\gamma}$ for the ${[4,1,4]}_{0}$ code on the {S6 device}.} The optimal $\gamma$ increases proportionally to the problem scale $\alpha$. A higher $\gamma_{\textrm{opt}}$ is required for the EP case, where we perform no decoding. The success probability depends strongly on $\overline{N}$, $\gamma$, and $\alpha$.}
  \label{fig:S6-squareCode-opt-beta}
\end{figure*}

\begin{figure*}
  \centering
  \begin{subfigure}[t]{0.28\textwidth}
      \includegraphics[width=\textwidth]{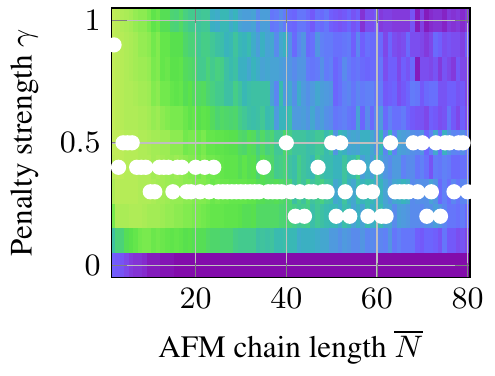}
      \caption{$\alpha=0.4$ EP}
  \end{subfigure}\;\;
  \begin{subfigure}[t]{0.28\textwidth}
      \includegraphics[width=\textwidth]{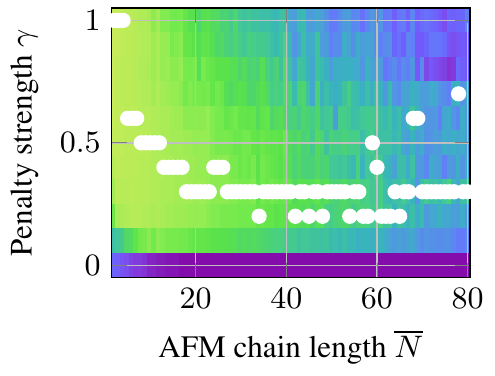}
      \caption{$\alpha=0.5$ EP}
  \end{subfigure}\;\;
  \begin{subfigure}[t]{0.28\textwidth}
      \includegraphics[width=\textwidth]{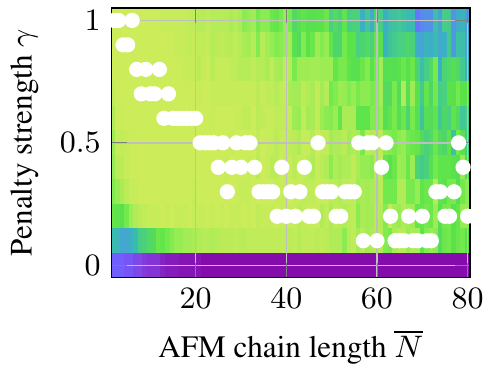}
      \caption{$\alpha=1.0$ EP}
  \end{subfigure}
  \begin{subfigure}[t]{0.28\textwidth}
      \includegraphics[width=\textwidth]{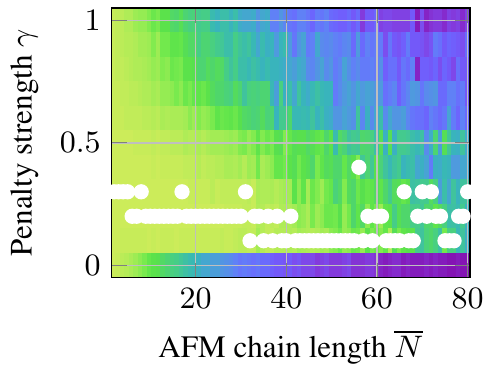}
      \caption{$\alpha=0.4$ QAC}
  \end{subfigure}\;\;
  \begin{subfigure}[t]{0.28\textwidth}
      \includegraphics[width=\textwidth]{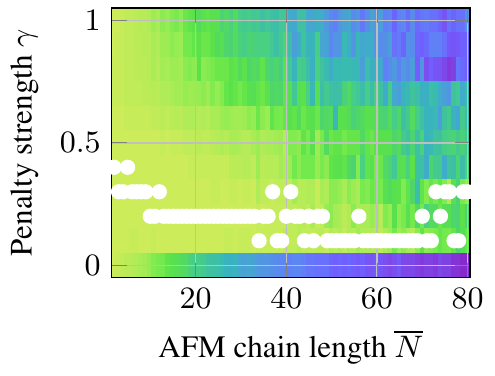}
      \caption{$\alpha=0.5$ QAC}
  \end{subfigure}\;\;
  \begin{subfigure}[t]{0.28\textwidth}
      \includegraphics[width=\textwidth]{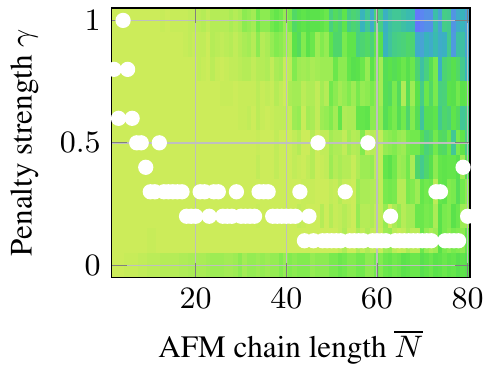}
      \caption{$\alpha=1.0$ QAC}
  \end{subfigure}
  \caption{\textbf{Optimal $\gamma$ on the S6 device for the ${[3,1,3]}_{1}$ code.} The optimal $\gamma$ increases proportionally to the problem scale $\alpha$ in the EP case, but remains fairly constant for the QAC case, again in agreement with Ref.~\cite{PAL:13} (though note that the latter used the DW2-ISI device). The success probability depends strongly on $\overline{N}$ and $\alpha$, but not as strongly on $\gamma$.}
  \label{fig:S6-pudenzCode-opt-beta}
\end{figure*}

For each chain instance, we identified the optimal penalty coupling strength $\gamma$ by varying it in increments of $0.1$ in the range $[0,1]$. This is shown in Figs.~\ref{fig:ISI-squareCode-opt-beta}-\ref{fig:S6-pudenzCode-opt-beta} where we plot the success probability as a function of $\gamma$ and $\overline{N}$. We note that for the ${[4,1,4]}_{0}$ code the optimal penalty scales with $\alpha$, i.e., $\gamma_{\text{opt}}\propto \alpha$. Lower values of $\gamma_{\text{opt}}$ are observed on the {S6 device}. For the  ${[3,1,3]}_{1}$ code, the optimal $\gamma$ is around $\gamma\approx0.2-0.3$ for all $\alpha$ values studied, and the optimal values are unchanged across the two devices.
\section{Comparing decoding strategies}
\label{sec:comp-decod-strat}
\begin{figure*}
  \centering
  \begin{subfigure}[t]{\textwidth}
      \includegraphics[width=\textwidth]{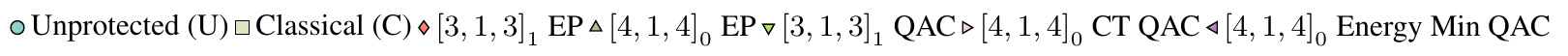}
  \end{subfigure}
  \begin{subfigure}[t]{0.45\textwidth}
      \includegraphics[width=\textwidth]{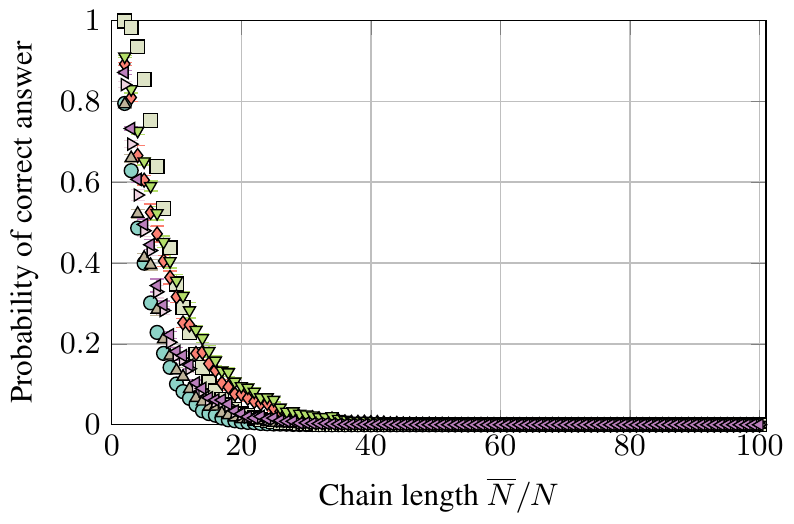}
      \caption{$\alpha=0.1$}
  \end{subfigure}\;\;
  \begin{subfigure}[t]{0.45\textwidth}
      \includegraphics[width=\textwidth]{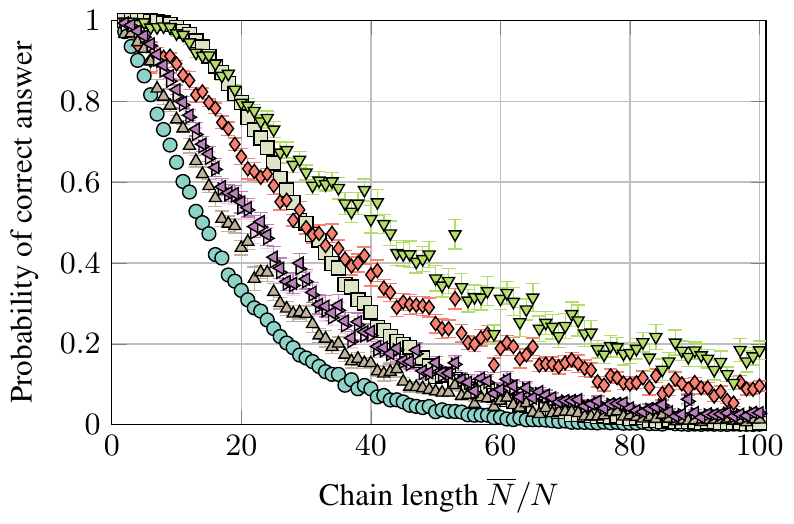}
      \caption{$\alpha=0.3$}
  \end{subfigure}
  \begin{subfigure}[t]{0.45\textwidth}
      \includegraphics[width=\textwidth]{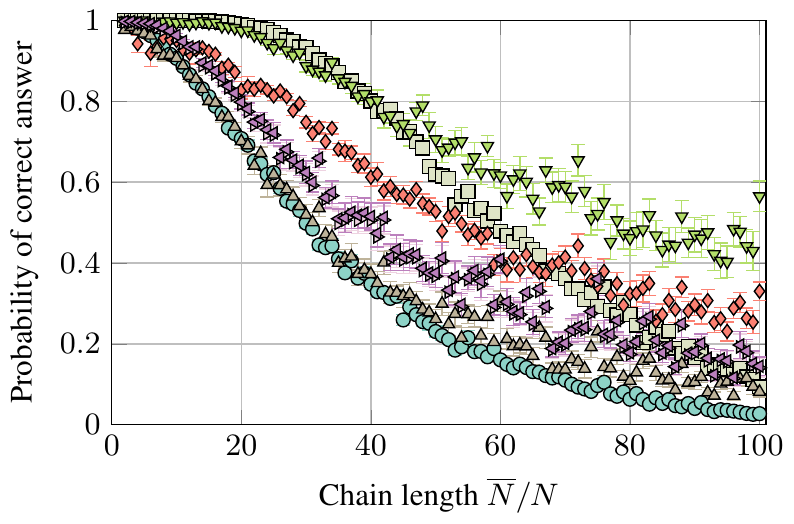}
      \caption{$\alpha=0.5$}
  \end{subfigure}\;\;
  \begin{subfigure}[t]{0.45\textwidth}
      \includegraphics[width=\textwidth]{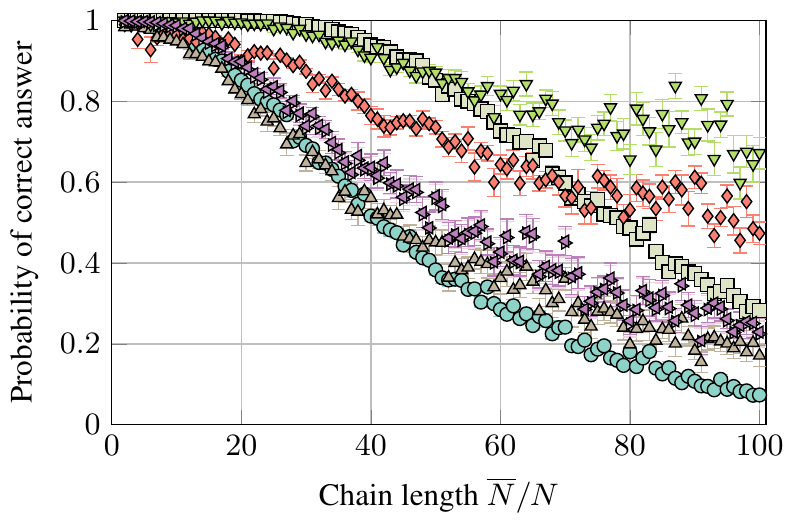}
      \caption{$\alpha=0.7$}
  \end{subfigure}
  \begin{subfigure}[t]{0.45\textwidth}
      \includegraphics[width=\textwidth]{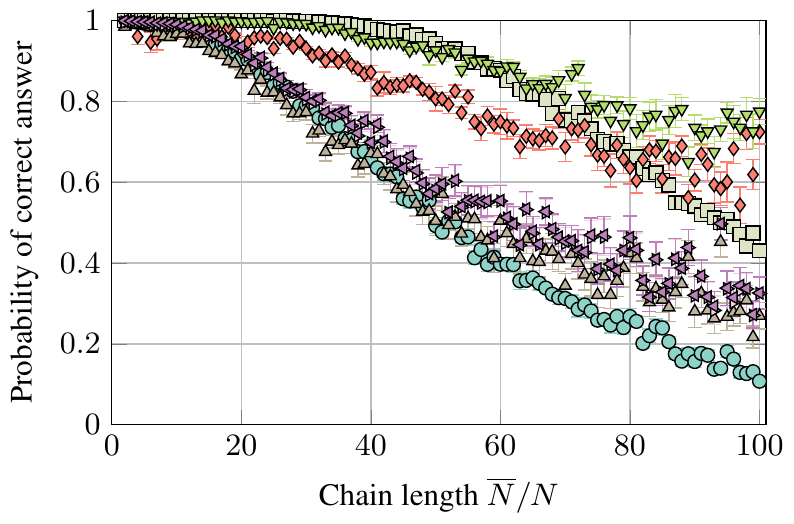}
      \caption{$\alpha=0.9$}
  \end{subfigure}\;\;
  \begin{subfigure}[t]{0.45\textwidth}
      \includegraphics[width=\textwidth]{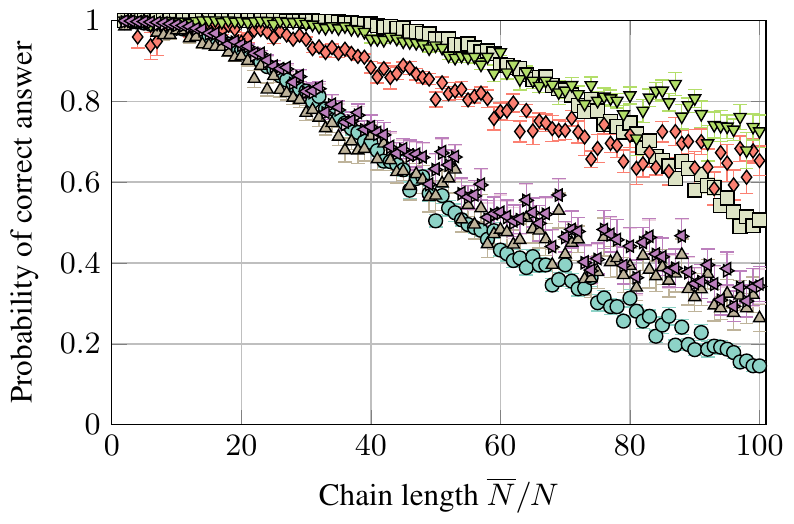}
      \caption{$\alpha=1.0$}
  \end{subfigure}
  \caption{\textbf{Performance comparison for the DW2-ISI device.} The ${[4,1,4]}_{0}$ code starts to outperform the C strategy below $\alpha\approx0.5$ and for sufficiently long chains.  }
  \label{fig:ISI-decod-strat}
\end{figure*}

\begin{figure*}
  \centering
  \begin{subfigure}[t]{\textwidth}
      \includegraphics[width=\textwidth]{allPlotsStrat-Legend.pdf}
  \end{subfigure}
  \begin{subfigure}[t]{0.45\textwidth}
      \includegraphics[width=\textwidth]{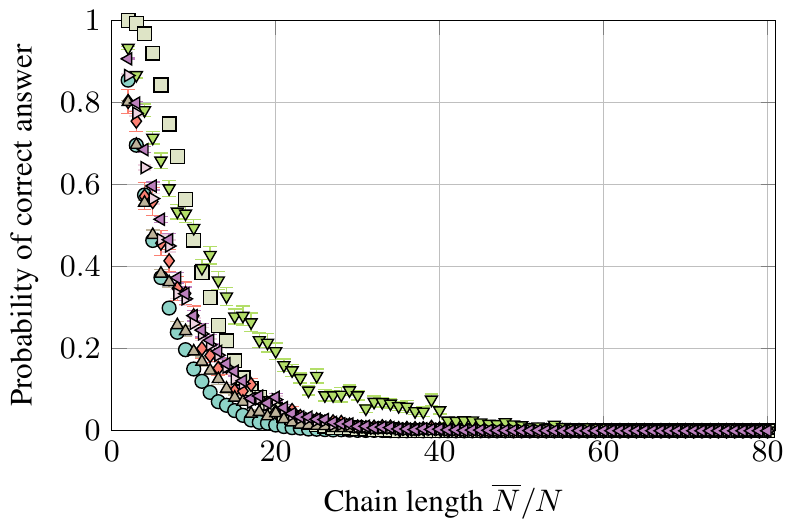}
      \caption{$\alpha=0.1$}
  \end{subfigure}\;\;
  \begin{subfigure}[t]{0.45\textwidth}
      \includegraphics[width=\textwidth]{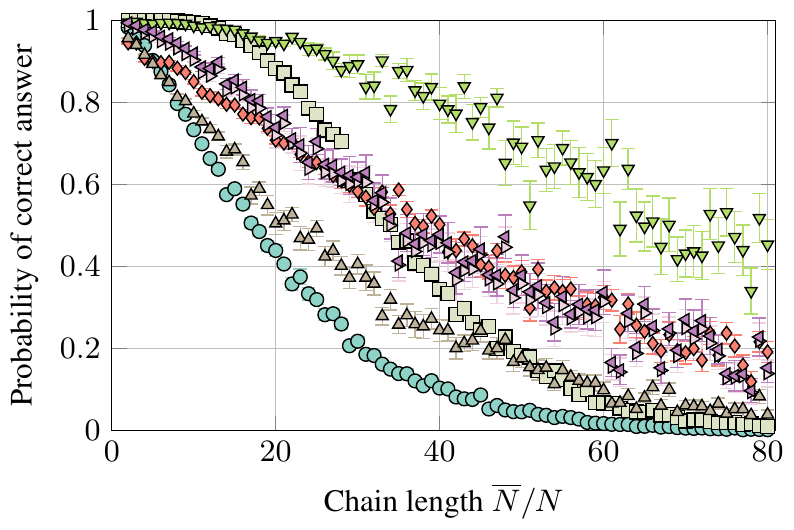}
      \caption{$\alpha=0.3$}
  \end{subfigure}
  \begin{subfigure}[t]{0.45\textwidth}
      \includegraphics[width=\textwidth]{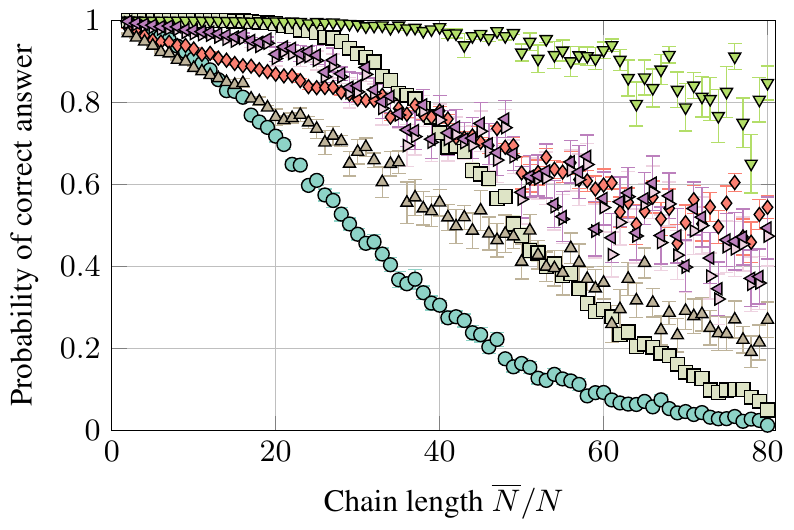}
      \caption{$\alpha=0.5$}
  \end{subfigure}\;\;
  \begin{subfigure}[t]{0.45\textwidth}
      \includegraphics[width=\textwidth]{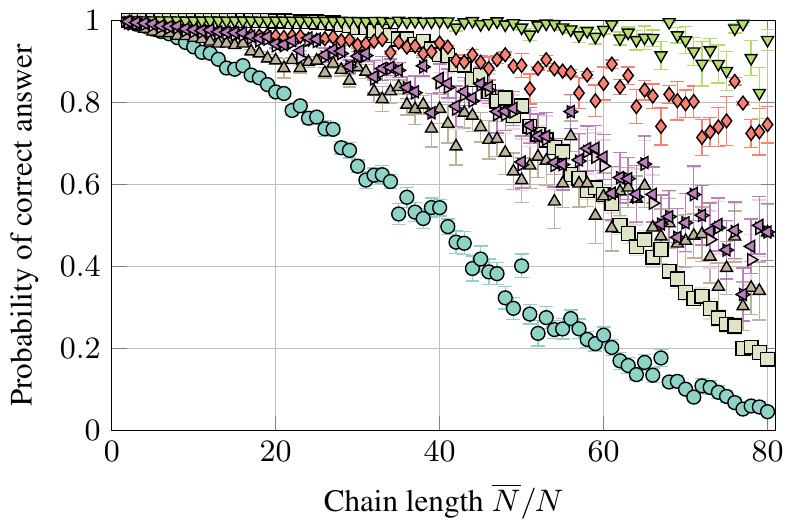}
      \caption{$\alpha=0.7$}
  \end{subfigure}
  \begin{subfigure}[t]{0.45\textwidth}
      \includegraphics[width=\textwidth]{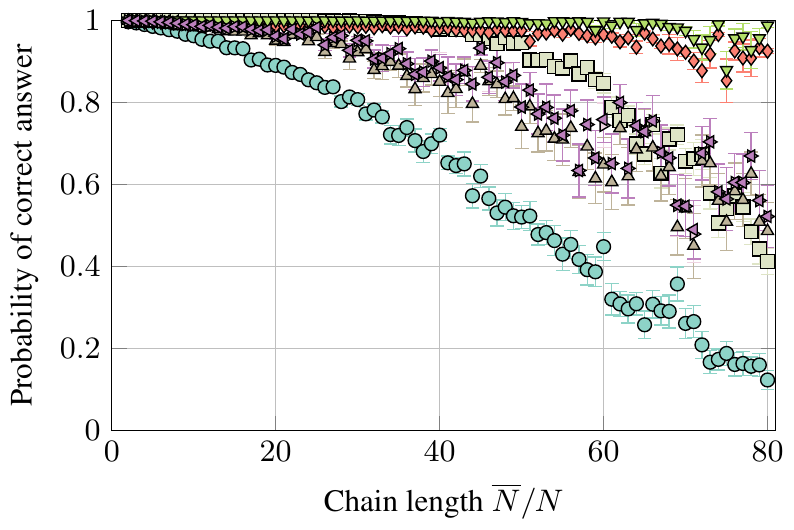}
      \caption{$\alpha=0.9$}
  \end{subfigure}\;\;
  \begin{subfigure}[t]{0.45\textwidth}
      \includegraphics[width=\textwidth]{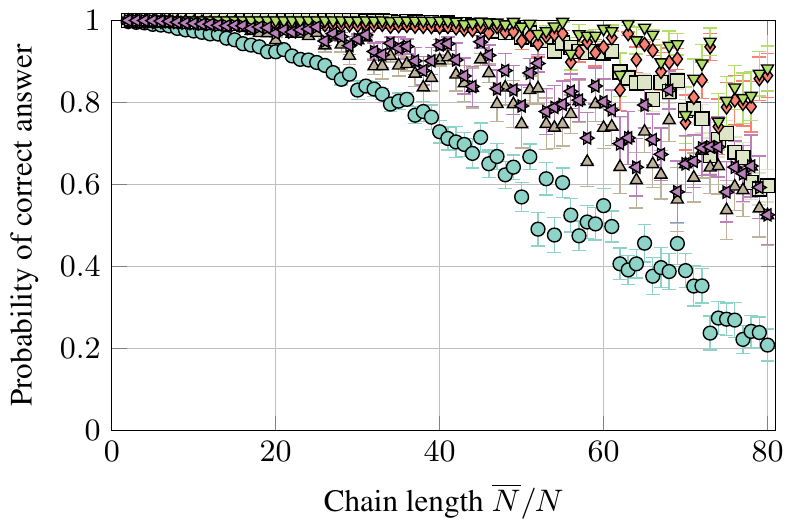}
      \caption{$\alpha=1.0$}
  \end{subfigure}
  \caption{\textbf{Performance comparison for the S6 device.} The ${[4,1,4]}_{0}$ code starts to outperform the C strategy below $\alpha\approx0.9$ and for sufficiently long chains. }
  \label{fig:S6-decod-strat}
\end{figure*}

In the main text we compared four strategies: U, C, the ${[4,1,4]}_{0}$ code, and ${[3,1,3]}_{1}$ code.  We also used different decoding strategies: EM, EP, and CT. Figure~\ref{fig:ISI-decod-strat} and Fig.~\ref{fig:S6-decod-strat} show all these strategies for a few chosen values of the scaling parameter $\alpha$ for the {DW2-ISI} and {S6} devices, respectively. The U strategy is always worst. The ${[3,1,3]}_{1}$ code can be seen to outperform all other strategies at each $\alpha$ value for sufficiently long chains. The ${[4,1,4]}_{0}$ code outperforms the C strategy below a device-dependent $\alpha$ value and for sufficiently long chains. The fact that the success probabilities of the CT and EM strategies are nearly equal suggests that there are very few tied qubits in the ${[4,1,4]}_{0}$-encoded chains, an observation that holds for both devices.

\begin{figure*}
\centering
\begin{subfigure}[t]{\textwidth}
\centering
\includegraphics[width=.5\textwidth]{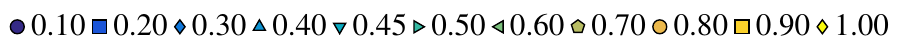}
\end{subfigure}\\
\begin{subfigure}[t]{0.5\textwidth}
\includegraphics[width=\textwidth]{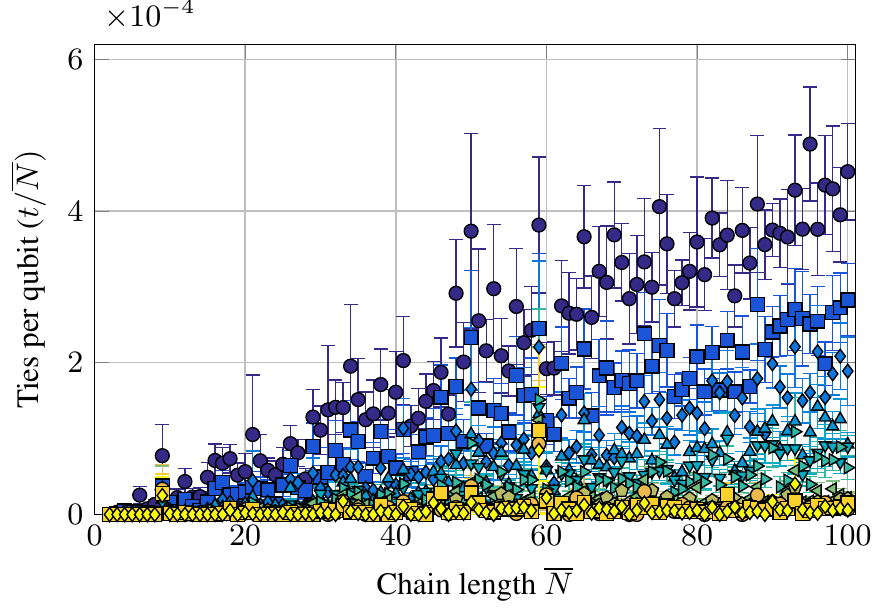}
\end{subfigure}
\caption{\textbf{Ties in the ${[4,1,4]}_{0}$ code.} The number of ties per qubit is shown for each value of $\alpha \in [0.1,1.0]$, minimized over the penalty strength $\gamma$. The number of ties generally increases with  the chain length $\overline{N}$ and decreases with $\alpha$. The largest number of ties per qubit is $\sim 5\times 10^{-4}$.}
\label{fig:ties}
\end{figure*}

In the main text we also presented indirect evidence for the small number of ties in the the ${[4,1,4]}_{0}$ code. Figure~\ref{fig:ties} shows this directly.

%%Extra figures
\balancecolsandclearpage
%Annealer comparison table
%Chimera graphs
\begin{table*}[b]
\begin{tabular}{|c|c|c|c|c|}
\hline
Annealer & Number of working qubits & Temperature(mK) & $M_{\text{AFM}}$ (pH) & $1/f$ Amplitude \\
\hline
DW2-ISI & $504$ & $16\pm1$ & $1.33$ & $7.5\pm1$ \\
S6 & $476$ & $11\pm1$ & $1.92$ & $4.1\pm0.3$ \\
\hline
\end{tabular}
\caption{Physical parameter of the two quantum annealing devices used in our study. Both devices belong to the same Vesuvius generation, with the major difference being a lower temperature and lower noise on the S6 device, and a higher qubit yield on the DW2-ISI device. $M_{\text{AFM}}$ is the inter-qubit coupling energy when a coupler is set to provide the maximum antiferromagnetic (AFM) coupling. $1/f$ is the low frequency flux noise in units of flux quanta, $\Phi_{0}={h}/{2e}$, where $h$ is the Planck constant and $e$ is the electron charge. Details about these physical parameters can be found in Ref.~\cite{Harris:2010kx}.}
\label{tab:annealerComp}
\end{table*}
%Square code construction
%Square code encoded ISI graph
%Pudenz code construction
%Pudenz code encoded ISI graph
%
%
%DW2-ISI results
%
%System 6 results
%
%Figure, S6/ISI comparison.
%
%Figure, thermodynamical comparison of two codes.

\begin{thebibliography}{48}%
\makeatletter
\providecommand \@ifxundefined [1]{%
 \@ifx{#1\undefined}
}%
\providecommand \@ifnum [1]{%
 \ifnum #1\expandafter \@firstoftwo
 \else \expandafter \@secondoftwo
 \fi
}%
\providecommand \@ifx [1]{%
 \ifx #1\expandafter \@firstoftwo
 \else \expandafter \@secondoftwo
 \fi
}%
\providecommand \natexlab [1]{#1}%
\providecommand \enquote  [1]{``#1''}%
\providecommand \bibnamefont  [1]{#1}%
\providecommand \bibfnamefont [1]{#1}%
\providecommand \citenamefont [1]{#1}%
\providecommand \href@noop [0]{\@secondoftwo}%
\providecommand \href [0]{\begingroup \@sanitize@url \@href}%
\providecommand \@href[1]{\@@startlink{#1}\@@href}%
\providecommand \@@href[1]{\endgroup#1\@@endlink}%
\providecommand \@sanitize@url [0]{\catcode `\\12\catcode `\$12\catcode
  `\&12\catcode `\#12\catcode `\^12\catcode `\_12\catcode `\%12\relax}%
\providecommand \@@startlink[1]{}%
\providecommand \@@endlink[0]{}%
\providecommand \url  [0]{\begingroup\@sanitize@url \@url }%
\providecommand \@url [1]{\endgroup\@href {#1}{\urlprefix }}%
\providecommand \urlprefix  [0]{URL }%
\providecommand \Eprint [0]{\href }%
\providecommand \doibase [0]{http://dx.doi.org/}%
\providecommand \selectlanguage [0]{\@gobble}%
\providecommand \bibinfo  [0]{\@secondoftwo}%
\providecommand \bibfield  [0]{\@secondoftwo}%
\providecommand \translation [1]{[#1]}%
\providecommand \BibitemOpen [0]{}%
\providecommand \bibitemStop [0]{}%
\providecommand \bibitemNoStop [0]{.\EOS\space}%
\providecommand \EOS [0]{\spacefactor3000\relax}%
\providecommand \BibitemShut  [1]{\csname bibitem#1\endcsname}%
\let\auto@bib@innerbib\@empty
%</preamble>
\bibitem [{\citenamefont {Kelly}\ \emph {et~al.}(2015)\citenamefont {Kelly},
  \citenamefont {Barends}, \citenamefont {Fowler}, \citenamefont {Megrant},
  \citenamefont {Jeffrey}, \citenamefont {White}, \citenamefont {Sank},
  \citenamefont {Mutus}, \citenamefont {Campbell}, \citenamefont {Chen},
  \citenamefont {Chen}, \citenamefont {Chiaro}, \citenamefont {Dunsworth},
  \citenamefont {Hoi}, \citenamefont {Neill}, \citenamefont {O/'Malley},
  \citenamefont {Quintana}, \citenamefont {Roushan}, \citenamefont
  {Vainsencher}, \citenamefont {Wenner}, \citenamefont {Cleland},\ and\
  \citenamefont {Martinis}}]{Kelly:2015fv}%
  \BibitemOpen
  \bibfield  {author} {\bibinfo {author} {\bibfnamefont {J.}~\bibnamefont
  {Kelly}}, \bibinfo {author} {\bibfnamefont {R.}~\bibnamefont {Barends}},
  \bibinfo {author} {\bibfnamefont {A.~G.}\ \bibnamefont {Fowler}}, \bibinfo
  {author} {\bibfnamefont {A.}~\bibnamefont {Megrant}}, \bibinfo {author}
  {\bibfnamefont {E.}~\bibnamefont {Jeffrey}}, \bibinfo {author} {\bibfnamefont
  {T.~C.}\ \bibnamefont {White}}, \bibinfo {author} {\bibfnamefont
  {D.}~\bibnamefont {Sank}}, \bibinfo {author} {\bibfnamefont {J.~Y.}\
  \bibnamefont {Mutus}}, \bibinfo {author} {\bibfnamefont {B.}~\bibnamefont
  {Campbell}}, \bibinfo {author} {\bibfnamefont {Y.}~\bibnamefont {Chen}},
  \bibinfo {author} {\bibfnamefont {Z.}~\bibnamefont {Chen}}, \bibinfo {author}
  {\bibfnamefont {B.}~\bibnamefont {Chiaro}}, \bibinfo {author} {\bibfnamefont
  {A.}~\bibnamefont {Dunsworth}}, \bibinfo {author} {\bibfnamefont {I.~C.}\
  \bibnamefont {Hoi}}, \bibinfo {author} {\bibfnamefont {C.}~\bibnamefont
  {Neill}}, \bibinfo {author} {\bibfnamefont {P.~J.~J.}\ \bibnamefont
  {O/'Malley}}, \bibinfo {author} {\bibfnamefont {C.}~\bibnamefont {Quintana}},
  \bibinfo {author} {\bibfnamefont {P.}~\bibnamefont {Roushan}}, \bibinfo
  {author} {\bibfnamefont {A.}~\bibnamefont {Vainsencher}}, \bibinfo {author}
  {\bibfnamefont {J.}~\bibnamefont {Wenner}}, \bibinfo {author} {\bibfnamefont
  {A.~N.}\ \bibnamefont {Cleland}}, \ and\ \bibinfo {author} {\bibfnamefont
  {J.~M.}\ \bibnamefont {Martinis}},\ }\href
  {http://dx.doi.org/10.1038/nature14270} {\bibfield  {journal} {\bibinfo
  {journal} {Nature}\ }\textbf {\bibinfo {volume} {519}},\ \bibinfo {pages}
  {66} (\bibinfo {year} {2015})}\BibitemShut {NoStop}%
\bibitem [{\citenamefont {Corcoles}\ \emph {et~al.}(2015)\citenamefont
  {Corcoles}, \citenamefont {Magesan}, \citenamefont {Srinivasan},
  \citenamefont {Cross}, \citenamefont {Steffen}, \citenamefont {Gambetta},\
  and\ \citenamefont {Chow}}]{Corcoles:2015zl}%
  \BibitemOpen
  \bibfield  {author} {\bibinfo {author} {\bibfnamefont {A.~D.}\ \bibnamefont
  {Corcoles}}, \bibinfo {author} {\bibfnamefont {E.}~\bibnamefont {Magesan}},
  \bibinfo {author} {\bibfnamefont {S.~J.}\ \bibnamefont {Srinivasan}},
  \bibinfo {author} {\bibfnamefont {A.~W.}\ \bibnamefont {Cross}}, \bibinfo
  {author} {\bibfnamefont {M.}~\bibnamefont {Steffen}}, \bibinfo {author}
  {\bibfnamefont {J.~M.}\ \bibnamefont {Gambetta}}, \ and\ \bibinfo {author}
  {\bibfnamefont {J.~M.}\ \bibnamefont {Chow}},\ }\href
  {http://dx.doi.org/10.1038/ncomms7979} {\bibfield  {journal} {\bibinfo
  {journal} {Nat Commun}\ }\textbf {\bibinfo {volume} {6}} (\bibinfo {year}
  {2015})}\BibitemShut {NoStop}%
\bibitem [{\citenamefont {Lloyd}(1996)}]{Lloyd:96}%
  \BibitemOpen
  \bibfield  {author} {\bibinfo {author} {\bibfnamefont {S.}~\bibnamefont
  {Lloyd}},\ }\href {http://www.sciencemag.org/content/273/5278/1073.abstract
  N2 - Feynman's 1982 conjecture, that quantum computers can be programmed to
  simulate any local quantum system, is shown to be correct.} {\bibfield
  {journal} {\bibinfo  {journal} {Science}\ }\textbf {\bibinfo {volume}
  {273}},\ \bibinfo {pages} {1073} (\bibinfo {year} {1996})}\BibitemShut
  {NoStop}%
\bibitem [{\citenamefont {Buluta}\ and\ \citenamefont
  {Nori}(2009)}]{Buluta:2009fk}%
  \BibitemOpen
  \bibfield  {author} {\bibinfo {author} {\bibfnamefont {I.}~\bibnamefont
  {Buluta}}\ and\ \bibinfo {author} {\bibfnamefont {F.}~\bibnamefont {Nori}},\
  }\href {http://www.sciencemag.org/content/326/5949/108.abstract N2 - Quantum
  simulators are controllable quantum systems that can be used to simulate
  other quantum systems. Being able to tackle problems that are intractable on
  classical computers, quantum simulators would provide a means of exploring
  new physical phenomena. We present an overview of how quantum simulators may
  become a reality in the near future as the required technologies are now
  within reach. Quantum simulators, relying on the coherent control of neutral
  atoms, ions, photons, or electrons, would allow studying problems in various
  fields including condensed-matter physics, high-energy physics, cosmology,
  atomic physics, and quantum chemistry.} {\bibfield  {journal} {\bibinfo
  {journal} {Science}\ }\textbf {\bibinfo {volume} {326}},\ \bibinfo {pages}
  {108} (\bibinfo {year} {2009})}\BibitemShut {NoStop}%
\bibitem [{\citenamefont {Barreiro}\ \emph {et~al.}(2011)\citenamefont
  {Barreiro}, \citenamefont {Muller}, \citenamefont {Schindler}, \citenamefont
  {Nigg}, \citenamefont {Monz}, \citenamefont {Chwalla}, \citenamefont
  {Hennrich}, \citenamefont {Roos}, \citenamefont {Zoller},\ and\ \citenamefont
  {Blatt}}]{barreiro2011open}%
  \BibitemOpen
  \bibfield  {author} {\bibinfo {author} {\bibfnamefont {J.~T.}\ \bibnamefont
  {Barreiro}}, \bibinfo {author} {\bibfnamefont {M.}~\bibnamefont {Muller}},
  \bibinfo {author} {\bibfnamefont {P.}~\bibnamefont {Schindler}}, \bibinfo
  {author} {\bibfnamefont {D.}~\bibnamefont {Nigg}}, \bibinfo {author}
  {\bibfnamefont {T.}~\bibnamefont {Monz}}, \bibinfo {author} {\bibfnamefont
  {M.}~\bibnamefont {Chwalla}}, \bibinfo {author} {\bibfnamefont
  {M.}~\bibnamefont {Hennrich}}, \bibinfo {author} {\bibfnamefont {C.~F.}\
  \bibnamefont {Roos}}, \bibinfo {author} {\bibfnamefont {P.}~\bibnamefont
  {Zoller}}, \ and\ \bibinfo {author} {\bibfnamefont {R.}~\bibnamefont
  {Blatt}},\ }\href {dx.doi.org/10.1038/nature09801} {\bibfield  {journal}
  {\bibinfo  {journal} {Nature}\ }\textbf {\bibinfo {volume} {470}},\ \bibinfo
  {pages} {486} (\bibinfo {year} {2011})}\BibitemShut {NoStop}%
\bibitem [{\citenamefont {Finnila}\ \emph {et~al.}(1994)\citenamefont
  {Finnila}, \citenamefont {Gomez}, \citenamefont {Sebenik}, \citenamefont
  {Stenson},\ and\ \citenamefont {Doll}}]{finnila_quantum_1994}%
  \BibitemOpen
  \bibfield  {author} {\bibinfo {author} {\bibfnamefont {A.~B.}\ \bibnamefont
  {Finnila}}, \bibinfo {author} {\bibfnamefont {M.~A.}\ \bibnamefont {Gomez}},
  \bibinfo {author} {\bibfnamefont {C.}~\bibnamefont {Sebenik}}, \bibinfo
  {author} {\bibfnamefont {C.}~\bibnamefont {Stenson}}, \ and\ \bibinfo
  {author} {\bibfnamefont {J.~D.}\ \bibnamefont {Doll}},\ }\href {\doibase
  10.1016/0009-2614(94)00117-0} {\bibfield  {journal} {\bibinfo  {journal}
  {Chemical Physics Letters}\ }\textbf {\bibinfo {volume} {219}},\ \bibinfo
  {pages} {343} (\bibinfo {year} {1994})}\BibitemShut {NoStop}%
\bibitem [{\citenamefont {Kadowaki}\ and\ \citenamefont
  {Nishimori}(1998)}]{kadowaki_quantum_1998}%
  \BibitemOpen
  \bibfield  {author} {\bibinfo {author} {\bibfnamefont {T.}~\bibnamefont
  {Kadowaki}}\ and\ \bibinfo {author} {\bibfnamefont {H.}~\bibnamefont
  {Nishimori}},\ }\href {\doibase 10.1103/PhysRevE.58.5355} {\bibfield
  {journal} {\bibinfo  {journal} {Phys. Rev. E}\ }\textbf {\bibinfo {volume}
  {58}},\ \bibinfo {pages} {5355} (\bibinfo {year} {1998})}\BibitemShut
  {NoStop}%
\bibitem [{\citenamefont {Brooke}\ \emph {et~al.}(1999)\citenamefont {Brooke},
  \citenamefont {Bitko}, \citenamefont {F.}, \citenamefont {Rosenbaum},\ and\
  \citenamefont {Aeppli}}]{Brooke1999}%
  \BibitemOpen
  \bibfield  {author} {\bibinfo {author} {\bibfnamefont {J.}~\bibnamefont
  {Brooke}}, \bibinfo {author} {\bibfnamefont {D.}~\bibnamefont {Bitko}},
  \bibinfo {author} {\bibfnamefont {T.}~\bibnamefont {F.}}, \bibinfo {author}
  {\bibnamefont {Rosenbaum}}, \ and\ \bibinfo {author} {\bibfnamefont
  {G.}~\bibnamefont {Aeppli}},\ }\href {\doibase 10.1126/science.284.5415.779}
  {\bibfield  {journal} {\bibinfo  {journal} {Science}\ }\textbf {\bibinfo
  {volume} {284}},\ \bibinfo {pages} {779} (\bibinfo {year}
  {1999})}\BibitemShut {NoStop}%
\bibitem [{\citenamefont {Brooke}\ \emph {et~al.}(2001)\citenamefont {Brooke},
  \citenamefont {Rosenbaum},\ and\ \citenamefont
  {Aeppli}}]{brooke_tunable_2001}%
  \BibitemOpen
  \bibfield  {author} {\bibinfo {author} {\bibfnamefont {J.}~\bibnamefont
  {Brooke}}, \bibinfo {author} {\bibfnamefont {T.~F.}\ \bibnamefont
  {Rosenbaum}}, \ and\ \bibinfo {author} {\bibfnamefont {G.}~\bibnamefont
  {Aeppli}},\ }\href {\doibase 10.1038/35098037} {\bibfield  {journal}
  {\bibinfo  {journal} {Nature}\ }\textbf {\bibinfo {volume} {413}},\ \bibinfo
  {pages} {610} (\bibinfo {year} {2001})}\BibitemShut {NoStop}%
\bibitem [{\citenamefont {{Kaminsky}}\ and\ \citenamefont
  {{Lloyd}}(2004)}]{2002quant.ph.11152K}%
  \BibitemOpen
  \bibfield  {author} {\bibinfo {author} {\bibfnamefont {W.~M.}\ \bibnamefont
  {{Kaminsky}}}\ and\ \bibinfo {author} {\bibfnamefont {S.}~\bibnamefont
  {{Lloyd}}},\ }in\ \href {http://arxiv.org/abs/quant-ph/0211152} {\emph
  {\bibinfo {booktitle} {Quantum Computing and Quantum Bits in Mesoscopic
  Systems}}},\ \bibinfo {editor} {edited by\ \bibinfo {editor} {\bibfnamefont
  {A.}~\bibnamefont {Leggett}}, \bibinfo {editor} {\bibfnamefont
  {B.}~\bibnamefont {Ruggiero}}, \ and\ \bibinfo {editor} {\bibfnamefont
  {P.}~\bibnamefont {Silvestrini}}}\ (\bibinfo  {publisher} {Kluwer
  Academic/Plenum Publ.},\ \bibinfo {year} {2004})\ \Eprint
  {http://arxiv.org/abs/arXiv:quant-ph/0211152} {arXiv:quant-ph/0211152}
  \BibitemShut {NoStop}%
\bibitem [{\citenamefont {Johnson}\ \emph {et~al.}(2011)\citenamefont
  {Johnson}, \citenamefont {Amin}, \citenamefont {Gildert}, \citenamefont
  {Lanting}, \citenamefont {Hamze}, \citenamefont {Dickson}, \citenamefont
  {Harris}, \citenamefont {Berkley}, \citenamefont {Johansson}, \citenamefont
  {Bunyk}, \citenamefont {Chapple}, \citenamefont {Enderud}, \citenamefont
  {Hilton}, \citenamefont {Karimi}, \citenamefont {Ladizinsky}, \citenamefont
  {Ladizinsky}, \citenamefont {Oh}, \citenamefont {Perminov}, \citenamefont
  {Rich}, \citenamefont {Thom}, \citenamefont {Tolkacheva}, \citenamefont
  {Truncik}, \citenamefont {Uchaikin}, \citenamefont {Wang}, \citenamefont
  {Wilson},\ and\ \citenamefont {Rose}}]{Dwave}%
  \BibitemOpen
  \bibfield  {author} {\bibinfo {author} {\bibfnamefont {M.~W.}\ \bibnamefont
  {Johnson}}, \bibinfo {author} {\bibfnamefont {M.~H.~S.}\ \bibnamefont
  {Amin}}, \bibinfo {author} {\bibfnamefont {S.}~\bibnamefont {Gildert}},
  \bibinfo {author} {\bibfnamefont {T.}~\bibnamefont {Lanting}}, \bibinfo
  {author} {\bibfnamefont {F.}~\bibnamefont {Hamze}}, \bibinfo {author}
  {\bibfnamefont {N.}~\bibnamefont {Dickson}}, \bibinfo {author} {\bibfnamefont
  {R.}~\bibnamefont {Harris}}, \bibinfo {author} {\bibfnamefont {A.~J.}\
  \bibnamefont {Berkley}}, \bibinfo {author} {\bibfnamefont {J.}~\bibnamefont
  {Johansson}}, \bibinfo {author} {\bibfnamefont {P.}~\bibnamefont {Bunyk}},
  \bibinfo {author} {\bibfnamefont {E.~M.}\ \bibnamefont {Chapple}}, \bibinfo
  {author} {\bibfnamefont {C.}~\bibnamefont {Enderud}}, \bibinfo {author}
  {\bibfnamefont {J.~P.}\ \bibnamefont {Hilton}}, \bibinfo {author}
  {\bibfnamefont {K.}~\bibnamefont {Karimi}}, \bibinfo {author} {\bibfnamefont
  {E.}~\bibnamefont {Ladizinsky}}, \bibinfo {author} {\bibfnamefont
  {N.}~\bibnamefont {Ladizinsky}}, \bibinfo {author} {\bibfnamefont
  {T.}~\bibnamefont {Oh}}, \bibinfo {author} {\bibfnamefont {I.}~\bibnamefont
  {Perminov}}, \bibinfo {author} {\bibfnamefont {C.}~\bibnamefont {Rich}},
  \bibinfo {author} {\bibfnamefont {M.~C.}\ \bibnamefont {Thom}}, \bibinfo
  {author} {\bibfnamefont {E.}~\bibnamefont {Tolkacheva}}, \bibinfo {author}
  {\bibfnamefont {C.~J.~S.}\ \bibnamefont {Truncik}}, \bibinfo {author}
  {\bibfnamefont {S.}~\bibnamefont {Uchaikin}}, \bibinfo {author}
  {\bibfnamefont {J.}~\bibnamefont {Wang}}, \bibinfo {author} {\bibfnamefont
  {B.}~\bibnamefont {Wilson}}, \ and\ \bibinfo {author} {\bibfnamefont
  {G.}~\bibnamefont {Rose}},\ }\href {\doibase 10.1038/nature10012} {\bibfield
  {journal} {\bibinfo  {journal} {Nature}\ }\textbf {\bibinfo {volume} {473}},\
  \bibinfo {pages} {194} (\bibinfo {year} {2011})}\BibitemShut {NoStop}%
\bibitem [{\citenamefont {Farhi}\ \emph {et~al.}(2000)\citenamefont {Farhi},
  \citenamefont {Goldstone}, \citenamefont {Gutmann},\ and\ \citenamefont
  {Sipser}}]{farhi_quantum_2000}%
  \BibitemOpen
  \bibfield  {author} {\bibinfo {author} {\bibfnamefont {E.}~\bibnamefont
  {Farhi}}, \bibinfo {author} {\bibfnamefont {J.}~\bibnamefont {Goldstone}},
  \bibinfo {author} {\bibfnamefont {S.}~\bibnamefont {Gutmann}}, \ and\
  \bibinfo {author} {\bibfnamefont {M.}~\bibnamefont {Sipser}},\ }\href
  {http://arxiv.org/abs/quant-ph/0001106} {\bibfield  {journal} {\bibinfo
  {journal} {arXiv:quant-ph/0001106}\ } (\bibinfo {year} {2000})}\BibitemShut
  {NoStop}%
\bibitem [{\citenamefont {Farhi}\ \emph {et~al.}(2001)\citenamefont {Farhi},
  \citenamefont {Goldstone}, \citenamefont {Gutmann}, \citenamefont {Lapan},
  \citenamefont {Lundgren},\ and\ \citenamefont {Preda}}]{farhi_quantum_2001}%
  \BibitemOpen
  \bibfield  {author} {\bibinfo {author} {\bibfnamefont {E.}~\bibnamefont
  {Farhi}}, \bibinfo {author} {\bibfnamefont {J.}~\bibnamefont {Goldstone}},
  \bibinfo {author} {\bibfnamefont {S.}~\bibnamefont {Gutmann}}, \bibinfo
  {author} {\bibfnamefont {J.}~\bibnamefont {Lapan}}, \bibinfo {author}
  {\bibfnamefont {A.}~\bibnamefont {Lundgren}}, \ and\ \bibinfo {author}
  {\bibfnamefont {D.}~\bibnamefont {Preda}},\ }\href {\doibase
  10.1126/science.1057726} {\bibfield  {journal} {\bibinfo  {journal}
  {Science}\ }\textbf {\bibinfo {volume} {292}},\ \bibinfo {pages} {472}
  (\bibinfo {year} {2001})}\BibitemShut {NoStop}%
\bibitem [{\citenamefont {Barahona}(1982)}]{Barahona1982}%
  \BibitemOpen
  \bibfield  {author} {\bibinfo {author} {\bibfnamefont {F.}~\bibnamefont
  {Barahona}},\ }\href {http://stacks.iop.org/0305-4470/15/i=10/a=028}
  {\bibfield  {journal} {\bibinfo  {journal} {J. Phys. A: Math. Gen}\ }\textbf
  {\bibinfo {volume} {15}},\ \bibinfo {pages} {3241} (\bibinfo {year}
  {1982})}\BibitemShut {NoStop}%
\bibitem [{\citenamefont {Kato}(1950)}]{Kato:50}%
  \BibitemOpen
  \bibfield  {author} {\bibinfo {author} {\bibfnamefont {T.}~\bibnamefont
  {Kato}},\ }\href {\doibase 10.1143/JPSJ.5.435} {\bibfield  {journal}
  {\bibinfo  {journal} {J. Phys. Soc. Jap.}\ }\textbf {\bibinfo {volume} {5}},\
  \bibinfo {pages} {435} (\bibinfo {year} {1950})}\BibitemShut {NoStop}%
\bibitem [{\citenamefont {Jansen}\ \emph {et~al.}(2007)\citenamefont {Jansen},
  \citenamefont {Ruskai},\ and\ \citenamefont {Seiler}}]{Jansen:07}%
  \BibitemOpen
  \bibfield  {author} {\bibinfo {author} {\bibfnamefont {S.}~\bibnamefont
  {Jansen}}, \bibinfo {author} {\bibfnamefont {M.-B.}\ \bibnamefont {Ruskai}},
  \ and\ \bibinfo {author} {\bibfnamefont {R.}~\bibnamefont {Seiler}},\ }\href
  {http://scitation.aip.org/content/aip/journal/jmp/48/10/10.1063/1.2798382}
  {\bibfield  {journal} {\bibinfo  {journal} {J. Math. Phys.}\ }\textbf
  {\bibinfo {volume} {48}},\  (\bibinfo {year} {2007})}\BibitemShut {NoStop}%
\bibitem [{\citenamefont {Lidar}\ \emph {et~al.}(2009)\citenamefont {Lidar},
  \citenamefont {Rezakhani},\ and\ \citenamefont {Hamma}}]{lidar:102106}%
  \BibitemOpen
  \bibfield  {author} {\bibinfo {author} {\bibfnamefont {D.~A.}\ \bibnamefont
  {Lidar}}, \bibinfo {author} {\bibfnamefont {A.~T.}\ \bibnamefont
  {Rezakhani}}, \ and\ \bibinfo {author} {\bibfnamefont {A.}~\bibnamefont
  {Hamma}},\ }\href
  {http://scitation.aip.org/content/aip/journal/jmp/50/10/10.1063/1.3236685}
  {\bibfield  {journal} {\bibinfo  {journal} {J. Math. Phys.}\ }\textbf
  {\bibinfo {volume} {50}},\  (\bibinfo {year} {2009})}\BibitemShut {NoStop}%
\bibitem [{\citenamefont {Childs}\ \emph {et~al.}(2001)\citenamefont {Childs},
  \citenamefont {Farhi},\ and\ \citenamefont
  {Preskill}}]{childs_robustness_2001}%
  \BibitemOpen
  \bibfield  {author} {\bibinfo {author} {\bibfnamefont {A.~M.}\ \bibnamefont
  {Childs}}, \bibinfo {author} {\bibfnamefont {E.}~\bibnamefont {Farhi}}, \
  and\ \bibinfo {author} {\bibfnamefont {J.}~\bibnamefont {Preskill}},\ }\href
  {\doibase 10.1103/PhysRevA.65.012322} {\bibfield  {journal} {\bibinfo
  {journal} {Phys. Rev. A}\ }\textbf {\bibinfo {volume} {65}},\ \bibinfo
  {pages} {012322} (\bibinfo {year} {2001})}\BibitemShut {NoStop}%
\bibitem [{\citenamefont {Sarandy}\ and\ \citenamefont
  {Lidar}(2005)}]{PhysRevLett.95.250503}%
  \BibitemOpen
  \bibfield  {author} {\bibinfo {author} {\bibfnamefont {M.~S.}\ \bibnamefont
  {Sarandy}}\ and\ \bibinfo {author} {\bibfnamefont {D.~A.}\ \bibnamefont
  {Lidar}},\ }\href {http://link.aps.org/doi/10.1103/PhysRevLett.95.250503}
  {\bibfield  {journal} {\bibinfo  {journal} {Phys. Rev. Lett.}\ }\textbf
  {\bibinfo {volume} {95}},\ \bibinfo {pages} {250503} (\bibinfo {year}
  {2005})}\BibitemShut {NoStop}%
\bibitem [{\citenamefont {Aberg}\ \emph {et~al.}(2005)\citenamefont {Aberg},
  \citenamefont {Kult},\ and\ \citenamefont {Sj\"oqvist}}]{Aberg:2005rt}%
  \BibitemOpen
  \bibfield  {author} {\bibinfo {author} {\bibfnamefont {J.}~\bibnamefont
  {Aberg}}, \bibinfo {author} {\bibfnamefont {D.}~\bibnamefont {Kult}}, \ and\
  \bibinfo {author} {\bibfnamefont {E.}~\bibnamefont {Sj\"oqvist}},\ }\href
  {http://link.aps.org/doi/10.1103/PhysRevA.72.042317} {\bibfield  {journal}
  {\bibinfo  {journal} {Phys. Rev. A}\ }\textbf {\bibinfo {volume} {72}},\
  \bibinfo {pages} {042317} (\bibinfo {year} {2005})}\BibitemShut {NoStop}%
\bibitem [{\citenamefont {Roland}\ and\ \citenamefont
  {Cerf}(2005)}]{PhysRevA.71.032330}%
  \BibitemOpen
  \bibfield  {author} {\bibinfo {author} {\bibfnamefont {J.}~\bibnamefont
  {Roland}}\ and\ \bibinfo {author} {\bibfnamefont {N.~J.}\ \bibnamefont
  {Cerf}},\ }\href {\doibase 10.1103/PhysRevA.71.032330} {\bibfield  {journal}
  {\bibinfo  {journal} {Phys. Rev. A}\ }\textbf {\bibinfo {volume} {71}},\
  \bibinfo {pages} {032330} (\bibinfo {year} {2005})}\BibitemShut {NoStop}%
\bibitem [{\citenamefont {Amin}\ \emph {et~al.}(2009)\citenamefont {Amin},
  \citenamefont {Averin},\ and\ \citenamefont
  {Nesteroff}}]{amin_decoherence_2009}%
  \BibitemOpen
  \bibfield  {author} {\bibinfo {author} {\bibfnamefont {M.~H.~S.}\
  \bibnamefont {Amin}}, \bibinfo {author} {\bibfnamefont {D.~V.}\ \bibnamefont
  {Averin}}, \ and\ \bibinfo {author} {\bibfnamefont {J.~A.}\ \bibnamefont
  {Nesteroff}},\ }\href {\doibase 10.1103/PhysRevA.79.022107} {\bibfield
  {journal} {\bibinfo  {journal} {Phys. Rev. A}\ }\textbf {\bibinfo {volume}
  {79}},\ \bibinfo {pages} {022107} (\bibinfo {year} {2009})}\BibitemShut
  {NoStop}%
\bibitem [{\citenamefont {Albash}\ and\ \citenamefont
  {Lidar}(2015)}]{Albash:2015nx}%
  \BibitemOpen
  \bibfield  {author} {\bibinfo {author} {\bibfnamefont {T.}~\bibnamefont
  {Albash}}\ and\ \bibinfo {author} {\bibfnamefont {D.~A.}\ \bibnamefont
  {Lidar}},\ }\href {http://link.aps.org/doi/10.1103/PhysRevA.91.062320}
  {\bibfield  {journal} {\bibinfo  {journal} {Phys. Rev. A}\ }\textbf {\bibinfo
  {volume} {91}},\ \bibinfo {pages} {062320} (\bibinfo {year}
  {2015})}\BibitemShut {NoStop}%
\bibitem [{\citenamefont {Young}\ \emph
  {et~al.}(2013{\natexlab{a}})\citenamefont {Young}, \citenamefont
  {Blume-Kohout},\ and\ \citenamefont {Lidar}}]{Young:2013fk}%
  \BibitemOpen
  \bibfield  {author} {\bibinfo {author} {\bibfnamefont {K.~C.}\ \bibnamefont
  {Young}}, \bibinfo {author} {\bibfnamefont {R.}~\bibnamefont {Blume-Kohout}},
  \ and\ \bibinfo {author} {\bibfnamefont {D.~A.}\ \bibnamefont {Lidar}},\
  }\href {http://link.aps.org/doi/10.1103/PhysRevA.88.062314} {\bibfield
  {journal} {\bibinfo  {journal} {Phys. Rev. A}\ }\textbf {\bibinfo {volume}
  {88}},\ \bibinfo {pages} {062314} (\bibinfo {year}
  {2013}{\natexlab{a}})}\BibitemShut {NoStop}%
\bibitem [{\citenamefont {Lidar}\ and\ \citenamefont
  {Brun}(2013)}]{Lidar-Brun:book}%
  \BibitemOpen
  \bibinfo {editor} {\bibfnamefont {D.}~\bibnamefont {Lidar}}\ and\ \bibinfo
  {editor} {\bibfnamefont {T.}~\bibnamefont {Brun}},\ eds.,\ \href
  {http://www.cambridge.org/9780521897877} {\emph {\bibinfo {title} {Quantum
  Error Correction}}}\ (\bibinfo  {publisher} {Cambridge University Press},\
  \bibinfo {address} {{Cambridge, UK}},\ \bibinfo {year} {2013})\BibitemShut
  {NoStop}%
\bibitem [{\citenamefont {Jordan}\ \emph {et~al.}(2006)\citenamefont {Jordan},
  \citenamefont {Farhi},\ and\ \citenamefont {Shor}}]{jordan2006error}%
  \BibitemOpen
  \bibfield  {author} {\bibinfo {author} {\bibfnamefont {S.~P.}\ \bibnamefont
  {Jordan}}, \bibinfo {author} {\bibfnamefont {E.}~\bibnamefont {Farhi}}, \
  and\ \bibinfo {author} {\bibfnamefont {P.~W.}\ \bibnamefont {Shor}},\ }\href
  {http://link.aps.org/doi/10.1103/PhysRevA.74.052322} {\bibfield  {journal}
  {\bibinfo  {journal} {{Phys. Rev. A}}\ }\textbf {\bibinfo {volume} {74}},\
  \bibinfo {pages} {052322} (\bibinfo {year} {2006})}\BibitemShut {NoStop}%
\bibitem [{\citenamefont {Lidar}(2008)}]{PhysRevLett.100.160506}%
  \BibitemOpen
  \bibfield  {author} {\bibinfo {author} {\bibfnamefont {D.~A.}\ \bibnamefont
  {Lidar}},\ }\href {http://link.aps.org/doi/10.1103/PhysRevLett.100.160506}
  {\bibfield  {journal} {\bibinfo  {journal} {{Phys.~Rev.~Lett.}}\ }\textbf
  {\bibinfo {volume} {100}},\ \bibinfo {pages} {160506} (\bibinfo {year}
  {2008})}\BibitemShut {NoStop}%
\bibitem [{\citenamefont {Quiroz}\ and\ \citenamefont
  {Lidar}(2012)}]{PhysRevA.86.042333}%
  \BibitemOpen
  \bibfield  {author} {\bibinfo {author} {\bibfnamefont {G.}~\bibnamefont
  {Quiroz}}\ and\ \bibinfo {author} {\bibfnamefont {D.~A.}\ \bibnamefont
  {Lidar}},\ }\href {\doibase 10.1103/PhysRevA.86.042333} {\bibfield  {journal}
  {\bibinfo  {journal} {Phys. Rev. A}\ }\textbf {\bibinfo {volume} {86}},\
  \bibinfo {pages} {042333} (\bibinfo {year} {2012})}\BibitemShut {NoStop}%
\bibitem [{\citenamefont {Ganti}\ \emph {et~al.}(2014)\citenamefont {Ganti},
  \citenamefont {Onunkwo},\ and\ \citenamefont {Young}}]{Ganti:13}%
  \BibitemOpen
  \bibfield  {author} {\bibinfo {author} {\bibfnamefont {A.}~\bibnamefont
  {Ganti}}, \bibinfo {author} {\bibfnamefont {U.}~\bibnamefont {Onunkwo}}, \
  and\ \bibinfo {author} {\bibfnamefont {K.}~\bibnamefont {Young}},\ }\href
  {http://link.aps.org/doi/10.1103/PhysRevA.89.042313} {\bibfield  {journal}
  {\bibinfo  {journal} {Phys. Rev. A}\ }\textbf {\bibinfo {volume} {89}},\
  \bibinfo {pages} {042313} (\bibinfo {year} {2014})}\BibitemShut {NoStop}%
\bibitem [{\citenamefont {Bookatz}\ \emph {et~al.}(2015)\citenamefont
  {Bookatz}, \citenamefont {Farhi},\ and\ \citenamefont
  {Zhou}}]{Bookatz:2014uq}%
  \BibitemOpen
  \bibfield  {author} {\bibinfo {author} {\bibfnamefont {A.~D.}\ \bibnamefont
  {Bookatz}}, \bibinfo {author} {\bibfnamefont {E.}~\bibnamefont {Farhi}}, \
  and\ \bibinfo {author} {\bibfnamefont {L.}~\bibnamefont {Zhou}},\ }\href
  {http://link.aps.org/doi/10.1103/PhysRevA.92.022317} {\bibfield  {journal}
  {\bibinfo  {journal} {Physical Review A}\ }\textbf {\bibinfo {volume} {92}},\
  \bibinfo {pages} {022317} (\bibinfo {year} {2015})}\BibitemShut {NoStop}%
\bibitem [{\citenamefont {Young}\ \emph
  {et~al.}(2013{\natexlab{b}})\citenamefont {Young}, \citenamefont {Sarovar},\
  and\ \citenamefont {Blume-Kohout}}]{Young:13}%
  \BibitemOpen
  \bibfield  {author} {\bibinfo {author} {\bibfnamefont {K.~C.}\ \bibnamefont
  {Young}}, \bibinfo {author} {\bibfnamefont {M.}~\bibnamefont {Sarovar}}, \
  and\ \bibinfo {author} {\bibfnamefont {R.}~\bibnamefont {Blume-Kohout}},\
  }\href {http://link.aps.org/doi/10.1103/PhysRevX.3.041013} {\bibfield
  {journal} {\bibinfo  {journal} {Phys. Rev. X}\ }\textbf {\bibinfo {volume}
  {3}},\ \bibinfo {pages} {041013} (\bibinfo {year}
  {2013}{\natexlab{b}})}\BibitemShut {NoStop}%
\bibitem [{\citenamefont {Sarovar}\ and\ \citenamefont
  {Young}(2013)}]{Sarovar:2013kx}%
  \BibitemOpen
  \bibfield  {author} {\bibinfo {author} {\bibfnamefont {M.}~\bibnamefont
  {Sarovar}}\ and\ \bibinfo {author} {\bibfnamefont {K.~C.}\ \bibnamefont
  {Young}},\ }\href {http://stacks.iop.org/1367-2630/15/i=12/a=125032}
  {\bibfield  {journal} {\bibinfo  {journal} {New J. of Phys.}\ }\textbf
  {\bibinfo {volume} {15}},\ \bibinfo {pages} {125032} (\bibinfo {year}
  {2013})}\BibitemShut {NoStop}%
\bibitem [{\citenamefont {Marvian}\ and\ \citenamefont
  {Lidar}(2014)}]{Marvian:2014nr}%
  \BibitemOpen
  \bibfield  {author} {\bibinfo {author} {\bibfnamefont {I.}~\bibnamefont
  {Marvian}}\ and\ \bibinfo {author} {\bibfnamefont {D.~A.}\ \bibnamefont
  {Lidar}},\ }\href {http://link.aps.org/doi/10.1103/PhysRevLett.113.260504}
  {\bibfield  {journal} {\bibinfo  {journal} {Phys. Rev. Lett.}\ }\textbf
  {\bibinfo {volume} {113}},\ \bibinfo {pages} {260504} (\bibinfo {year}
  {2014})}\BibitemShut {NoStop}%
\bibitem [{\citenamefont {Aliferis}\ \emph {et~al.}(2006)\citenamefont
  {Aliferis}, \citenamefont {Gottesman},\ and\ \citenamefont
  {Preskill}}]{Aliferis:05}%
  \BibitemOpen
  \bibfield  {author} {\bibinfo {author} {\bibfnamefont {P.}~\bibnamefont
  {Aliferis}}, \bibinfo {author} {\bibfnamefont {D.}~\bibnamefont {Gottesman}},
  \ and\ \bibinfo {author} {\bibfnamefont {J.}~\bibnamefont {Preskill}},\
  }\href {http://www.rintonpress.com/xqic6/qic-6-2/097-165.pdf} {\bibfield
  {journal} {\bibinfo  {journal} {Quantum Inf. Comput.}\ }\textbf {\bibinfo
  {volume} {6}},\ \bibinfo {pages} {97} (\bibinfo {year} {2006})}\BibitemShut
  {NoStop}%
\bibitem [{\citenamefont {Knill}(2005)}]{Knill:05}%
  \BibitemOpen
  \bibfield  {author} {\bibinfo {author} {\bibfnamefont {E.}~\bibnamefont
  {Knill}},\ }\href {http://dx.doi.org/10.1038/nature03350} {\bibfield
  {journal} {\bibinfo  {journal} {Nature}\ }\textbf {\bibinfo {volume} {434}},\
  \bibinfo {pages} {39} (\bibinfo {year} {2005})}\BibitemShut {NoStop}%
\bibitem [{\citenamefont {Mizel}(2014)}]{Mizel:2014sp}%
  \BibitemOpen
  \bibfield  {author} {\bibinfo {author} {\bibfnamefont {A.}~\bibnamefont
  {Mizel}},\ }\href {http://arXiv.org/abs/1403.7694} {\bibfield  {journal}
  {\bibinfo  {journal} {arXiv:1403.7694}\ } (\bibinfo {year}
  {2014})}\BibitemShut {NoStop}%
\bibitem [{\citenamefont {Johnson}\ \emph {et~al.}(2010)\citenamefont
  {Johnson}, \citenamefont {Bunyk}, \citenamefont {Maibaum}, \citenamefont
  {Tolkacheva}, \citenamefont {Berkley}, \citenamefont {Chapple}, \citenamefont
  {Harris}, \citenamefont {Johansson}, \citenamefont {Lanting}, \citenamefont
  {Perminov}, \citenamefont {Ladizinsky}, \citenamefont {Oh},\ and\
  \citenamefont {Rose}}]{Johnson:2010ys}%
  \BibitemOpen
  \bibfield  {author} {\bibinfo {author} {\bibfnamefont {M.~W.}\ \bibnamefont
  {Johnson}}, \bibinfo {author} {\bibfnamefont {P.}~\bibnamefont {Bunyk}},
  \bibinfo {author} {\bibfnamefont {F.}~\bibnamefont {Maibaum}}, \bibinfo
  {author} {\bibfnamefont {E.}~\bibnamefont {Tolkacheva}}, \bibinfo {author}
  {\bibfnamefont {A.~J.}\ \bibnamefont {Berkley}}, \bibinfo {author}
  {\bibfnamefont {E.~M.}\ \bibnamefont {Chapple}}, \bibinfo {author}
  {\bibfnamefont {R.}~\bibnamefont {Harris}}, \bibinfo {author} {\bibfnamefont
  {J.}~\bibnamefont {Johansson}}, \bibinfo {author} {\bibfnamefont
  {T.}~\bibnamefont {Lanting}}, \bibinfo {author} {\bibfnamefont
  {I.}~\bibnamefont {Perminov}}, \bibinfo {author} {\bibfnamefont
  {E.}~\bibnamefont {Ladizinsky}}, \bibinfo {author} {\bibfnamefont
  {T.}~\bibnamefont {Oh}}, \ and\ \bibinfo {author} {\bibfnamefont
  {G.}~\bibnamefont {Rose}},\ }\href
  {http://stacks.iop.org/0953-2048/23/i=6/a=065004} {\bibfield  {journal}
  {\bibinfo  {journal} {Superconductor Science and Technology}\ }\textbf
  {\bibinfo {volume} {23}},\ \bibinfo {pages} {065004} (\bibinfo {year}
  {2010})}\BibitemShut {NoStop}%
\bibitem [{\citenamefont {Berkley}\ \emph {et~al.}(2010)\citenamefont
  {Berkley}, \citenamefont {Johnson}, \citenamefont {Bunyk}, \citenamefont
  {Harris}, \citenamefont {Johansson}, \citenamefont {Lanting}, \citenamefont
  {Ladizinsky}, \citenamefont {Tolkacheva}, \citenamefont {Amin},\ and\
  \citenamefont {Rose}}]{Berkley:2010zr}%
  \BibitemOpen
  \bibfield  {author} {\bibinfo {author} {\bibfnamefont {A.~J.}\ \bibnamefont
  {Berkley}}, \bibinfo {author} {\bibfnamefont {M.~W.}\ \bibnamefont
  {Johnson}}, \bibinfo {author} {\bibfnamefont {P.}~\bibnamefont {Bunyk}},
  \bibinfo {author} {\bibfnamefont {R.}~\bibnamefont {Harris}}, \bibinfo
  {author} {\bibfnamefont {J.}~\bibnamefont {Johansson}}, \bibinfo {author}
  {\bibfnamefont {T.}~\bibnamefont {Lanting}}, \bibinfo {author} {\bibfnamefont
  {E.}~\bibnamefont {Ladizinsky}}, \bibinfo {author} {\bibfnamefont
  {E.}~\bibnamefont {Tolkacheva}}, \bibinfo {author} {\bibfnamefont {M.~H.~S.}\
  \bibnamefont {Amin}}, \ and\ \bibinfo {author} {\bibfnamefont
  {G.}~\bibnamefont {Rose}},\ }\href
  {http://stacks.iop.org/0953-2048/23/i=10/a=105014} {\bibfield  {journal}
  {\bibinfo  {journal} {Superconductor Science and Technology}\ }\textbf
  {\bibinfo {volume} {23}},\ \bibinfo {pages} {105014} (\bibinfo {year}
  {2010})}\BibitemShut {NoStop}%
\bibitem [{\citenamefont {Harris}\ \emph {et~al.}(2010)\citenamefont {Harris},
  \citenamefont {Johnson}, \citenamefont {Lanting}, \citenamefont {Berkley},
  \citenamefont {Johansson}, \citenamefont {Bunyk}, \citenamefont {Tolkacheva},
  \citenamefont {Ladizinsky}, \citenamefont {Ladizinsky}, \citenamefont {Oh},
  \citenamefont {Cioata}, \citenamefont {Perminov}, \citenamefont {Spear},
  \citenamefont {Enderud}, \citenamefont {Rich}, \citenamefont {Uchaikin},
  \citenamefont {Thom}, \citenamefont {Chapple}, \citenamefont {Wang},
  \citenamefont {Wilson}, \citenamefont {Amin}, \citenamefont {Dickson},
  \citenamefont {Karimi}, \citenamefont {Macready}, \citenamefont {Truncik},\
  and\ \citenamefont {Rose}}]{Harris:2010kx}%
  \BibitemOpen
  \bibfield  {author} {\bibinfo {author} {\bibfnamefont {R.}~\bibnamefont
  {Harris}}, \bibinfo {author} {\bibfnamefont {M.~W.}\ \bibnamefont {Johnson}},
  \bibinfo {author} {\bibfnamefont {T.}~\bibnamefont {Lanting}}, \bibinfo
  {author} {\bibfnamefont {A.~J.}\ \bibnamefont {Berkley}}, \bibinfo {author}
  {\bibfnamefont {J.}~\bibnamefont {Johansson}}, \bibinfo {author}
  {\bibfnamefont {P.}~\bibnamefont {Bunyk}}, \bibinfo {author} {\bibfnamefont
  {E.}~\bibnamefont {Tolkacheva}}, \bibinfo {author} {\bibfnamefont
  {E.}~\bibnamefont {Ladizinsky}}, \bibinfo {author} {\bibfnamefont
  {N.}~\bibnamefont {Ladizinsky}}, \bibinfo {author} {\bibfnamefont
  {T.}~\bibnamefont {Oh}}, \bibinfo {author} {\bibfnamefont {F.}~\bibnamefont
  {Cioata}}, \bibinfo {author} {\bibfnamefont {I.}~\bibnamefont {Perminov}},
  \bibinfo {author} {\bibfnamefont {P.}~\bibnamefont {Spear}}, \bibinfo
  {author} {\bibfnamefont {C.}~\bibnamefont {Enderud}}, \bibinfo {author}
  {\bibfnamefont {C.}~\bibnamefont {Rich}}, \bibinfo {author} {\bibfnamefont
  {S.}~\bibnamefont {Uchaikin}}, \bibinfo {author} {\bibfnamefont {M.~C.}\
  \bibnamefont {Thom}}, \bibinfo {author} {\bibfnamefont {E.~M.}\ \bibnamefont
  {Chapple}}, \bibinfo {author} {\bibfnamefont {J.}~\bibnamefont {Wang}},
  \bibinfo {author} {\bibfnamefont {B.}~\bibnamefont {Wilson}}, \bibinfo
  {author} {\bibfnamefont {M.~H.~S.}\ \bibnamefont {Amin}}, \bibinfo {author}
  {\bibfnamefont {N.}~\bibnamefont {Dickson}}, \bibinfo {author} {\bibfnamefont
  {K.}~\bibnamefont {Karimi}}, \bibinfo {author} {\bibfnamefont
  {B.}~\bibnamefont {Macready}}, \bibinfo {author} {\bibfnamefont {C.~J.~S.}\
  \bibnamefont {Truncik}}, \ and\ \bibinfo {author} {\bibfnamefont
  {G.}~\bibnamefont {Rose}},\ }\href {\doibase 10.1103/PhysRevB.82.024511}
  {\bibfield  {journal} {\bibinfo  {journal} {Phys. Rev. B}\ }\textbf {\bibinfo
  {volume} {82}},\ \bibinfo {pages} {024511} (\bibinfo {year}
  {2010})}\BibitemShut {NoStop}%
\bibitem [{\citenamefont {Pudenz}\ \emph {et~al.}(2014)\citenamefont {Pudenz},
  \citenamefont {Albash},\ and\ \citenamefont {Lidar}}]{PAL:13}%
  \BibitemOpen
  \bibfield  {author} {\bibinfo {author} {\bibfnamefont {K.~L.}\ \bibnamefont
  {Pudenz}}, \bibinfo {author} {\bibfnamefont {T.}~\bibnamefont {Albash}}, \
  and\ \bibinfo {author} {\bibfnamefont {D.~A.}\ \bibnamefont {Lidar}},\ }\href
  {\doibase 10.1038/ncomms4243} {\bibfield  {journal} {\bibinfo  {journal}
  {Nat. Commun.}\ }\textbf {\bibinfo {volume} {5}},\ \bibinfo {pages} {3243}
  (\bibinfo {year} {2014})}\BibitemShut {NoStop}%
\bibitem [{\citenamefont {Pudenz}\ \emph {et~al.}(2015)\citenamefont {Pudenz},
  \citenamefont {Albash},\ and\ \citenamefont {Lidar}}]{PAL:14}%
  \BibitemOpen
  \bibfield  {author} {\bibinfo {author} {\bibfnamefont {K.~L.}\ \bibnamefont
  {Pudenz}}, \bibinfo {author} {\bibfnamefont {T.}~\bibnamefont {Albash}}, \
  and\ \bibinfo {author} {\bibfnamefont {D.~A.}\ \bibnamefont {Lidar}},\ }\href
  {http://link.aps.org/doi/10.1103/PhysRevA.91.042302} {\bibfield  {journal}
  {\bibinfo  {journal} {{Phys. Rev. A}}\ }\textbf {\bibinfo {volume} {91}},\
  \bibinfo {pages} {042302} (\bibinfo {year} {2015})}\BibitemShut {NoStop}%
\bibitem [{\citenamefont {Vinci}\ \emph {et~al.}(2015)\citenamefont {Vinci},
  \citenamefont {Albash}, \citenamefont {Paz-Silva}, \citenamefont {Hen},\ and\
  \citenamefont {Lidar}}]{Vinci:2015jt}%
  \BibitemOpen
  \bibfield  {author} {\bibinfo {author} {\bibfnamefont {W.}~\bibnamefont
  {Vinci}}, \bibinfo {author} {\bibfnamefont {T.}~\bibnamefont {Albash}},
  \bibinfo {author} {\bibfnamefont {G.}~\bibnamefont {Paz-Silva}}, \bibinfo
  {author} {\bibfnamefont {I.}~\bibnamefont {Hen}}, \ and\ \bibinfo {author}
  {\bibfnamefont {D.~A.}\ \bibnamefont {Lidar}},\ }\href
  {http://link.aps.org/doi/10.1103/PhysRevA.92.042310} {\bibfield  {journal}
  {\bibinfo  {journal} {{Phys. Rev. A}}\ }\textbf {\bibinfo {volume} {92}},\
  \bibinfo {pages} {042310} (\bibinfo {year} {2015})}\BibitemShut {NoStop}%
\bibitem [{\citenamefont {Lucas}(2014)}]{2013arXiv1302.5843L}%
  \BibitemOpen
  \bibfield  {author} {\bibinfo {author} {\bibfnamefont {A.}~\bibnamefont
  {Lucas}},\ }\href {\doibase 10.3389/fphy.2014.00005} {\bibfield  {journal}
  {\bibinfo  {journal} {Front. Phys.}\ }\textbf {\bibinfo {volume} {2}},\
  \bibinfo {pages} {5} (\bibinfo {year} {2014})}\BibitemShut {NoStop}%
\bibitem [{\citenamefont {R{\o}nnow}\ \emph {et~al.}(2014)\citenamefont
  {R{\o}nnow}, \citenamefont {Wang}, \citenamefont {Job}, \citenamefont
  {Boixo}, \citenamefont {Isakov}, \citenamefont {Wecker}, \citenamefont
  {Martinis}, \citenamefont {Lidar},\ and\ \citenamefont {Troyer}}]{speedup}%
  \BibitemOpen
  \bibfield  {author} {\bibinfo {author} {\bibfnamefont {T.~F.}\ \bibnamefont
  {R{\o}nnow}}, \bibinfo {author} {\bibfnamefont {Z.}~\bibnamefont {Wang}},
  \bibinfo {author} {\bibfnamefont {J.}~\bibnamefont {Job}}, \bibinfo {author}
  {\bibfnamefont {S.}~\bibnamefont {Boixo}}, \bibinfo {author} {\bibfnamefont
  {S.~V.}\ \bibnamefont {Isakov}}, \bibinfo {author} {\bibfnamefont
  {D.}~\bibnamefont {Wecker}}, \bibinfo {author} {\bibfnamefont {J.~M.}\
  \bibnamefont {Martinis}}, \bibinfo {author} {\bibfnamefont {D.~A.}\
  \bibnamefont {Lidar}}, \ and\ \bibinfo {author} {\bibfnamefont
  {M.}~\bibnamefont {Troyer}},\ }\href {\doibase 10.1126/science.1252319}
  {\bibfield  {journal} {\bibinfo  {journal} {Science}\ }\textbf {\bibinfo
  {volume} {345}},\ \bibinfo {pages} {420} (\bibinfo {year}
  {2014})}\BibitemShut {NoStop}%
\bibitem [{\citenamefont {Matsuura}\ \emph {et~al.}(2015)\citenamefont
  {Matsuura}, \citenamefont {Nishimori}, \citenamefont {Albash},\ and\
  \citenamefont {Lidar}}]{matsuura_mean_2015}%
  \BibitemOpen
  \bibfield  {author} {\bibinfo {author} {\bibfnamefont {S.}~\bibnamefont
  {Matsuura}}, \bibinfo {author} {\bibfnamefont {H.}~\bibnamefont {Nishimori}},
  \bibinfo {author} {\bibfnamefont {T.}~\bibnamefont {Albash}}, \ and\ \bibinfo
  {author} {\bibfnamefont {D.~A.}\ \bibnamefont {Lidar}},\ }\href
  {http://arxiv.org/abs/1510.07709} {\bibfield  {journal} {\bibinfo  {journal}
  {arXiv:1510.07709 [quant-ph]}\ } (\bibinfo {year} {2015})}\BibitemShut {NoStop}%
\bibitem [{\citenamefont {Albash}\ \emph {et~al.}(2012)\citenamefont {Albash},
  \citenamefont {Boixo}, \citenamefont {Lidar},\ and\ \citenamefont
  {Zanardi}}]{ABLZ:12-SI}%
  \BibitemOpen
  \bibfield  {author} {\bibinfo {author} {\bibfnamefont {T.}~\bibnamefont
  {Albash}}, \bibinfo {author} {\bibfnamefont {S.}~\bibnamefont {Boixo}},
  \bibinfo {author} {\bibfnamefont {D.~A.}\ \bibnamefont {Lidar}}, \ and\
  \bibinfo {author} {\bibfnamefont {P.}~\bibnamefont {Zanardi}},\ }\href
  {\doibase 10.1088/1367-2630/14/12/123016} {\bibfield  {journal} {\bibinfo
  {journal} {New J. of Phys.}\ }\textbf {\bibinfo {volume} {14}},\ \bibinfo
  {pages} {123016} (\bibinfo {year} {2012})}\BibitemShut {NoStop}%
\bibitem [{\citenamefont {{Reed}}\ \emph {et~al.}(2012)\citenamefont {{Reed}},
  \citenamefont {{Dicarlo}}, \citenamefont {{Nigg}}, \citenamefont {{Sun}},
  \citenamefont {{Frunzio}}, \citenamefont {{Girvin}},\ and\ \citenamefont
  {{Schoelkopf}}}]{2012Natur.482..382R}%
  \BibitemOpen
  \bibfield  {author} {\bibinfo {author} {\bibfnamefont {M.~D.}\ \bibnamefont
  {{Reed}}}, \bibinfo {author} {\bibfnamefont {L.}~\bibnamefont {{Dicarlo}}},
  \bibinfo {author} {\bibfnamefont {S.~E.}\ \bibnamefont {{Nigg}}}, \bibinfo
  {author} {\bibfnamefont {L.}~\bibnamefont {{Sun}}}, \bibinfo {author}
  {\bibfnamefont {L.}~\bibnamefont {{Frunzio}}}, \bibinfo {author}
  {\bibfnamefont {S.~M.}\ \bibnamefont {{Girvin}}}, \ and\ \bibinfo {author}
  {\bibfnamefont {R.~J.}\ \bibnamefont {{Schoelkopf}}},\ }\href {\doibase
  10.1038/nature10786} {\bibfield  {journal} {\bibinfo  {journal} {Nature}\
  }\textbf {\bibinfo {volume} {482}},\ \bibinfo {pages} {382} (\bibinfo {year}
  {2012})}\BibitemShut {NoStop}%
\bibitem [{\citenamefont {Amin}\ \emph {et~al.}(2008)\citenamefont {Amin},
  \citenamefont {Love},\ and\ \citenamefont {Truncik}}]{TAQC}%
  \BibitemOpen
  \bibfield  {author} {\bibinfo {author} {\bibfnamefont {M.~H.~S.}\
  \bibnamefont {Amin}}, \bibinfo {author} {\bibfnamefont {P.~J.}\ \bibnamefont
  {Love}}, \ and\ \bibinfo {author} {\bibfnamefont {C.~J.~S.}\ \bibnamefont
  {Truncik}},\ }\href {\doibase 10.1103/PhysRevLett.100.060503} {\bibfield
  {journal} {\bibinfo  {journal} {Phys. Rev. Lett.}\ }\textbf {\bibinfo
  {volume} {100}},\ \bibinfo {pages} {060503} (\bibinfo {year}
  {2008})}\BibitemShut {NoStop}%
\end{thebibliography}
\end{document}